\documentclass[11pt,a4paper]{article}

\usepackage{epsfig}
\usepackage{amsmath}
\usepackage{amssymb}
\usepackage[footnotesize]{caption}
\usepackage{subfigure}
\usepackage{cancel}
\usepackage{cite}
\usepackage{textcomp}
\usepackage{calc}
\usepackage{geometry}
\usepackage{graphicx}
\usepackage{dcolumn}
\usepackage{hyperref}
\usepackage{color}
\usepackage{bbm}
\usepackage{fourier-orns}

\newcommand{\drawsquare}[2]{\hbox{%
\rule{#2pt}{#1pt}\hskip-#2pt
\rule{#1pt}{#2pt}\hskip-#1pt
\rule[#1pt]{#1pt}{#2pt}}\rule[#1pt]{#2pt}{#2pt}\hskip-#2pt
\rule{#2pt}{#1pt}}

\newcommand{\Yfund}{\raisebox{-.5pt}{\drawsquare{6.5}{0.4}}}

\begin{document}


\noindent July 2014
\hfill\begin{tabular}{ll}
IPMU & \hspace{-0.3cm}14-0156 \\
ICRR & \hspace{-0.3cm}687-2014-14
\end{tabular}

\vskip 2.25cm

\begin{center}
{\LARGE\bf Dynamical Fractional Chaotic Inflation}

\smallskip

{\small\bf \decothreeleft\:\:Dynamical Generation of a Fractional Power-Law Potential for
Chaotic Inflation\:\:\decothreeright}

\vskip 1.5cm

{\large Keisuke~Harigaya$^a$, Masahiro~Ibe$^{a,b}$,
Kai~Schmitz$^a$, Tsutomu~T.~Yanagida$^a$}\\[3mm]
{\it{
a Kavli IPMU (WPI), TODIAS, University of Tokyo, Kashiwa, 277-8583, Japan\\
b ICRR, University of Tokyo, Kashiwa, 277-8582, Japan}}
\end{center}

\vskip 1cm


\begin{abstract}

\noindent Chaotic inflation based on a simple monomial scalar potential,
$V(\phi)\propto\phi^p$, is an attractive large-field model of 
inflation capable of generating a sizable tensor-to-scalar ratio $r$.
Therefore, assuming that future CMB observations will confirm the large $r$ value
reported by BICEP2, it is important to determine what kind of dynamical mechanism
could possibly endow the inflaton field with such a simple effective potential.
In this paper, we answer this question in the context of field theory,
i.e.\ in the framework of dynamical chaotic inflation (DCI), where
strongly interacting supersymmetric gauge dynamics around the scale
of grand unification dynamically generate a fractional power-law potential
via the quantum effect of dimensional transmutation.
In constructing explicit models, we significantly extend our previous
work, as we now consider a large variety of possible underlying gauge dynamics
and relax our conditions on the field content of the model.
This allows us to realize almost arbitrary rational values for the power $p$
in the inflaton potential.
The present paper may hence be regarded as a first step towards a more complete
theory of dynamical chaotic inflation.
\end{abstract}

\newpage


\tableofcontents
\newpage


\section{Introduction}


Recently, the BICEP2 collaboration has reported on the observation
of a B-mode polarization signal in the cosmic microwave background (CMB)
radiation~\cite{Ade:2014xna} which might indicate the presence of primordial tensor
perturbations generated during the inflationary phase in the very early universe.
Neglecting any possible foreground contributions due to polarized dust
in our own galaxy, the BICEP2 data is well fit by the standard cosmological
model including a remarkably large tensor-to-scalar ratio,
$r=0.20^{+0.07}_{-0.05}$ (at $68\,\%$\,CL)~\cite{Ade:2014xna}.
Once the effect of polarized dust on the B-mode signal is taken into account,
the corresponding best-fit value for $r$ decreases.
At present, no conclusive dust polarization data for the relevant
patch in the sky is available, which is why it remains to be seen
whether $r$ is indeed of $\mathcal{O}(0.1)$ or in fact much smaller.
In this paper, we shall take the attitude that the large tensor-to-scalar ratio
reported by the BICEP2 collaboration will in fact survive further scrutiny.
The natural question which then immediately arises is which inflationary
dynamics may be responsible for an $r$ value as large as $r\simeq 0.2$.
A particularly attractive scenario in this context is chaotic
inflation~\cite{Linde:1983gd}, which can begin at
Planck-scale energies without running into problems related to
the initial conditions at the onset of inflation and which,
at the same time, easily yields tensor perturbations with large amplitude.
A further virtue of chaotic inflation is the simple form of the
potential $V$ for the inflaton field $\phi$.
In order to realize chaotic inflation, one merely has to assume
that, effectively, the scalar potential is nothing but a monomial,
i.e.\ simply proportional to a single power $\phi^p$ of the inflaton field $\phi$,
\begin{align}
V(\phi) \propto M^4 \left(\frac{\left|\phi\right|}{M}\right)^p \,,
\quad \mathbb{Q}\ni p > 0 \,, \label{eq:powerlawV}
\end{align}
for some appropriate mass scale $M$ and some power $p$, which does not
necessarily need to be an integer, but which can very well be a rational number.
Assuming that the true value of $r$ lies somewhere around $r\simeq0.1..0.2$,
powers between $p\simeq 1$ and $p\simeq 2$ look particularly promising at the moment.
But in the end, it is still too early to draw any definite conclusions
and we just have to wait for additional experimental data to 
shed more light on the range of viable $p$ values.


Given the preeminent role of chaotic inflation within the class of models
predicting large~$r$, it is important to address the question
as to the dynamical origin of the inflaton potential in chaotic inflation.
Why would the effective potential driving inflation just be a simple
monomial?
In string theory, the mechanism of axion monodromy~\cite{Silverstein:2008sg}
has proven to be a successful means to realize large inflaton field excursions
without spoiling the flatness of the potential.
Different variants of this mechanism result in effective
potentials such as the one in Eq.~\eqref{eq:powerlawV} with
a fractional power $p=2/3$ or $p=2/5$~\cite{Silverstein:2008sg},
in a linear potential~\cite{McAllister:2008hb},
quadratic potential~\cite{Kaloper:2011jz} or
a power series in $\phi$ with a leading quadratic term~\cite{Marchesano:2014mla}.
At the same time, it has been shown that chaotic inflation can also be
consistently embedded into supergravity (SUGRA).
The first such embedding was accomplished in Ref.~\cite{Kawasaki:2000yn},
which presented a model of chaotic inflation featuring a quadratic potential.
Subsequently, the results of Ref.~\cite{Kawasaki:2000yn} were substantially
generalized in Refs.~\cite{Kallosh:2010ug,Kallosh:2010xz,Kallosh:2011qk},
where a general prescription was given for how to construct models
of chaotic inflation based on an almost arbitrary potential.%
\footnote{For a recent review of inflation in general and
chaotic inflation in supergravity in particular,
cf.\ Ref.~\cite{Linde:2014nna}.\smallskip}
Meanwhile, it was demonstrated in Ref.~\cite{Takahashi:2010ky}
that a fractional power-law potential may also result
from a running kinetic term for the inflaton.
In contrast to the stringy constructions relying on the idea
of axion monodromy, all these field-theoretic models however
lack a dynamical origin for the inflaton potential. 
Virtually all SUGRA models of chaotic inflation make \textit{ad hoc}
choices for the super- as well as the K\"ahler potential, which may or may
not be well motivated by more fundamental principles, and then merely
deduce the scalar potential from these input functions.
Indeed, concepts like superconformal symmetry~\cite{Kallosh:2010ug}
and no-scale supergravity~\cite{Ellis:2014rxa} are sensible
guiding principles in the construction of SUGRA models of
inflation.
But still, it would be desirable to have a dynamical mechanism
at hand that allows one to dynamically generate
the potential of chaotic inflation within field theory.


A first and promising approach in this respect seems to be the
class of models originally proposed in Ref.\,\cite{Harigaya:2012pg}, in which
the inflaton field couples to a strongly interacting supersymmetric
gauge theory in such  way that it dynamically acquires a fractional
power-law potential via the effect of dimensional transmutation.
As these models provide a \textit{dynamical} explanation for the origin of the
inflaton potential in chaotic inflation, we collectively dubbed them
``dynamical chaotic inflation'' (DCI).%
\footnote{The authors of Ref.~\cite{Harigaya:2014eta}
and Ref.~\cite{Yonekura:2014oja} recently pointed out how the inflaton potential of
monodromy inflation may be dynamically generated in supersymmetric QCD-like theories
as well as in (not necessarily supersymmetric) pure Yang-Mills theories, respectively.
For other models in which the scale of inflation is
generated dynamically, including hybrid and new inflation models,
cf.\ for example Ref.~\cite{Dimopoulos:1997fv}.}
This class of models explains the occurrence of the fractional power $p$
in the inflaton potential and provides a dynamical origin for the energy
scale of inflation.
In its simplest realization, dynamical chaotic inflation 
is conformally invariant at the classical level---it does not
involve any dimensionful input scales,
but rather generates the scale of inflation from quantum effects.
It hence naturally explains why the scale of inflation
lies roughly two orders of magnitude below the Planck scale.
On the other hand, it is clear that also dynamical chaotic inflation
cannot be an ultraviolet (UV)-complete theory of inflation as it requires the introduction
of a softly broken shift symmetry in the direction of the inflaton
field, a property which it shares with many other realizations of chaotic inflation
in supergravity.


In Ref.\,\cite{Harigaya:2012pg}, we introduced the idea of dynamical chaotic
inflation and presented first explicit models based on $SP(N_c)$ gauge dynamics
in combination with $N_f = 2\left(N_c + 2\right)$ flavors.
The largest power $p$ we could obtain in this way was $p=1$,
which allowed us to predict $r$ values as large as $r\simeq 0.08$.
In this paper, we will now (i) review our earlier work, (ii) further develop
the theoretical framework of dynamical chaotic inflation and (iii) give a
comprehensive discussion of its phenomenological implications.
In doing so, we will significantly extend the range of possible gauge dynamics
underlying dynamical chaotic inflation from $SP(N_c)$ theories only
to theories either based on $SP(N_c)$, $SO(10)$, $SU(5)$, $SU(3)\times SU(2)$ or $SU(N_c)$.
At the same time, we will also relax our conditions on the field content of our models.
This will provide us with a wealth of consistent DCI scenarios, allowing us to
approximatively realize any arbitrary power $p$.


The organization of this paper is as follows.
In the next section, Sec.~\ref{sec:idea}, we first outline the general recipe according
to which all DCI models are to be constructed. 
In Secs.~\ref{sec:SPN}, \ref{sec:SO10} and
\ref{sec:32}, we then implement this algorithm, respectively basing our
construction on $SP(N_c)$ vector gauge theories, $SO(10)$ and $SU(5)$ chiral gauge theories
as well as on a chiral $SU(3) \times SU(2)$ model.
Subsequently, we comment on the embedding of dynamical
chaotic inflation into supergravity in Sec.~\ref{sec:SUGRA}
and discuss its phenomenological consequences in Sec.~\ref{sec:phenomenology}.
In the latter section, we derive in particular constraints on the
parameters space of our model, discuss the implications for the preheating
and reheating processes after inflation and summarize the
predictions for the inflationary CMB observables.
Finally, we present our conclusions and give a brief outlook,
cf.\ Sec.~\ref{sec:conclusions}.
In the two appendices, we construct additional DCI models based on
$SU(N_c)$ dynamics, which turn out to come with certain theoretical limitations,
cf.\ App.~\ref{app:SUN}, and illustrate how our general DCI recipe
may be altered, so as to connect dynamical chaotic inflation to dynamical
supersymmetry (SUSY) breaking in the true vacuum after the end of inflation,
cf.\ App.~\ref{app:dynSUSY}.


\section{Dynamical generation of the inflaton potential}
\label{sec:idea}


To illustrate the basic idea of how the inflation potential is
dynamically generated in general DCI models, let us consider
as an example a strongly interacting non-supersymmetric $SU(N_c)$
gauge theory with $N_f$ pairs of ``quarks" and ``antiquarks",
i.e.\ fermions respectively transforming in the fundamental and
antifundamental representation of $SU(N_c)$, similarly as in
quantum chromodynamics (QCD).
Although the precise vacuum structure of such QCD-like theories is not
yet fully understood theoretically, it is generally believed that,
depending on the numbers of colors and flavors, $N_c$ and $N_f$,
they often times reach a confined phase at low energies,
thereby giving rise to a vacuum
energy density of $\mathcal{O}\left(\Lambda^4\right)$.
Here, $\Lambda$ denotes the dynamical scale of the strong interactions,
at which the canonical gauge coupling constant formally diverges.%
\footnote{When explicitly constructing models, we will actually rely on supersymmetric
gauge dynamics, in which the vacuum energy is effectively obtained as a
SUSY-breaking effect.
Therefore, even if the above intuitive discussion of the QCD contribution
to the vacuum energy should not hold in the non-supersymmetric
version of a given theory~\cite{Brodsky:2009zd}, our arguments
pertaining the supersymmetric variant of this theory will still apply.
\label{fn:vacuumenergySUSY}}


Now, let us imagine that a number of $N_d$ quarks are coupled to the inflaton
field $\phi$ via Yukawa interaction terms in the Lagrangian,
\begin{eqnarray}
{\cal L} = \lambda_i \, \phi\,q_L^i\bar{q}_R^i +
\textrm{h.c.} \,,\quad i = 1,2,..,N_d \,, \quad 0 < N_d \leq N_f \,,
\label{eq:phiqqLag}
\end{eqnarray}
where the $\lambda_i$ denote dimensionless coupling constants, which
we assume to be equal here for simplicity,
$\lambda_i \equiv \lambda$.
Next, suppose that the inflaton field has a very large field value,
$\left|\phi\right| \gg \Lambda/\lambda$.
In this situation, the effective masses of the quarks coupled to
the inflaton become much larger than the dynamical scale, so that
the $N_d$ quark flavors in Eq.~\eqref{eq:phiqqLag} decouple perturbatively.
At energy scales $\mu$ below the heavy quark mass threshold, i.e.\ below the
scale $\lambda\left|\phi\right|$, the original gauge theory therefore reduces to a gauge
theory with a smaller number of flavors, $N_f^{\textrm{eff}} = N_f - N_d$.
Above and below the heavy quark mass threshold, the running of the gauge
coupling constant $g$ is therefore determined by two different beta functions,
\begin{align}
\frac{d}{d\ln\mu} \frac{8\pi^2}{g^2(\mu)} =
\begin{cases}
b & ; \:\: \mu \gg \lambda \left|\phi\right| \\
b_{\textrm{eff}} & ; \:\: \mu \ll \lambda \left|\phi\right|
\end{cases} \,, 
\label{eq:gRGE}
\end{align}
with $b$ and $b_{\rm eff}$ respectively denoting the
beta-function coefficients of the gauge coupling constant of the original
as well as of the effective reduced gauge theory.
At the one-loop level, we have
\begin{align}
b = \frac{11}{3}N_c-\frac{2}{3}N_f \,, \quad
b_{\textrm{eff}} = \frac{11}{3}N_c-\frac{2}{3}N_f^{\textrm{eff}} =
\frac{11}{3}N_c-\frac{2}{3}\left(N_f - N_d\right) \,.
\end{align}
Solving the renormalization group equation (RGE) in Eq.~\eqref{eq:gRGE}
in the high-energy as well as in the low-energy regime yields two
expressions for the gauge coupling
constant, $g_{\textrm{HE}}$ and $g_{\textrm{LH}}$,
\begin{align}
\frac{8\pi^2}{g_{\textrm{HE}}^2(\mu)} = b \, \ln\frac{\mu}{\Lambda} \,, \quad
\frac{8\pi^2}{g_{\textrm{LE}}^2(\mu)} = b_{\textrm{eff}} \, \ln\frac{\mu}{\Lambda_{\textrm{eff}}} \,,
\label{eq:gHELE}
\end{align}
which have to match at the heavy quark mass threshold.
Evaluating both equations in Eq.~\eqref{eq:gHELE} at $\mu = \lambda \left|\phi\right|$ and equating
their respective right-hand sides, we hence find the effective, field-dependent scale
$\Lambda_{\textrm{eff}}$ in terms of the fundamental scale $\Lambda$ and the inflaton
field value $\phi$,
\begin{eqnarray}
\Lambda_{\rm eff}\left(\phi\right) \simeq \Lambda \left(\frac{\lambda\left|\phi\right|}
{\Lambda}\right)^{(b_{\rm eff}-b)/b_{\rm eff}} \,.
\label{eq:Lambdaeff}
\end{eqnarray}
Given this effective, field-dependent dynamical scale,
the strong dynamics now lead to an effective, field-dependent vacuum energy
density of $\mathcal{O}\left(\Lambda_{\textrm{eff}}^4\right)$,
\begin{eqnarray}
V\left(\phi\right) \sim \Lambda_{\rm eff}^4\left(\phi\right) \simeq
\Lambda^4 \left(\frac{\lambda\left|\phi\right|}{\Lambda}
\right)^{4(b_{\rm eff}-b)/b_{\rm eff}} \,.
\label{eq:dyn1}
\end{eqnarray}
This is nothing but a monomial scalar potential for the
inflaton field $\phi$ as in Eq.~\eqref{eq:powerlawV} featuring a
fractional power
\begin{eqnarray}
p = \frac{4\,(b_{\rm eff}-b)}{b_{\rm eff}} =
\frac{8(N_f-N_f^{\textrm{eff}})}{11 N_c - 2 N_f^{\textrm{eff}}} =
\frac{8\,N_d}{11 N_c - 2 \left(N_f-N_d\right)} \,.
\label{eq:pbb}
\end{eqnarray}


To sum up, the monomial inflaton potential is generated via dimensional
transmutation in consequence of the coupling between the inflaton field
and a subset of quarks in the strongly interacting sector.
Here, the mass scale $M$ in Eq.~\eqref{eq:powerlawV} is in particular
identified as the dynamical scale $\Lambda$ associated with the strong
dynamics.
Moreover, the power $p$ turns out to be solely determined
by the numbers of colors and flavors of the strong gauge theory,
$N_c$, $N_f$ and $N_f^{\textrm{eff}}$.
For these reasons, we shall refer to the inflationary scenario described
by the dynamically generated potential in Eq.~\eqref{eq:powerlawV} as
``dynamical chaotic inflation'' (DCI).
We expect the scale $\Lambda$ to be parametrically suppressed compared to the
reduced Planck scale, $M_{\textrm{Pl}}\simeq 2.44 \times 10^{18}\,\textrm{GeV}$.
In fact, as we will see in Sec.~\ref{subsec:observables}, the scale $\Lambda$ is
typically required to take a value close or shortly below the scale of grand
unification (GUT), $\Lambda_{\textrm{GUT}} \simeq 2\times 10^{16}\,\textrm{GeV}$.
Dynamical chaotic inflation thus provides an elegant explanation for why inflation
takes place at sub-Planckian energies.
Besides that, it is an exciting prospect that even more precise measurements of
the inflationary CMB observables
would allow to narrow down the range of viable $p$ values and hence
to infer information regarding the gauge dynamics underlying
the epoch of cosmic inflation.


\begin{figure}
\centering
\includegraphics[width=0.54\textwidth]{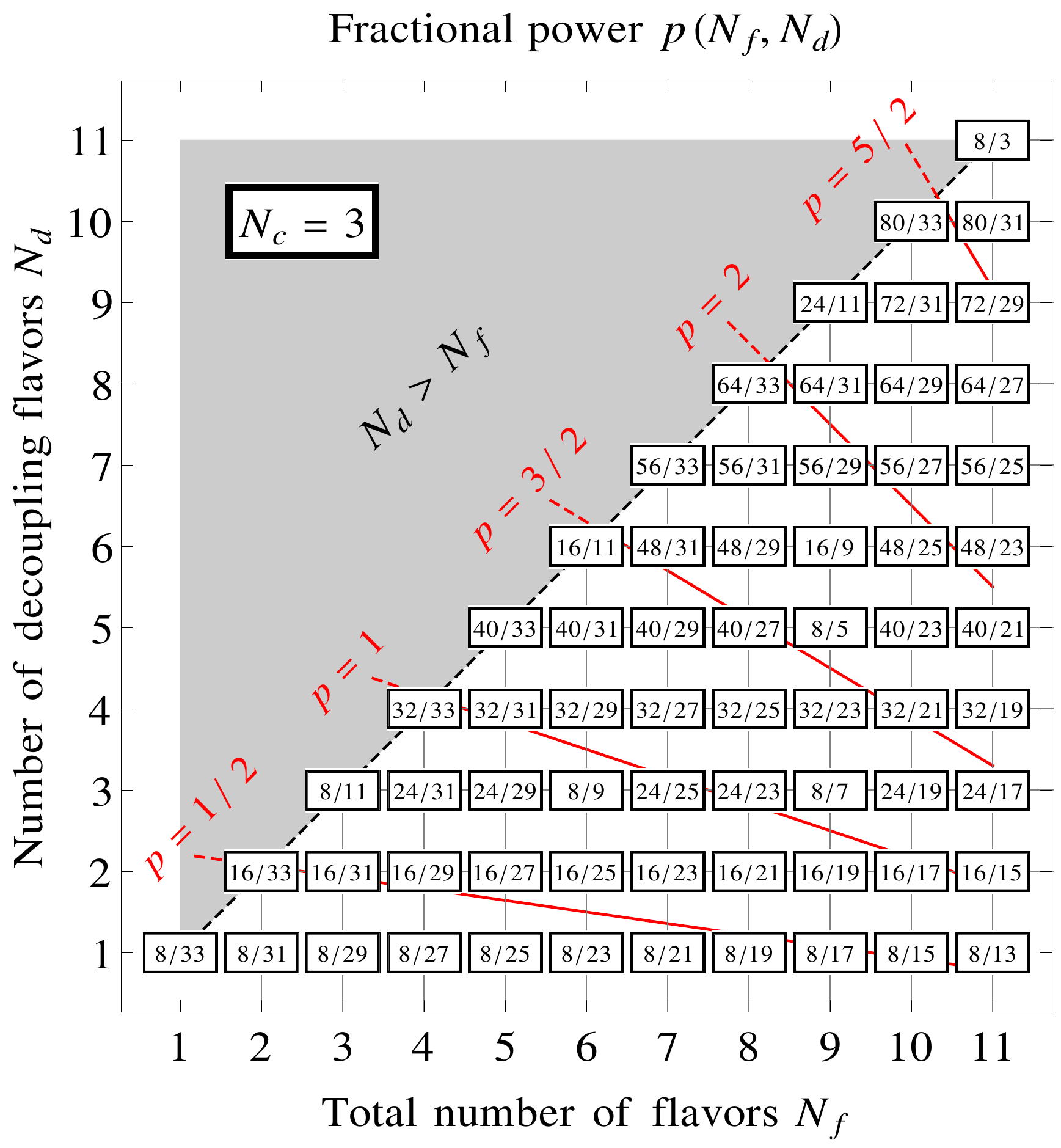}
\caption{Power $p$ in the inflaton potential as function
of $N_f$ and $N_d$ for $N_c = 3$.
Here, the gray-shaded region is excluded by construction, as
$N_d = N_f - N_f^{\textrm{eff}}$ must satisfy $0 < N_d \leq N_f$,
i.e.\ as $N_f^{\textrm{eff}}$ must satisfy $0 \leq N_f^{\textrm{eff}} <N_f$.}
\label{fig:nonSUSY}
\end{figure}


In Fig.\,\ref{fig:nonSUSY}, we show the possible values for the
power in the inflaton potential for $N_c = 3$ as an example.
Here, we restrict ourselves to $N_f \le 11$, for which 
the infrared (IR) behavior of the $SU(3)$ gauge theory is 
expected to be in the confinement phase\,\cite{Appelquist:1996dq}.
The figure demonstrates that, depending on the values chosen for $N_f$
and $N_d$, the power $p$ ranges from very small
values, $p\ll 1$, to values as large as $p \simeq 2.5$.
This illustrates that the DCI mechanism outlined above
allows to construct monomial potentials with a multitude
of different possible $p$ values.


So far, we considered a non-supersymmetric theory
and assumed a QCD-like contribution to the vacuum energy density of
$\mathcal{O}\left(\Lambda_{\textrm{eff}}^4\right)$ at low energies.
As we merely intended to convey our main idea, our reasoning has, however,
been far from quantitative.
In fact, our argumentation up to this point does not rest on a sound
theoretical footing for three reasons.
(i)~We contemporarily lack a precise understanding of the vacuum
structure of non-supersymmetric gauge theories
and hence we are unable to describe the low-energy
behavior of the inflaton sector with certainty.
(ii)~Besides that, since we have introduced the inflaton field as an elementary
scalar field, it is also difficult to suppress its mass term,
which is at least generated radiatively and which might easily spoil
the inflationary dynamics.
(iii)~Last but not least, in our derivation of the inflaton potential in Eq.~\eqref{eq:dyn1},
we relied on the validity of perturbation theory during inflation.
Depending on the number of flavors $N_f$ in relation to the number of colors $N_c$,
this assumption may however not be justified, requiring nonperturbative
techniques to compute the inflaton potential.
Due to those obstructions in the non-supersymmetric case, we shall exclusively
focus on DCI models based on supersymmetric gauge theories in the following.
As we shall see, supersymmetry allows us to remedy or at least alleviate all
of the three previously mentioned shortcomings in a simple manner.
(i)~First of all, it guarantees us good control over the gauge
dynamics in the infrared.
Here, we shall rely in particular on the supersymmetric gauge theory of smooth
confinement or \textit{s-confinement}\,\cite{Seiberg:1994bz,Intriligator:1995ne,Csaki:1996sm},
which is characterized by the fact that it
is not accompanied by chiral symmetry breaking, but, at the same time,
responsible for the nonperturbative
generation of a non-vanishing confining effective superpotential.
In the s-confined phase of a strongly interacting supersymmetric gauge
theory, the degrees of freedom in the infrared correspond to gauge-invariant
composites of the fundamental fields, which are described by a smooth effective
theory spanning over the entire quantum moduli space and in particular including its origin.
A particular virtue of s-confinement is that it allows to easily stabilize
the flat directions of the moduli space through couplings between the fundamental
matter fields and an appropriate number of gauge singlets.
(ii) Moreover, supersymmetry eliminates the radiative corrections to the
inflaton mass and hence stabilizes the inflationary dynamics.
(iii) Finally, in the context of supersymmetry, we no longer need to rely
on perturbation theory when examining the gauge interactions of the strongly
interacting quarks.
Now, we may simply deduce the superpotential for the corresponding chiral superfields
from the running of the holomorphic gauge coupling or, alternatively,
from the requirement of consistency with $R$ symmetry.%
\footnote{This procedure does not apply to the interactions
encoded in the K\"ahler potential.
Also in the supersymmetric case, a complete departure from
perturbation theory is hence not feasible.}
We emphasize that it is for these three virtues of supersymmetry that we turn away
from non-supersymmetric theories and concentrate on supersymmetric DCI models instead.


One important implication of our decision to solely work with supersymmetric
theories is that now, in contrast to the non-supersymmetric case,
the mere confinement due to the strong dynamics no longer automatically leads to
a nonzero vacuum energy density since
the vacuum energy density is protected by supersymmetry.
Therefore, as we intend to identify the inflaton potential as an
effective vacuum energy density, we are led to constructing models in which
supersymmetry is spontaneously broken once the inflaton field takes a large
field value.
All in all, the basic recipe to cook up a working DCI
model hence consists of two ingredients:
\begin{enumerate}
\item Construct a supersymmetric s-confining gauge theory which 
can be mutated into a model of dynamical supersymmetry breaking (DSB)
by means of appropriate mass deformations.
\item Couple the inflaton to some of the matter fields to
give them an effective mass, such that for nonzero inflaton field values
the s-confining theory transforms into a DSB model, i.e.\ identify
the mass deformations in step 1 as the turning-on of the
inflaton field value.
\end{enumerate}
In any successful implementation of this mechanism,
a field other than the inflaton field is bound to obtain
a non-vanishing $F$-term during inflation.
This is an important detail ensuring that we obtain stable inflationary dynamics.
If the inflaton itself was responsible for spontaneous supersymmetry breaking,
i.e.\ if it acquired a nonzero $F$-term itself during inflation, its potential
would receive a large negative contribution from SUGRA corrections at super-Planckian
inflaton field values, which would ultimately ruin chaotic inflation~\cite{Kawasaki:2000yn}.


In the subsequent three sections, we now follow the above outlined procedure and
construct DCI models based on $SP(N_c)$, $SO(10)$, $SU(5)$ and
$SU(3)\times SU(2)$ gauge dynamics, respectively.


\section{Models based on \texorpdfstring{\boldmath{$SP(N_{\lowercase{c}})$}}{SP(Nc)} dynamics}
\label{sec:SPN}


\subsection[Massless matter fields only (fractional powers \texorpdfstring{$0 < p\leq1$}{0 < p <= 1})]
{Massless matter fields only (fractional powers \texorpdfstring{\boldmath{$0 < p\leq1$}}{0 < p <= 1})}
\label{subsec:SPNcmassless}


We begin by reviewing our earlier work presented in Ref.~\cite{Harigaya:2012pg}.
Let us consider an $SP(N_c)$ gauge theory%
\footnote{We choose a convention such that $SP(1)$ is equivalent to $SU(2)$.}
with $2N_f = 2(N_c+2)$ vector-like quark fields $Q^I$, where $I=1,2,..,2N_f$,
transforming in the fundamental representation of $SP(N_c)$.
In the infrared, below the dynamical scale $\Lambda$,
this theory exhibits an s-confinement phase, which is well described
by a smooth effective theory in terms of $(N_c+2)(2N_c+3)$
gauge-invariant meson fields, $M^{IJ} = Q^I Q^J / \Lambda = - M^{JI}$.
This model can be mutated into the so-called IYIT model of dynamical
supersymmetry breaking~\cite{Izawa:1996pk} by making one pair of
quarks very heavy, while at the same time stabilizing all
other matter fields through couplings to an appropriate number
of singlet fields in the superpotential.
To do so, we introduce a corresponding singlet field $Z_{IJ}$
for each quark bilinear $M^{IJ}$, where we identify w.l.o.g.\
the field $Z_{2N_c+3,2N_c+4}$
as the chiral inflaton superfield $\Phi$, which contains the complex
scalar inflaton field $\phi$.
We can then write down the following tree-level superpotential
\begin{align}
W_{\textrm{tree}} = \frac{1}{2} \lambda_{IJ} Z_{IJ} Q^I Q^J = W_{\textrm{IYIT}} + \Delta W \,,
\label{eq:tree1}
\end{align}
where the $\lambda_{IJ} = \lambda_{JI}$ denote real dimensionless coupling constants.
This superpotential may be decomposed into the superpotential of the IYIT
model
\begin{align}
W_{\textrm{IYIT}} = \frac{1}{2}\lambda_{ij} Z_{ij}Q^iQ^j \,,
\quad i,j = 1,2,..,2(N_c+1) \,.
\label{eq:SPN1}
\end{align}
as well as into an additional piece involving the $\Phi$-dependent
mass term of the heavy quark pair,
\begin{align}
\Delta W = \lambda_{ik}Z_{ik}Q^i Q^k + \lambda \,\Phi\, Q^{2N_c+3}Q^{2N_c+4}
\,, \quad i = 1,2,..,2(N_c+1) \,, \:\: k = 2N_c+3,2N_c+4 \,.
\label{eq:DeltaW}
\end{align}


Notice that a large inflaton field value provides exactly the
mass deformation of our s-confining theory that is necessary to
transform it into the IYIT model of dynamical supersymmetry breaking.
The above construction therefore represents a direct implementation
of the general DCI mechanism described in Sec.~\ref{sec:idea}.
We point out that a similar approach is also conceivable
for strong gauge dynamics based on the group $SU(N_c)$ in combination with
$N_f = N_c + 1$ quark flavors.
However, in this case, the massless spectrum in the s-confined phase
also includes baryon and antibaryon fields,
$B \propto \epsilon_{i_1 i_2 .. i_{N_f}} Q^{i_1} Q^{i_2} .. Q^{i_{N_c}}$
and $\bar{B} \propto \epsilon_{\bar{\imath}_1 \bar{\imath}_2 ..
\bar{\imath}_{N_f}} \bar{Q}^{\bar{\imath}_1} \bar{Q}^{\bar{\imath}_2} ..
\bar{Q}^{\bar{\imath}_{N_c}}$, which cannot be as easily stabilized
as the meson fields $M^{IJ}$ in our $SP(N_c)$ theory.
As for the $SP(N_c)$ case, it is enough to introduce ordinary Yukawa couplings
between the quark fields and a sufficiently large number of singlet fields,
cf.\ Eq.~\eqref{eq:tree1}.
The stabilization of the baryon and antibaryon fields requires, by contrast,
the introduction of higher-dimensional operators, which complicates the analysis
of the entire model.
It is for this reason that we focus on $SP(N_c)$ gauge theories for now.
DCI models based on $SU(N_c)$ dynamics are discussed in App.~\ref{app:SUN}.


In the low-energy effective theory, where the $(Q^{2N_c+3},Q^{2N_c+4})$ quark
pair is perturbatively decoupled, the IYIT superpotential in Eq.~\eqref{eq:SPN1}
gives rise to spontaneous supersymmetry breaking associated with
a non-vanishing vacuum energy density~\cite{Izawa:1996pk}.
For completeness and convenience, let us now briefly summarize the computation
of the effective potential in our manifestation of the IYIT model.
In the effective $SP(N_c)$ gauge theory with $N_f = N_c + 1$ flavors below the heavy
quark mass threshold, the vacuum expectation values (VEVs) of the $(N_c + 1)(2N_c + 1)$
meson fields $M^{ij}$ obey the following, so-called deformed quantum moduli
constraint~\cite{Seiberg:1994bz},
\begin{align}
\textrm{Pf}^{(N_c + 1)}\left(M^{ij}\right)
= \Lambda_{\textrm{eff}}^{N_c + 1} \,,
\label{eq:dqmconstraint}
\end{align}
where the left-hand side of this relation denotes the Pfaffian%
\footnote{Here and in the following, the Pfaffian $\textrm{Pf}^{(n)}(M)$ of a
$2n\times2n$ antisymmetric matrix $M$ is defined such that the symplectic form
$J = \mathbbm{1}_n\otimes i\sigma_2$ with $\mathbbm{1}_n$ being the $n$-dimensional
unit matrix and $\sigma_2$ denoting the second Pauli matrix,
satisfies $\textrm{Pf}^{(n)}(J) = 1$.}
of the antisymmetric meson matrix $M^{ij}$.
This moduli constraint gives rise to a dynamically generated superpotential
featuring a new chiral multiplet $Y$, which may be regarded as a Lagrange
multiplier enforcing the moduli constraint by means of its $F$-term condition,
\begin{align}
W_{\textrm{dyn}} = \frac{Y}{\Lambda_{\textrm{eff}}^{N_c - 1}}
\left[\Lambda_{\textrm{eff}}^{N_c + 1} - \textrm{Pf}^{(N_c + 1)}\left(M^{ij}\right)\right] \,.
\end{align}
This $F$-term condition conflicts with the $F$-term conditions for the singlet fields
$Z_{ij}$, cf.\ Eq.~\eqref{eq:SPN1},
\begin{align}
W_{\textrm{IYIT}} \simeq \frac{1}{2}\lambda_{ij}\,\Lambda_{\textrm{eff}} \,Z_{ij} M^{ij}
\,, \quad M^{ij} = 0 \,,
\label{eq:WIYITconf}
\end{align}
and hence supersymmetry is spontaneously broken.
Here, in writing down the IYIT superpotential in the confined phase, we have assumed
that the coupling constants $\lambda_{ij}$ receive at most $\mathcal{O}(1)$ corrections due
to nonperturbative effects as well as that the meson fields $M^{ij}$ are canonically
normalized to good approximation.
The $F$-term scalar potential induced by the singlet fields
$Z_{ij}$ is minimized in accordance with the moduli constraint in Eq.~\eqref{eq:dqmconstraint}
once the $N_c+1$ meson fields $M^{12}$, $M^{34}$, ..., $M^{2N_c+1,2N_c+2}$
are all set to a VEV of $\mathcal{O}\left(\Lambda_{\textrm{eff}}\right)$,
\begin{align}
M^{ij} = \frac{\lambda'}{\lambda_{ij}}\Lambda_{\textrm{eff}} \times J^{ij} \, \quad
\textrm{(no summation over $i,j$),}
\label{eq:Mijsol}
\end{align}
with $\lambda'$ denoting the geometric mean of the coupling constants
$\lambda_{12}$, $\lambda_{34}$, ..., $\lambda_{2N_c+1,2N_c+2}$,
\begin{align}
\lambda' = \left(\prod_{i=1}^{N_c+1} \lambda_{2i-1,2i}\right)^{1/(N_c+1)} \,.
\end{align}
Inserting this solution into Eq.~\eqref{eq:WIYITconf},
i.e.\ integrating out the meson fields, results in the following effective superpotential
describing the vacuum manifold of the IYIT model at low energies,
\begin{align}
W_{\textrm{eff}} \simeq \lambda' \sqrt{N_c + 1} \,\Lambda_{\textrm{eff}}^2 \,X \,,
\label{eq:Weff}
\end{align}
where the chiral superfield $X$ is the canonically normalized goldstino superfield,
\begin{align}
X = \frac{1}{2\sqrt{N_c + 1}} J^{ij} Z_{ij}
= \frac{1}{\sqrt{N_c+1}} \sum_{i=1}^{N_c+1} Z_{2i-1,2i} \,.
\label{eq:goldstino}
\end{align}
The vacuum energy density associated with the spontaneous breaking of supersymmetry
corresponds to the $F$-term scalar potential resulting from the field $X$,
\begin{align}
V \simeq \lambda'^2 (N_c + 1) \,\Lambda_{\textrm{eff}}^4 \,.
\label{eq:XFpot}
\end{align}
As anticipated, it is of $\mathcal{O}\left(\Lambda_{\textrm{eff}}^4\right)$,
which illustrates that in the context of supersymmetry, in contrast to the
non-supersymmetric case, we are indeed able to engineer the vacuum energy density
in a controlled fashion, cf.\ our discussion in the previous section and in
particular Fn.~\ref{fn:vacuumenergySUSY}.


The effective scale $\Lambda_{\textrm{eff}}$ now depends on the mass
of the heavy quark pair $(Q^{2N_c+3},Q^{2N_c+4})$ and thus on the inflaton field
value, cf.\ Eq~\eqref{eq:Lambdaeff}.
In order to relate it to the fundamental dynamical scale $\Lambda$,
we need the beta-function coefficients above and below the
heavy quark mass threshold,
\begin{align}
b & \: = 3(N_c+1) - N_f = 3(N_c+1) - (N_c+2) = 2 N_c + 1 \,, \label{eq:bbSPNc}\\
b_{\textrm{eff}} & \: = 3(N_c+1) - N_f^{\textrm{eff}} = 3(N_c+1) - (N_c+1) = 2 N_c + 2 \,.
\nonumber
\end{align}
Combining Eqs.~\eqref{eq:Lambdaeff}, \eqref{eq:XFpot} and \eqref{eq:bbSPNc},
we then find the potential for the complex inflaton
field $\phi$,
\begin{align}
\label{eq:spot2}
V \simeq \lambda'^2 (N_c + 1) \,\Lambda^4
\left(\frac{\lambda\left|\phi\right|}{\Lambda}\right)^{2/(N_c+1)} \,,
\end{align}
which represents the main result of our analysis in Ref.~\cite{Harigaya:2012pg}.
In summary, we conclude that in the above described class of DCI models
we obtain monomial inflaton potentials featuring fractional powers $p$
that are fully determined by the size of the gauge group,
\begin{eqnarray}
p = \frac{2}{N_c+1} \,.
\label{eq:pSPNc}
\end{eqnarray}


We note that our derivation of the potential in Eq.~\eqref{eq:spot2} was based on
the assumption that the heavy quark pair $(Q^{2N_c+3},Q^{2N_c+4})$ always decouples
perturbatively.
During inflation, the inflaton field must therefore satisfy the condition
$\lambda\left|\phi\right| \gg \Lambda$ at all times.
After the end of inflation, this conditions however becomes violated and we need
to invoke different properties of our strongly coupled gauge theory in order to
compute the scalar potential.
At intermediate field values, $\lambda\left|\phi\right| \sim \Lambda$,
nonperturbative effects become important, which unfortunately deprives
us of the ability to accurately calculate the inflaton potential.
In fact, we do not know how the transition from the large-field
to the small-field regime takes place and under unfortunate circumstances
it might be that the inflaton becomes trapped in a metastable local
minimum at a field value close to $\Lambda/\lambda$.
This is a potential problem that appears in every DCI scenario
and we have no other choice but to presume that no peculiar
feature such as a metastable vacuum is present in the inflaton potential
at values of $\lambda\left|\phi\right|$ close to the dynamical scale.


Around the origin in field space, where $Z_{IJ} \ll \Lambda/\lambda_{IJ}$ for all $I,J$,
the situation is fortunately much simpler.
In that situation, all quarks are light and dynamical.
Consequently, no quark flavor decouples and the gauge dynamics are that
of the full $SP(N_c)$ gauge theory with $2(N_c+2)$ quark fields.
It is known that this theory exhibits a phase of a s-confinement
close to the origin in field space, which is well described in terms
of the composite meson fields $M^{IJ} \sim Q^I Q^J/\Lambda$.
The strong gauge dynamics in the confined phase nonperturbatively
generate a dynamical superpotential, $W_{\textrm{dyn}}$, which combines
with our tree-level superpotential in Eq.~\eqref{eq:tree1} to yield
\begin{align}
W = W_{\textrm{dyn}} + W_{\textrm{tree}}  \,, \quad
W_{\textrm{tree}} \simeq \frac{1}{2}\lambda_{IJ} \Lambda \,Z_{IJ}M^{IJ} \,, \quad
W_{\textrm{dyn}} = -\frac{\textrm{Pf}^{(N_c + 2)}\left(M^{IJ}\right)}{\Lambda^{N_c-1}} \,.
\label{eq:Wtotsconf}
\end{align}
From this superpotential, it is evident that our $SP(N_c)$ theory
possesses a supersymmetric vacuum at the origin in field space, in which all
fields are massive.
After the end of inflation, all fields settle in this vacuum, thereby
restoring supersymmetry in the strongly interacting sector.
Meanwhile, it is also conceivable that upon some small
modification of our set-up a small amount of SUSY breaking
may remain in the true vacuum, which is then transmitted to the
visible sector and hence responsible for the soft SUSY
breaking in the supersymmetric standard model.
In App.~\ref{app:dynSUSY}, we present a toy model exemplifying
how such a modification may look like and discuss the possible
connection between our DCI mechanism and soft SUSY
breaking in more detail.


Now, in order to compute the effective superpotential for inflaton field values
$\left|\phi\right| \ll \Lambda / \lambda$, it is convenient to decompose the tree-level
superpotential in a similar fashion as in Eq.~\eqref{eq:tree1},
\begin{align}
W_{\textrm{tree}} & \: \simeq \frac{1}{2} \lambda_{ij}\, \Lambda\, Z_{ij} M^{ij} +
\lambda_{ik}\, \Lambda\, Z_{ik} M^{ik} + \lambda\,\Lambda\,\Phi\, Y \,, \quad
Y \equiv M^{2N_c+3,2N_c+4} \,, \\ \nonumber
i,j & \: = 1,2,..,2(N_c+1) \,, \:\: k = 2N_c+3,2N_c+4 \,, 
\end{align}
The total superpotential in Eq.~\eqref{eq:Wtotsconf} may then be rewritten as,
\begin{align}
W \simeq \frac{Y}{\Lambda^{N_c-1}} \left[\lambda\, \Lambda^{N_c} \,\Phi
- \textrm{Pf}^{(N_c+1)} \left(M^{ij}\right)\right] +
\frac{1}{2} \lambda_{ij}\, \Lambda\, Z_{ij} M^{ij} +
\lambda_{ik}\, \Lambda\, Z_{ik} M^{ik} .. \,,
\label{eq:WYPhi}
\end{align}
where the ellipses stands for a polynomial only in the
$M^{ij}$ and $M^{ik}$, but not involving the field $Y$.
The first two terms in Eq.~\eqref{eq:WYPhi} just represent the
total superpotential of the IYIT model in the peculiar case of
a field-dependent moduli constraint.
The solution to this constraint which also minimizes the
$F$-term potential induced by the singlet fields $Z_{ij}$
is given by, cf.\ Eq.~\eqref{eq:Mijsol},
\begin{align}
M^{ij} = \frac{\lambda'}{\lambda_{ij}}\left(\lambda\,\Lambda^{N_c}\,\Phi\right)^{1/(N_c+1)}
\times J^{ij} \, \quad \textrm{(no summation over $i,j$).}
\label{eq:MijPhisol}
\end{align}
Close to the origin in field space, the interaction between the goldstino field $X$
and the inflaton field $\Phi$ is therefore described by the same effective superpotential
as during inflation, cf.\ Eq.~\eqref{eq:Weff},
\begin{align}
W_{\textrm{eff}} \simeq \lambda'
\sqrt{N_c+1} \,\Lambda_{\textrm{eff}}^2 \,X + .. \,, \quad
\Lambda_{\textrm{eff}}^2 \simeq \Lambda^2 \left(\frac{\lambda\,\Phi}{\Lambda}\right)^{1/(N_c+1)} \,,
\label{eq:Weff2}
\end{align}
where we have omitted all terms that do not involve the fields $X$ and $\Phi$.
The fact that our results in the large-field as well as in the
small-field regime, cf.\ Eqs.~\eqref{eq:Weff} and \eqref{eq:Weff2},
coincide with each other is all but a surprise, but rather expected.
The reason for this coincidence is that, as a matter of fact, the form of the effective
superpotential is uniquely determined by holomorphicity, dimensional analysis as well as
by the requirement of consistency with an (anomalous) $R$ symmetry.
We will return to this argument further below in Sec.~\ref{subsec:SPNWeff}.


In the small-field regime, we can now not deduce the inflaton potential
from the effective superpotential as easily as in the large-field regime.
The naive $F$-term potential induced by the field $X$ may be the same
as for large inflaton field values, cf.\ Eq.~\eqref{eq:spot2},
\begin{align}
V \simeq \lambda'^2 (N_c + 1) \,\Lambda^4
\left(\frac{\lambda\left|\phi\right|}{\Lambda}\right)^p \,, \quad
p = \frac{2}{N_c+1} \,.
\label{eq:Vphismall}
\end{align}
This potential, however, lacks a mass term for the inflaton
and hence conflicts with our above statement that all fields
are expected to be massive in the true vacuum.
Moreover, for $p<1$, it even implies a singular first derivative at the origin.
All these discrepancies are due to the fact that, in consequence of non-minimal
corrections to the K\"ahler potential, $\delta K$, the complex field $\phi$ now no longer
corresponds to the canonically normalized inflaton.
To see this, note that, upon eliminating the meson fields $M^{ij}$ according to
Eq.~\eqref{eq:MijPhisol}, the canonical K\"ahler potential for the fields $M^{ij}$
turns into a non-canonical K\"ahler potential for the inflaton,
\begin{align}
\delta K = \frac{1}{2} M_{ij}^\dagger M^{ij} = \frac{\lambda'^2}{\overline{\lambda^2}}
\left(N_c+1\right) \,\Lambda^2
\left(\frac{\lambda\left|\phi\right|}{\Lambda}\right)^p \,,
\end{align}
where $1/\overline{\lambda^2}$ denotes the arithmetic mean of the inverse
Yukawa coupling constants squared,
\begin{align}
1/\overline{\lambda^2} =  \frac{1}{2\left(N_c+1\right)}
\frac{J_{ij}^\dagger J^{ij}}{\lambda_{ij}^2}
= \frac{1}{\left(N_c+1\right)} \sum_{i=1}^{N_c+1} \frac{1}{\lambda_{ij}^2} \,.
\end{align}
At small inflaton field values, $\lambda\left|\phi\right|\ll \Lambda$, this term
in the K\"ahler potential dominates over the canonical K\"ahler potential
for the inflaton, $K = \left|\phi\right|^2$, such that the canonically normalized
inflaton must eventually be identified as the reparametrized field
$\tilde{\phi} = \lambda'/\bar{\lambda} \sqrt{N_c+1}\,\Lambda
\left(\lambda\,\phi/\Lambda\right)^{p/2}$, where $\bar{\lambda} =
\big(\overline{\lambda^2}\big)^{1/2}$.
The scalar potential for the field $\tilde{\phi}$ is then quadratic as expected,
\begin{align}
V \simeq \bar{\lambda}^2 \Lambda^2 \big|\tilde{\phi}\big|^2 \,,
\end{align}
such that, close to the origin, the inflaton indeed possesses a
mass, $m_\phi \simeq \bar{\lambda}\Lambda$.


Before closing this section, we finally remark that it would have
been less favorable to start out with the following, seemingly simpler
superpotential to arrive at the above results,
\begin{align}
W_{\textrm{tree}} = W_{\textrm{IYIT}} + \lambda \,\Phi\, Q^{2N_c+3}Q^{2N_c+4} \,.
\end{align}
Indeed, our conclusions pertaining the inflationary dynamics
would have been the same in the case of this superpotential;
we would have still been led to the fractional power-law potential
in Eq.~\eqref{eq:spot2}.
But, without the first term on the right-hand side of Eq.~\eqref{eq:DeltaW},
the low-energy effective theory well after the end of inflation
features $4(N_c+1)$ meson fields $M^{ik}$ lacking a singlet partner $Z_{ik}$.
Therefore, once the VEV of the inflaton field has
decreased to zero, these mesons are completely massless,%
\footnote{The dynamically generated superpotential for the meson
fields does not induce any meson mass terms.}
which has great potential to jeopardize the reheating process after inflation.
We would, for instance, expect that the inflaton, due its direct coupling
to the strongly interacting sector, then predominantly decays
into the massless mesons and hardly into standard model (SM) radiation.
In this situation, we would thus fail to ignite the hot big bang
after the end of inflation.
On the other hand, the generation of a subdominant abundance
of dark radiation in the form of massless mesons might also be advantageous.
The question under which conditions a small number of massless mesons
would allow for a viable reheating process requires, however, a
more detailed study, which is beyond the scope of this paper.


\subsection[Massless and massive matter fields
(fractional powers \texorpdfstring{$p > 0$}{p > 0})]
{Massless and massive matter fields
(fractional powers \texorpdfstring{\boldmath{$p > 0$}}{p > 0})}
\label{subsec:SPNcmassive}


It is now straightforward to extend the above class of models along
the lines of our general discussion in Sec.~\ref{sec:idea}.
The largest power $p$ we managed to generate in the previous section
was $p=1$, corresponding to the simplest possible group,
$SP(1) \cong SU(2)$, among all gauge groups under consideration.
In order to generate larger powers than $p=1$, it is necessary to increase
the difference between the beta-function coefficients $b$ and $b_{\textrm{eff}}$,
cf.\ Eq.~\eqref{eq:pbb}.
At first sight, this is easily accomplished by relaxing the relation between $N_f$
and $N_c$ imposed in the previous section, $N_f = N_c + 2$.
It appears as if we only needed to couple the inflaton to not
only one, but several quark flavors in the superpotential,
such that simply a larger number of quarks perturbatively decouples during inflation.
The only problem with this approach however is that then, at low energies, our $SP(N_c)$
gauge theory will feature more than $N_c+2$ massless quark flavors.
Hence, we will no longer reach a phase of s-confinement, which conflicts with
the second ingredient of our DCI recipe outlined in Sec.~\ref{sec:idea} and
thus causes us to loose control over the IR dynamics of our theory.
Instead, our $SP(N_c)$ gauge theory will then be in a non-Abelian Coulomb phase, which,
depending on the exact number of flavors $N_f$, will be either very strongly coupled,
$N_c+3\leq N_f \leq \frac{3}{2}(N_c+1)$, in the conformal window,
$\frac{3}{2}(N_c+1)< N_f < 3(N_c+1)$, or free, $3(N_c+1)\leq N_f$,
in the infrared~\cite{Intriligator:1995ne,Seiberg:1994pq}.
In particular the second of these three cases might lead to very exotic
dynamics after the end of inflation, because it would render the
inflaton an unparticle~\cite{Banks:1981nn} at low energies, which would interact
with the SM sector in a very unconventional way.
While such a possibility is certainly intriguing, we shall not further
consider it in this paper.
Instead, we shall merely extend our model just a little bit further,
so as to ensure that also in the presence of additional flavors
our theory reaches the s-confined phase at low energies.
This is best done by simply equipping the additional flavors
with large enough supersymmetric mass terms in the superpotential.
The additional heavy quark flavors are then guaranteed to perturbatively decouple
at low energies, independently of the inflaton field value, so that our
model always reduces to the s-confining $SP(N_c)$ theory with
$N_c + 2$ flavors as required by our DCI recipe.


To illustrate our idea more concretely, we shall now consider
$N_f = N_c + 2 + N_m$ quark flavors, out of which the first (old)
$N_c + 2$ flavors, $Q^I$ with $I=1,2,..,2(N_c+2)$, couple again to
the singlet fields $Z_{IJ}$, while the last (new) $N_m$ flavors,
$P^a$ and $\bar{P}^a$ with $a=1,2,..,N_m$, all couple to the
inflaton field $\Phi$.
According to our above consideration, we also endow the new quark flavors with
supersymmetric masses $M_a$, where w.l.o.g.\ $M_1 \leq M_2 \leq .. \leq M_{N_m}$,
such that our total tree-level superpotential now reads
\begin{align}
\label{eq:tree2}
W_{\textrm{tree}} = \frac{1}{2}\lambda_{IJ} Z_{IJ}Q^IQ^J +
\left(\lambda_a \Phi + M_a\right) P^a \bar{P}^a \,, \quad
\Phi \equiv Z_{2N_c+3,2N_c+4} \,.
\end{align}
Alternatively, we could also consider couplings between $\Phi$
and the quark pairs $P^a \bar{P}^{a+1}$,
which would allow us to distinguish between the $M_a$ terms and the inflaton
Yukawa interactions by means of discrete symmetries.
Besides that, we emphasize that we do not specify the origin of
the masses terms in the above superpotential.
They may equally be fundamental input scales
or originate from some other dynamical sector.
We shall merely assume that all of the masses $M_a$ in
Eq.~\eqref{eq:tree2} are at least slightly larger than the
dynamical scale, $M_a \gtrsim \Lambda$.
This guarantees that the massive quarks decouple perturbatively
at energies above the dynamical scale, such that at energies
around and below the dynamical scale only $N_c + 2$
quark flavors remain.
Thus, as anticipated, the theory reaches again the phase of s-confinement in the infrared.
Similarly as in the previous section, we still identify the chiral inflaton superfield
$\Phi$ as the $(I,J) = (2N_c+3,2N_c+4)$ component of the singlet field tensor $Z_{IJ}$.
The superpotential in Eq.~\eqref{eq:tree2} can therefore be expanded
in a similar manner as our first superpotential in Eq.~\eqref{eq:tree1},
\begin{align}
\label{eq:Wtreefull}
W_{\textrm{tree}} & \: = \frac{1}{2} \lambda_{ij} Z_{ij} Q^iQ^j +
\lambda_{ik} Z_{ik} Q^i Q^k +
\left(\lambda_0 \Phi + M_0\right) P^0 \bar{P}^0 +
\left(\lambda_a \Phi + M_a\right) P^a \bar{P}^a \,, \\ \nonumber
i,j & \: = 1,2,..,2(N_c+1) \,, \:\:
k = 2N_c+3,2N_c+4 \,, \:\: a = 1,2,..,N_m \,,
\end{align}
with $\lambda_0$ now playing the role of the Yukawa coupling $\lambda$
in Eq.~\eqref{eq:DeltaW} and where we have introduced
$M_0 \equiv 0$, $P^0\equiv Q^{2N_c+3}$ and $\bar{P}^0\equiv Q^{2N_c+4}$
for notational purposes.


During inflation, all $N_d = N_m + 1$ quark flavors coupling to the inflaton
field receive a large mass, such that the effective superpotential at energies
below all heavy quark mass thresholds is again the superpotential of the IYIT
model of dynamical supersymmetry breaking, cf.\ Eq.~\eqref{eq:Weff},
\begin{align}
W_{\textrm{eff}} \simeq \lambda' \sqrt{N_c + 1} \,\Lambda_{\textrm{eff}}^2 \,X \,.
\label{eq:Weff3}
\end{align}
Matching the running of the gauge coupling constant at each of the $N_d$
heavy quark mass thresholds, we are able to relate the effective dynamical
scale $\Lambda_{\textrm{eff}}$ to $\Lambda$, the fundamental scale of the
full $SP(N_c)$ gauge theory with $N_f = N_c+2+N_m$ flavors,
\begin{align}
\Lambda_{\textrm{eff}} \simeq \Lambda \prod_{n=0}^{N_m}
\left(\frac{M_n + \lambda_n\Phi}{\Lambda}\right)^{(b_n-b_{n+1})/b_{\textrm{eff}}} \,,
\label{eq:Lambdaeffprod}
\end{align}
where the $b_n$ denote the beta-function coefficients that are to be respectively
employed at energies at which already $N_d - n$ heavy quarks have perturbatively
decoupled.
Thus, we have
\begin{align}
b_n = & \: 3(N_c+1) - (N_c+1) - n = 2(N_c+1) - n \,, \label{eq:bn} \\
b_{N_m+1} \equiv b = & \: 3(N_c+1) - (N_c+1) - (N_m+1) = 2N_c + 1 - N_m \,, \nonumber \\
b_0 \equiv b_{\textrm{eff}} = & \: 3(N_c+1) - (N_c+1) = 2(N_c+1) \,. \nonumber
\end{align}
For not too large masses $M_n$ and not too small Yukawa couplings $\lambda_n$,
we expect $\lambda_n\left|\phi\right| \gg M_n$ for all $n=0,1,..,N_m$ during inflation.
In this case, the relation in Eq.~\eqref{eq:Lambdaeffprod} simplifies to
\begin{align}
\lambda_n \left|\phi\right| \gg M_n \,, \quad
\Lambda_{\textrm{eff}} \simeq \Lambda \left(\frac{\lambda\,\Phi}
{\Lambda}\right)^{(b_{\textrm{eff}}-b)/b_{\textrm{eff}}} \,, \quad
\lambda = \left(\,\prod_{n=0}^{N_m} \lambda_n\right)^{1/(N_m+1)} \,,
\label{eq:Lambdaefflimit}
\end{align}
where we have re-introduced the coupling constant $\lambda$, which is
now supposed to denote the geometric mean of the Yukawa couplings
$\lambda_0$, $\lambda_1$, .., and $\lambda_{N_m}$.


Combining Eqs.~\eqref{eq:Weff3}, \eqref{eq:bn}, and \eqref{eq:Lambdaefflimit},
we find the same monomial inflaton potential as in Eq.~\eqref{eq:spot2},
the only difference being that now the fractional power $p$ is given as
\begin{eqnarray}
p = \frac{2\,(1+N_m)}{N_c+1} \,,
\label{eq:pNm}
\end{eqnarray}
which of course reduces to the expression in Eq.~\eqref{eq:pSPNc} for $N_m = 0$.
As we will see in our discussion of the inflationary observables in
Sec.~\ref{subsec:observables}, the dynamical scale $\Lambda$ is typically
required to be rather large.
We therefore demand that our full gauge theory be (almost)
asymptotically free, $b \geq 0$.
Otherwise, the RGE running between the the heavy quark mass thresholds
and the Planck scale could result in too large values of the gauge coupling
constant at the Planck scale.
In addition to that, we note that a negative beta-function coefficient
$b$ could, under certain circumstances, also eventually lead to a Landau pole for
the gauge coupling constant above the Planck scale.
The condition that $b$ should not be negative then results in an upper bound on
the power $p$.
Setting $b$ to zero in Eq.~\eqref{eq:pbb}, we find that $p$
can become at most as large as $p = 4$,
\begin{align}
p = \frac{4\,\left(b_{\textrm{eff}}-b\right)}{b_{\textrm{eff}}} \leq 
\frac{4\,\left(b_{\textrm{eff}}- 0\right)}{b_{\textrm{eff}}} = 4 \,.
\label{eq:pboundSPN}
\end{align}
As evident from Eqs.~\eqref{eq:bn} and \eqref{eq:pNm},
this bound is saturated for a total of $N_m = 2 N_c + 1$ quark
flavors coupling the inflaton field;
more quark flavors would result in a negative beta-function coefficient $b$.
In addition to this rather technical bound on the power $p$,
we will also give a more physical argument why $p$ should not take
values larger than $4$ in Sec.~\ref{subsec:constraints}, cf.\ Eq.~\eqref{eq:pbound}.
Most importantly, however, it is well known that powers larger than $4$ are
in severe tension with the observational data in any case,
cf.\ our discussion of the inflationary observables in Sec.~\ref{subsec:observables}.


For now, let us derive the scalar potential for the inflaton field
after the end of inflation.
At energies around and below the dynamical scale, all of the $N_m$
additional quark flavors are perturbatively decoupled.
Our model thus corresponds once more to the $SP(N_c)$ gauge theory
with $N_c+2$ flavors.
The dynamical scale of this theory, $\Lambda_{N_c+2}$, however differs
from $\Lambda$, the fundamental scale of the full theory with $N_f = N_c+2+N_m$
flavors.
Analogously to the relation between $\Lambda_{\textrm{eff}}$ and $\Lambda$ in
Eq.~\eqref{eq:Lambdaeffprod}, those two scales are related to
each other as follows,
\begin{align}
\Lambda_{N_c+2} \simeq \Lambda \prod_{a=1}^{N_m}
\left(\frac{M_a + \lambda_a\Phi}{\Lambda}\right)^{(b_a-b_{a+1})/b_1}
\,, \quad b_1 = 2(N_c+1) - 1 = 2 N_c + 1 \,.
\label{eq:LambdaNc2}
\end{align}
In the limit of a small inflaton field value, we have $\lambda_a\left|\phi\right| \ll M_a$
for all $a=1,2,..,N_m$ and the dynamical scale $\Lambda_{N_c+2}$ in Eq.~\eqref{eq:LambdaNc2}
reduces to
\begin{align}
\Lambda_{N_c+2} \simeq \Lambda \left(\frac{M}{\Lambda}\right)^{N_m/b_1} \,, \quad
M = \left(\,\prod_{a=1}^{N_m} M_a\right)^{1/N_m} \,,
\end{align}
where $M$ is the geometric mean of the masses $M_1$, $M_2$, .., and $M_{N_m}$.
From this point on, the computation of the scalar potential proceeds exactly
as in the $N_m=0$ case.
The only change we have to perform is to replace every scale $\Lambda$ in our
previous calculation with the effective scale $\Lambda_{N_c+2}$.
We then find again the effective superpotential in Eq.~\eqref{eq:Weff2},
\begin{align}
W_{\textrm{eff}} \simeq \lambda' \sqrt{N_c+1} \,\Lambda_{\textrm{eff}}^2 \,X \,,
\label{eq:Weff4}
\end{align}
which now contains the following effective dynamical scale $\Lambda_{\textrm{eff}}$,
\begin{align}
\Lambda_{\textrm{eff}} \simeq \Lambda_{N_c+2}
\left(\frac{\lambda_0\Phi}{\Lambda_{N_c+2}}\right)^{1/b_{\textrm{eff}}}
\simeq \Lambda \left(\frac{M}{\Lambda}\right)^{N_m/b_{\textrm{eff}}}
\left(\frac{\lambda_0\Phi}{\Lambda}\right)^{1/b_{\textrm{eff}}} \,.
\label{eq:LambaLambdaNc2}
\end{align}
In the small-field regime, the scalar potential for the non-canonically normalized field
$\phi$ therefore features the same power $p$ as in the $N_m = 0$ case, cf.\ Eq.~\eqref{eq:pSPNc},
\begin{align}
p = \frac{2}{N_c+1} \,.
\label{eq:pLE}
\end{align}
As expected, the additional heavy quark flavors thus do not affect
the scaling behavior of the inflaton potential.
Instead, their only contribution to the inflaton potential
ends up being an overall prefactor, $(M/\Lambda)^{4N_m/b_{\textrm{eff}}}$,
which accounts for the change in the dynamical scale above
the heavy quark mass thresholds.
However, as long as one is only interested in the low-energy dynamics below
the dynamical scale, this change is of course irrelevant and
it is actually the scale $\Lambda_{N_c+2}$, which should be regarded as
the fundamental scale.
From this perspective, the effective potential in the small-field regime
then turns out to be independent of the number of extra heavy quark flavors,
$N_m$, and instead it always corresponds to the potential in Eq.~\eqref{eq:Vphismall}.


In the above analysis, we have separately discussed the large- as well as the
small-field regime, which led us to two different fractional powers in the 
inflation potential, cf.\ Eqs.~\eqref{eq:pNm} and \eqref{eq:pLE}.
These two results, however, only correspond to two limiting cases for the power $p$
which are in fact continuously connected.
In both regimes, we found the same effective scale,
\begin{align}
\Lambda_{\textrm{eff}} \simeq \Lambda \prod_{n=0}^{N_m}
\left(\frac{M_n + \lambda_n\Phi}{\Lambda}\right)^{(b_n-b_{n+1})/b_{\textrm{eff}}} \,, 
\label{eq:Lambdaefffull}
\end{align}
which we then simplified either assuming $\lambda_a\left|\phi\right| \gg M_a$
or $\lambda_a\left|\phi\right| \ll M_a$ for all $a=1,2,..,N_m$,
cf.\ Eqs.~\eqref{eq:Lambdaefflimit} and \eqref{eq:LambaLambdaNc2}.
Therefore, refraining from making any such assumption about the size of the
ratios $x_a = \lambda_a\left|\phi\right|/M_a$ puts us in the position
to study how our two expressions for $p$ in Eqs.~\eqref{eq:pNm}
and \eqref{eq:pLE} are actually linked to each other.
After some elementary algebra, one finds that the product on the right-hand side
of Eq.~\eqref{eq:Lambdaefffull} can be rewritten as follows,
\begin{align}
\Lambda_{\textrm{eff}} \simeq
\Lambda \left(\frac{M_{\textrm{eff}}}{\Lambda}
\right)^{(N_m - N_m^{\textrm{eff}})/b_{\textrm{eff}}}
\left(\frac{\lambda_{\textrm{eff}}\,\Phi}{\Lambda}\right)^{p_{\textrm{eff}}/4} \,,
\label{eq:Lambdaeffeff}
\end{align}
where $p_{\textrm{eff}}$, $N_m^{\textrm{eff}}$, $\lambda_{\textrm{eff}}$, and
$M_{\textrm{eff}}$ are all effective quantities that depend on the field value
of the inflaton field.
The effective power $p_{\textrm{eff}}$ turns out to be related to $N_m^{\textrm{eff}}$,
the effective number of undecoupled quark flavors $(P_a,\bar{P}_a)$, just in the same way
as $p$ is related to $N_m$ in Eq.~\eqref{eq:pNm},
\begin{align}
p_{\textrm{eff}}(\phi) = \frac{2(1+N_m^{\textrm{eff}}(\phi))}{N_c+1} \,, \quad
N_m^{\textrm{eff}}(\phi) = \sum_{a=1}^{N_m} \omega_a(\phi) \,.
\end{align}
Here, the $\omega_a$ represent ``weights'' for the individual heavy quark flavors
that approach $1$ for large values of the ratio $x_a$
as well as $0$ for small values of this ratio,%
\footnote{These weights diverge once the ratios $x_a$ approach a value of $1$.
The effective quantities $p_{\textrm{eff}}$, $N_m^{\textrm{eff}}$,
$\lambda_{\textrm{eff}}$, and $M_{\textrm{eff}}$ therefore only have a physically
meaningful interpretation as long as the inflaton field value is sufficiently far
away from any heavy quark mass threshold, i.e.\ as long as $x_a\not\simeq1$ for all
$a =1,2,..,N_m$.}
\begin{align}
\omega_a(\phi) = \log_{x_a(\phi)}\left[1+x_a(\phi)\right] \,, \quad
x_a(\phi) = \frac{\lambda_a\left|\phi\right|}{M_a} \,.
\end{align}
For completeness, we also define $\omega_0 = 1$ for the massless quark flavor
$(P_0,\bar{P}_0)$.
At the same time, the effective Yukawa coupling $\lambda_{\textrm{eff}}$ as well the
effective heavy quark mass scale $M_{\textrm{eff}}$ correspond to the weighted
geometric means of the fundamental input parameters $\lambda_n$ and $M_a$,
\begin{align}
\lambda_{\textrm{eff}}(\phi) = \left(\,\prod_{n=0}^{N_m}
\lambda_n^{\omega_n(\phi)}\right)^{1/(1+N_m^{\textrm{eff}}(\phi))} \,, \quad
M_{\textrm{eff}}(\phi) = \left(\,\prod_{a=1}^{N_m}
M_a^{1-\omega_a(\phi)}\right)^{1/(N_m-N_m^{\textrm{eff}}(\phi))} \,.
\end{align}
In combination with the effective IYIT superpotential in Eq.~\eqref{eq:Weff},
the effective scale in Eq.~\eqref{eq:Lambdaeffeff} gives rise to an inflaton potential
which is well approximated by a monomial as long as all of the weights $\omega_a$
are either close to $0$ or $1$,
\begin{align}
\label{eq:Vtot}
V \simeq & \: \lambda'^2 (N_c+1) \,\Lambda^4
\prod_{n=0}^{N_m}
\left|\frac{M_n + \lambda_n\phi}{\Lambda}\right|^{2/(N_c+1)} \\
\nonumber
= & \: \lambda'^2 (N_c+1) \,\Lambda^4 \left(\frac{M_{\textrm{eff}}}{\Lambda}
\right)^{2\,(N_m - N_m^{\textrm{eff}})/(N_c+1)}
\left(\frac{\lambda_{\textrm{eff}}\left|\phi\right|}{\Lambda}\right)^{p_{\textrm{eff}}}
\,, \quad p_{\textrm{eff}} = \frac{2(1+N_m^{\textrm{eff}})}{N_c+1} \,.
\end{align}


In our above discussion, we assumed $\omega_a \simeq 1$ for all $a=1,2,..,N_m$
in the large-field regime as well as $\omega_a \simeq 0$ for all $a=1,2,..,N_m$
in the small-field regime.
This implied $N_m^{\textrm{eff}} \simeq N_m$ during inflation as well as
$N_m^{\textrm{eff}} \simeq 0$ after the end of inflation, which led us to
our results in Eq.~\eqref{eq:pNm} and \eqref{eq:pLE}.
The scalar potential in Eq.~\eqref{eq:Vtot} now nicely illustrates how those
two asymptotic regimes are continuously connected to each other.
In addition, we note that the potential in Eq.~\eqref{eq:Vtot} is also suitable to
study inflation in the presence of very large quark masses,
$\mathcal{O}(\Lambda) \lesssim M_a \lesssim \mathcal{O}(M_{\textrm{Pl}})$.
In such a case, the heavy quark mass thresholds would not be passed after, but
during inflation, resulting in the power in the inflaton potential to successively
decrease while inflation takes place.
The predictions for the observables related to the primordial power spectrum
would then be determined by the momentary value of the effective power at the time when
the CMB scales exit the Hubble horizon, $p_{\textrm{eff}} = p_{\textrm{eff}}^*$.
A smooth variation of the heavy quark masses would then translate into
a smooth change of $p_{\textrm{eff}}^*$ and hence also a smooth change
of the inflationary observables.
Here, we have implicitly assumed that the masses $M_n$ as well as
the Yukawa couplings $\lambda_n$ are all real and positive.
But this is in fact a very special case, as the mass parameters $M_n$ and 
the coupling constants $\lambda_n$ are generally complex-valued.
In the general case, we therefore have to ensure that there are no accidental
cancellations between the complex phases in the effective masses of
the additional quark flavors.
Indeed, under unfortunate circumstances, some quark flavors might become
almost or even exactly massless during inflation because of a conspiracy among
these phases such that $\left|M_n + \lambda_n\phi\right| \ll \Lambda$
for some particular inflaton field value.
This endangers our entire inflationary scenario, as the scalar potential
for the inflaton field is based on the idea that there
are exactly $N_f = N_c + 1$ massless flavors present
in the low-energy effective theory.
Thus, in order to avoid such a situation, we either have to assume that the phases of the
masses and couplings are adjusted or even fine-tuned
in such a way that none of the additional quark flavors $\left(P^n,\bar{P}^n\right)$ becomes
massless during inflation---or we have to stick to very large ratios $x_a$
during inflation in the first place and forget about the possibility of heavy
quark masses $M_a$ in between the dynamical scale and the Planck scale.
In any case, while the regime of large masses $M_a$ 
clearly deserves a more detailed study, we shall simply assume in this paper
that all of the ratios $x_a$ are very large
during inflation and hence restrict ourselves to the case of a constant
power $p$ that only begins to vary after the end of inflation.


\subsection[Effective superpotential from \texorpdfstring{$R$}{R} symmetry]
{Effective superpotential from \texorpdfstring{\boldmath{$R$}}{R} symmetry}
\label{subsec:SPNWeff}


All of our calculations in the previous two sections have eventually resulted
in the same effective superpotential at energies below the dynamical scale,
cf.\ Eqs.~\eqref{eq:Weff}, \eqref{eq:Weff2}, \eqref{eq:Weff3}, and \eqref{eq:Weff4},
which, in its most general form, may be written as
\begin{align}
W_{\textrm{eff}} = \lambda' \sqrt{N_c + 1} \,\Lambda^2 X 
\prod_{n=0}^{N_m} \left(\frac{M_n + \lambda_n\Phi}
{\Lambda}\right)^{2/b_{\textrm{eff}}} \,.
\label{eq:Weff5}
\end{align}
As we will now demonstrate, this is not a coincidence, but a direct
consequence of holomorphicity, dimensional analysis and the requirement
of consistency with an (anomalous) $R$ symmetry.
We shall assume that all types of quark flavors, $Q^i$ and $(P^n,\bar{P}^n)$,
as well as all the various sets of singlet fields, $Z_{ij}$, $Z_{ik}$ and $\Phi$, 
share common $R$ charges, respectively,
\begin{align}
R[Q^i] = q \,, \quad R[P^n] = R[\bar{P}^n] = p \,, \quad
R[Z_{ij}] = z \,, \quad R[Z_{ik}] = y \,, \quad R[\Phi] = x \,.
\end{align}
The tree-level superpotential in Eq.~\eqref{eq:Wtreefull} induces
three conditions among these charges, such that eventually only two
of them remain linearly independent.
If we choose these free charges to be $z$ and $x$,
the three dependent charges, $q$, $p$ and $y$, can
be parametrized as follows,
\begin{align}
q = 1-\frac{z}{2} \,, \quad p = 1 - \frac{x}{2} \,, \quad y = \frac{1}{2}(x+z) \,.
\label{eq:Rchargesqpy}
\end{align}
Besides that, the superpotential in Eq.~\eqref{eq:Wtreefull} also illustrates
that, for $N_m > 0$ and $x\neq0$, the masses $M_a$ must carry a common
spurious nonzero $R$ charge $R[M] = m \equiv x$.
All in all, this charge assignment is consistent with
a classical $R$ symmetry.
However, at the quantum level, this $R$ symmetry is in general
nonperturbatively broken by the $SP(N_c)$ gauge anomaly.
The anomaly coefficient $\mathcal{A}_R$ for the mixed
$U(1)_R\left[SP(N_c)\right]^2$ is given by
\begin{align}
\label{eq:AR}
\mathcal{A}_R = & \: 2(N_c+1) + 2(N_c+1)(q - 1) + 2(1+N_m)(p - 1) \\ \nonumber
= & \: 2(N_c+1)\left(1-\frac{z}{2}\right) - (1+N_m)\,x \,,
\end{align}
which does not vanish for generic values of $x$ and $z$.
But, this is not a problem.
Formally, we can still construct an exact symmetry, a so-called anomalous
or spurious $R$ symmetry, by promoting the holomorphic
gauge coupling constant and thus the dynamical scale $\Lambda$ to chiral multiplets
which nontrivially shift and rotate under $U(1)_R$ transformations~\cite{Seiberg:1993vc}.
Performing a $U(1)_R$ rotation by an angle $\alpha$, we then have,
\begin{align}
\label{eq:LambdaRcharge}
\frac{8\pi^2}{g^2} \rightarrow \frac{8\pi^2}{g^2} - i\alpha \,b\,\ell \,, \quad
\Lambda \rightarrow \Lambda \,e^{i\alpha\,\ell} \,, \quad 
R[\Lambda] = \ell = \frac{\mathcal{A}_R}{b} \,, \quad b = 2N_c + 1 - N_m \,,
\end{align}
where $\ell$ denotes the spurious $R$ charge of the dynamical scale $\Lambda$.
Note that the such constructed anomalous $R$ symmetry also encompasses
the special case of an anomaly-free $R$ symmetry, which we obtain after
adjusting $x$ and $z$, such that the anomaly coefficient $\mathcal{A}_R$
in Eq.~\eqref{eq:AR} vanishes,
\begin{align}
z = 2 - \frac{1+N_m}{N_c+1} \,x \,, \quad \mathcal{A}_R = 0 \,, \quad \ell = 0 \,.
\end{align}
In this case, all $R$ charges can be parametrized solely in terms of 
the inflaton charge $x$,
\begin{align}
q = \frac{1+N_m}{N_c+1}\frac{x}{2} \,, \quad p = 1 - \frac{x}{2} \,, \quad
z = 2 - \frac{1+N_m}{N_c+1}\,x \,, \quad
y = 1 + \frac{N_c - N_m}{N_c+1} \frac{x}{2} \,, \quad \ell = 0 \,.
\label{eq:Rchargesx}
\end{align}


After these remarks regarding the assignment of $R$ charges to
the various fields of our theory, we shall now explicitly re-derive
the effective superpotential in Eq.~\eqref{eq:Weff5}.
In doing so, we shall consider the most general case and merely require
that the superpotential be consistent with the anomalous $R$
symmetry in Eqs.~\eqref{eq:Rchargesqpy} and \eqref{eq:LambdaRcharge}.
We first neglect the masses $M_a$.
According to the $R$ charges listed in these two equations, the only
gauge-invariant term that could possibly appear in the effective superpotential
must then have the following structure,
\begin{align}
W_{\textrm{eff}} \propto \left(\Lambda^b\,\Phi^{1+N_m}\right)^{1/(N_c+1)} Z_{ij} \,.
\label{eq:Weffterms}
\end{align}
This term is readily rendered neutral under the global flavor symmetry by contracting
the singlet fields $Z_{ij}$ with the antisymmetric tensor $J^{ij}/2$,
\begin{align}
W_{\textrm{eff}} \propto \sqrt{N_c+1}
\left(\Lambda^b\,\Phi^{1+N_m}\right)^{1/(N_c+1)} X \,, \quad
X = \frac{1}{2\sqrt{N_c+1}}J^{ij} Z_{ij} \,,
\end{align}
where we have re-introduced the canonically normalized field $X$.
Next, we recall that the inflaton couples to quark flavors via the
superpotential term $(\lambda_n \Phi + M_n)P^n \bar{P}^n$.
This requires us to shift each of the
$1+N_m$ factors of $\Phi$ contained in the product $\Phi^{1+N_m}$ by a
different mass parameter,
\begin{align}
\Phi^{1+N_m} \rightarrow \prod_{n=0}^{N_m} \left(M_n + \lambda_n \Phi\right) \,.
\label{eq:Phishift}
\end{align}
Making use of the fact that the beta-function coeffcient $b$ is
given as $b=2N_c+1-N_m$, we then find that
the effective superpotential must be of the following form,
\begin{align}
W_{\textrm{eff}} \simeq  \lambda' \sqrt{N_c + 1} \,\Lambda^2 X 
\prod_{n=0}^{N_m} \left(\frac{M_n + \lambda_n\Phi}
{\Lambda}\right)^{1/(N_c+1)} \,, \label{eq:Weff6}
\end{align}
where $\lambda'$ plays the role of an arbitrary numerical proportionality constant,
which we cannot further determine based on $R$ symmetry arguments.
This expression is nothing but the superpotential in Eq.~\eqref{eq:Weff5}!
As we now see, it indeed entirely follows from dimensional
analysis and the requirement of consistency with an anomalous
$R$ symmetry.
At the same time, the fact that the superpotential is a holomorphic
object guarantees that it does not change its functional form while the inflaton
field decreases from a large to a very small field value.
Instead, given the superpotential at a particular value of the inflaton field,
we are able to analytically continue it across the entire inflaton field
range without picking up any additional terms.
Its form hence always remains the same, regardless of the
magnitudes of the fields $\Phi$ and $X$.


We derived the effective superpotential in Eq.~\eqref{eq:Weff5} taking
into account the full field content of our theory.
Alternatively, we can also first integrate out all the heavy quark flavors
$(P^n,\bar{P}^n)$ and then construct the effective superpotential from the
requirement of consistency with $R$ symmetry.
We shall now show that this approach is equivalent to our first calculation,
allowing us to recover the superpotential in Eq.~\eqref{eq:Weff5} as well.
Upon integrating out all heavy quark flavors, our theory reduces to the
IYIT model of dynamical supersymmetry breaking with $N_f = N_c+1$ flavors
and a new effective dynamical scale, $\Lambda_{\textrm{eff}}$, which can be
obtained by matching the running gauge coupling constant at the various
heavy quark mass thresholds, cf.\ Eq.~\eqref{eq:Lambdaefffull},
\begin{align}
\Lambda_{\textrm{eff}} \simeq \Lambda \prod_{n=0}^{N_m}
\left(\frac{M_n + \lambda_n\Phi}{\Lambda}\right)^{1/(2N_c+2)} \,. 
\label{eq:Lambdaeff2}
\end{align}
The only fields left in our theory are then the quark fields $Q^i$
as well as the singlet fields $Z_{ij}$.
Their $R$ charges are constrained as in our previous calculation,
\begin{align}
R[Q^i] = q = 1 - \frac{z}{2} \,, \quad R[Z_{ij}] = z \,,
\end{align}
which results in general in a nonzero coefficient $\mathcal{A}_R^{\textrm{eff}}$
for the $U(1)_R\left[SP(N_c)\right]^2$ gauge anomaly,
\begin{align}
\mathcal{A}_R^{\textrm{eff}} = 2(N_c+1) + 2(N_c+1)(q-1) =
2(N_c+1)\left(1-\frac{z}{2}\right) \,.
\end{align}
Hence, the effective dynamical scale $\Lambda_{\textrm{eff}}$ must be
assigned a spurious $R$ charge $\ell_{\textrm{eff}}$,
\begin{align}
R\left[\Lambda_{\textrm{eff}}\right] = \ell_{\textrm{eff}} =
\frac{\mathcal{A}_R^{\textrm{eff}}}{b_{\textrm{eff}}} = 1-\frac{z}{2}\,, \quad
b_{\textrm{eff}} = 2(N_c+1) \,.
\end{align}
The only gauge-invariant term possibly appearing in $W_{\textrm{eff}}$
therefore corresponds to $\Lambda_{\textrm{eff}}^2 \,Z_{ij}$.
Again contracting the flavor index with the antisymmetric tensor
$J^{ij}/2$ and introducing the canonically normalized field $X$,
we then find for the effective superpotential of the IYIT model,
\begin{align}
W_{\textrm{eff}} \simeq \lambda' \sqrt{N_c + 1} \,\Lambda_{\textrm{eff}}^2 \,X \,,
\quad X = \frac{1}{2\sqrt{N_c+1}} J^{ij}Z_{ij} \,,
\end{align}
with $\lambda'$ being an undetermined numerical proportionality constant.
Inserting our expression for $\Lambda_{\textrm{eff}}$ in Eq.~\eqref{eq:Lambdaeff2}
finally leads us once more to the effective superpotential in Eq.~\eqref{eq:Weff5}.
No matter how we turn it or where we begin our calculation, we hence always find
the same superpotential for our $SP(N_c)$ gauge theory with $N_f = N_c+2+N_m$
flavors; and each time we arrive at the conclusion that the inflaton field possesses
a monomial potential featuring a fractional power $p$ the exact value of which
depends on the number of colors and flavors of our theory.


Finally, in order to conclude this section, we comment on the possibility
that an anomaly-free or anomalous $R$ symmetry is indeed realized
as an actual symmetry in our theory.
Note that, in the above derivation of the effective superpotential, we
merely employed $R$ symmetry as a tool and did not care about any specific
values of our $R$ charges, as they might result from the structure of a
possible UV completion of our theory.
Now we however turn to $R$ symmetry as an ingredient to model building and present
examples of $R$ charge assignments which may possibly be singled out
in a hypothetical high-energy theory.
An interesting special case of Eq.~\eqref{eq:Rchargesqpy}, for instance,
corresponds to equal $R$ charges for all singlet and quark fields, respectively,
\begin{align}
q = p = 1 - \frac{x}{2} \,, \quad x = y = z \,.
\end{align}
For given values of $N_c$ and $N_m$, there thus exists one unique nonanomalous
$R$ charge assignment such that all singlet and quark fields
share equal charges, respectively, cf.\ Eq.~\eqref{eq:Rchargesx},
\begin{align}
q = p = \frac{N_m + 1}{N_c+2+N_m} \,, \quad x = y = z = \frac{2(N_c+1)}{N_c+2+N_m}
\,, \quad \ell = 0 \,.
\end{align}
If this nonanomalous $R$ symmetry was indeed realized in our theory,
none of the singlet fields would carry $R$ charge $0$.
This would render us unable to endow the scalar potential with a shift symmetry
in the direction of any of the scalar fields, which would endanger inflation in the
context of supergravity, cf.\ Sec.~\ref{sec:SUGRA}.
Instead, assuming that our theory does preserve an anomaly-free $R$ symmetry,
we should better take the inflaton charge $x$ to be zero, so that
\begin{align}
q = 0 \,, \quad p = 1 \,, \quad z = 2 \,, \quad y = 1 \,, \quad x = 0 \,, \quad \ell = 0 \,.
\label{eq:SPNR0}
\end{align}
Then the inflaton ends up being the only singlet field carrying zero
$R$ charge, which would explain why only this field corresponds to a flat
direction in the scalar potential, i.e.\ why it is exactly the field $\Phi$
that plays the role of the inflaton.
Moreover, zero $R$ charge for the inflaton has the
further advantage that, simultaneously, the $R$ charges of
the masses $M_a$ vanish.
On the other hand, depending on the origin of these masses in the high-energy theory,
it might also be completely natural that the parameters $M_a$ are spurion fields
with nonzero $R$ charges.


\section{Models based on \texorpdfstring{\boldmath{$SO(10)$}}{SO(10)} dynamics}
\label{sec:SO10}


\subsection[Massless matter fields only (fractional power \texorpdfstring{$p=14/11$}{p = 14/11})]
{Massless matter fields only (fractional power \texorpdfstring{\boldmath{$p=14/11$}}{p = 14/11})}
\label{subsec:SO10massless}


The general DCI recipe can also be readily
applied to theories featuring gauge groups other than $SP(N_c)$.
A potential alternative are for instance models based on $SO(N_c)$
gauge dynamics including matter fields in the spinor, antispinor and vector
representations of $SO(N_c)$.
According to the first step of our general algorithm, it is important
that, in the low-energy effective regime of our theory, supersymmetry is
dynamically broken.
If we restrict ourselves to $SO(N_c)$ theories with only spinor matter
fields and vanishing superpotential, only one candidate DSB model remains,
namely the chiral $SO(10)$ theory with one spinor representation~\cite{Affleck:1984mf}.
All other $SO(N_c)$ theories of the same type can be shown to exhibit flat directions
in moduli space, along which supersymmetry is preserved.
The ground state of the SUSY-breaking $SO(10)$ theory is located
around the origin of field space.
In this vacuum, the theory is strongly coupled and does not admit a semiclassical
description in terms of an effective K\"ahler and an effective
superpotential.
By contrast to the IYIT model discussed in Sec.~\ref{subsec:SPNcmassless},
we are therefore unable to precisely calculate the vacuum energy
density at low energies.
This is, however, not a problem as we are mostly interested in the functional
dependence of the scalar potential on the inflaton field value and not as much on
the precise value of its magnitude.
In the context of nonsupersymmetric gauge theories, we could not even be
certain whether a nonzero vacuum energy density is generated at all.
Now, we know at least for sure that, in consequence of spontaneous $R$ symmetry breaking,
supersymmetry is broken at low energies, resulting in a vacuum energy density of 
$\mathcal{O}\left(\Lambda_{\textrm{eff}}^4\right)$,
\begin{align}
V \simeq C \, \Lambda_{\textrm{eff}}^4 \,,
\label{eq:SO10V}
\end{align}
where $C$ is some $\mathcal{O}(1)$ factor accounting for non-calculable
strong coupling effects.


Proceeding with implementing our general DCI recipe, we must now ask which
s-confining $SO(10)$ theories can be mutated into the 
DSB model with only one spinor representation.
According to the general analysis presented in Ref.~\cite{Csaki:1996sm},
there are exactly seven s-confining $SO(10)$ gauge theories,
differing from each other in terms of their matter content,
cf.\ Tab.~\ref{tab:SO10theories}.
In order to remove representations from these theories, we can always
equip spinor-antispinor pairs, $\mathbf{16}\,\overline{\mathbf{16}}$, as
well as vector squares, $\mathbf{10}\,\mathbf{10}$, with large mass terms.
Therefore, given that we ought to retain exactly one spinor representation,
only theories with one spinor more than antispinors come into question.
There is, however, only one such theory, namely theory \textnumero~5 in
Tab.~\ref{tab:SO10theories}.
This theory contains two spinors, $S_0$ and $S_1$, one antispinor
$\bar{S}_1$ as well as three vectors, $Q_1$, $Q_2$ and $Q_3$.
By providing the fields $S_1$, $\bar{S}_1$ and all of the $Q_i$ with mass terms,
this model reduces to the $SO(10)$ gauge theory with one spinor representation $S_0$,
which is nothing but the DSB model based on $SO(10)$.
It is therefore aptly suited to accommodate a realization of our DCI mechanism.


\begin{table}
\centering
\begin{tabular}{c||ccccccc}
$SO(10)$ theory & \textnumero~1 & \textnumero~2 & \textnumero~3 &
\textnumero~4 & \textnumero~5 & \textnumero~6 & \textnumero~7 \\\hline\hline
$\left(\#\,\mathbf{16},\#\,\overline{\mathbf{16}},\#\,\mathbf{10}\right)$ &
$(4,0,1)$ & $(3,0,3)$ & $(2,0,5)$ & $(3,1,1)$ & $(2,1,3)$ & $(1,1,5)$ & $(2,2,1)$
\end{tabular}
\caption{The complete set of s-confining $SO(10)$ theories according to
Ref.~\cite{Csaki:1996sm}.
As indicated, the different theories feature different numbers of matter fields
in the spinor ($\mathbf{16}$), antispinor ($\overline{\mathbf{16}}$) and
vector representation ($\mathbf{10}$).}
\label{tab:SO10theories}
\end{table}


According to the second step of our general algorithm, the mass deformations mutating
the s-confining $SO(10)$ theory into the DSB model must be associated
with large values of the inflaton field $\Phi$.
We therefore introduce the following tree-level superpotential,
\begin{eqnarray}
\label{eq:SO10}
W_{\textrm{tree}} = \Phi \left(\lambda_0\, S_1\bar{S}_1 +
\lambda_1\,Q_1 Q_1 + \lambda_2\,Q_2 Q_2 + \lambda_3\,Q_3 Q_3\right) \,,
\end{eqnarray}
which equips all fields except for the spinor $S_0$ with an effective
$\Phi$-dependent mass.
During inflation, at energies below the heavy quark mass thresholds, the
theory thus reduces to the DSB model with
vacuum energy density of $\mathcal{O}\left(\Lambda_{\textrm{eff}}^4\right)$.
Here, the effective dynamical scale $\Lambda_{\textrm{eff}}$ is related to
the fundamental dynamical scale $\Lambda$ just as in Eq.~\eqref{eq:Lambdaefffull},
the only difference being that now we do not have any explicit mass terms
for the matter fields,
\begin{align}
\Lambda_{\textrm{eff}} \simeq \Lambda \prod_{n=0}^{3}
\left(\frac{\lambda_n\Phi}{\Lambda}\right)^{(b_n-b_{n+1})/b_{\textrm{eff}}}
= \Lambda \left(\frac{\lambda\,\Phi}
{\Lambda}\right)^{(b_{\textrm{eff}}-b)/b_{\textrm{eff}}} \,, \quad
\lambda = \prod_{n=0}^{3}\lambda_n^{(b_n-b_{n+1})/(b_{\textrm{eff}}-b)} \,.
\label{eq:LambdaeffSO10}
\end{align}
For an $SO(2N)$ gauge theory with $N_s$ fields in the spinor \textit{or}
antispinor representation and $N_v$ fields in the vector representation,
the beta-function coefficient $b$ is given as
\begin{align}
b = 3\,(2N-2) - 2\,N_s - N_v \,.
\label{eq:SO10b}
\end{align}
In our specific case, $N=5$, $N_s = 3$ and $N_v = 3$, this translates into the
following coefficients $b_n$,
\begin{align}
b_4 \equiv b = 15 \,, \quad b_3 = 16 \,, \quad b_2 = 17 \,, \quad b_1 = 18 \,, \quad
b_0 \equiv b_{\textrm{eff}} = 22 \,.
\end{align}
The scalar potential of the low-energy effective theory is therefore once more
a monomial potential for the inflaton field featuring a fractional power,
cf.\ Eq.~\eqref{eq:SO10V},
\begin{align}
\label{eq:potSO(10)}
V \simeq C \,\Lambda_{\rm eff}^4 \simeq C\, \Lambda^4
\left(\frac{\lambda\left|\phi\right|}{\Lambda}\right)^{p} \,, \quad
p = \frac{4\left(b_{\textrm{eff}}-b\right)}{b_{\textrm{eff}}} = \frac{14}{11} \,.
\end{align}
Interestingly enough, this potential mimics the potential of the DCI model based
on $SP(10)$ dynamics with $1 + N_m = 7$ heavy quark flavors coupling
to the inflaton field, cf.\ Eq.~\eqref{eq:pSPNc}.


The scalar potential in Eq.~\eqref{eq:potSO(10)} is only valid at large values
of the inflaton field, where all matter fields except one spinor perturbatively
decouple.
At small inflaton field values, the dynamics of our model turn out to be more
complicated.
In contrast to the s-confining $SP(N_c)$ theories considered in Sec.~\ref{sec:SPN},
the effective degrees of freedom at low energies now not only correspond to mesons
($S\bar{S}$, $Q^2$), but also encompass baryons ($S^2 Q$, $\bar{S}^2 Q$) as well as
further exotic bound states ($S^2\bar{S}^2$, $S^4$, $S\bar{S}Q^2$, $S^3\bar{S}Q$, $S^2Q^3$,
$S^2\bar{S}^2Q^2$, $S^3\bar{S}Q^3$)~\cite{Csaki:1996sm}.
Therefore, in order to stabilize all composite fields at their origin,
we need to introduce a multitude of singlet fields coupling to the
effective degrees of freedom in the tree-level superpotential.
Schematically, we envision a superpotential of the following form,
\begin{align}
W_{\textrm{tree}} = \sum_{m=2}^7 \lambda_m
\left(\frac{\Lambda}{M_{\textrm{Pl}}}\right)^{m-2}
\Lambda\,Z_m \, \mathcal{O}_m \,,
\label{eq:WSO10Z}
\end{align}
where $\mathcal{O}_2$, $\mathcal{O}_3$, .. stand for the various meson operators,
baryon operators and so on.
The mesons are hence expected to have masses of $\mathcal{O}(\Lambda)$,
whereas the baryons and exotic bound states should have comparatively smaller
masses which are suppressed compared to the dynamical scale $\Lambda$
by at least one power of the ratio $\Lambda/M_{\textrm{Pl}}$.
With such light states in the inflaton sector, the thermal
history of the universe after the end of inflation can easily become very involved,
depending on how exactly the inflaton sector is coupled to the SM sector.
Because of this, a more detailed study of the reheating process after inflation is 
definitely imperative to fully capture the phenomenological implications of our
$SO(10)$ model. 
Such an analysis is, however, necessarily model-dependent and therefore
beyond the scope of this paper.
We leave a more thorough study of the low-energy dynamics of our $SO(10)$
scenario after inflation to future work.
In particular, we do not attempt to calculate the inflaton potential
in the small-field regime.
On general grounds, we expect it to be quadratic in the inflaton field
around the origin in field space.
Beyond that, any more concrete statement requires a further specification of
the superpotential in Eq.~\eqref{eq:WSO10Z} as well as a better understanding
of the K\"ahler potential.


\subsection[Massless and massive matter fields
(fractional powers \texorpdfstring{$p \geq 14/11$}{p >= 14/11})]
{Massless and massive matter fields
(fractional powers \texorpdfstring{\boldmath{$p \geq 14/11$}}{p >= 14/11})}


A power of $p=14/11$ is not the only fractional power we may generate in the
inflaton potential by means of $SO(10)$ dynamics.
Just as in the $SP(N_c)$ case, the above outlined DCI model based on $SO(10)$
can be easily generalized to larger values of $p$ 
by allowing for further matter fields with supersymmetric masses
$M_a$ above the dynamical scale, cf.\ Eq.~\eqref{eq:tree2}.
The discussion of this generalized $SO(10)$ scenario is completely
analogous to the discussion of our $SP(N_c)$ models with $N_m$
additional quark flavors, cf.\ Sec.~\ref{subsec:SPNcmassive},
which is why here we only state our final results.
Consider that on top of the matter content of theory
\textnumero~5 in Tab.~\ref{tab:SO10theories} we still have
$N_m^s/2$ spinors, $N_m^s/2$ antispinors and $N_m^v$ vectors,
where $N_m^s=0,2,4,..$ and $N_m^v= 0,1,2,..$, all of which possess
supersymmetric masses $M_a$ above the dynamical scale.
The beta-function coefficients of the high- and low-energy theories,
$b$ and $b_{\textrm{eff}}$, then read, cf.~\eqref{eq:SO10b},
\begin{align}
b = 3\, (2N-2) - 2 \left(N_s + N_m^s\right) - \left(N_v + N_m^v\right)
= 15 - 2 \, N_m^s - N_m^v \,, \quad b_{\textrm{eff}} = 22 \,,
\end{align}
which eventually leads to the following power in the inflaton potential
in the large-field regime,
\begin{align}
p = \frac{2\left(7 + 2\,N_m^s + N_m^v\right)}{11} =
\frac{2\left(7 + N_m^{\textrm{eff}}\right)}{11}\,,
\label{eq:pSO10}
\end{align}
where we have introduced $N_m^{\textrm{eff}} = 2N_m^s + N_m^v$ as the
effective number of additional matter fields.
Similarly as in the $SP(N_c)$ case, we require our $SO(10)$ theory
to be (almost) asymptotically free at high energies, which is why
we again restrict the coefficient $b$ to nonnegative values.
According to Eq.~\eqref{eq:pboundSPN}, the condition  $b \geq 0$ then
implies that the power $p$ cannot be larger than $4$.
This bound is now saturated for effectively $N_m^{\textrm{eff}} = 15$
additional flavors coupling to the inflaton field.
Furthermore, we point out that, by comparing Eq.~\eqref{eq:pSO10}
with Eq.~\eqref{eq:pNm}, one immediately finds that the DCI models based on
$S0(10)$ with $N_m^{\textrm{eff}}$ extra quark fields yield
the same predictions for the power $p$ as $SP(10)$ models
with $N_m' \equiv N_m^{\textrm{eff}} + 6$ extra quark fields.
The following two sets of DCI scenarios are hence equivalent
in the sense that they give rise to the same inflationary dynamics
at inflaton field values above all heavy quark mass thresholds,
\begin{align}
SP(10) \textrm{ with } N_f = N_c + 2 + N_m' \quad \leftrightsquigarrow \quad
SO(10) \textrm{ with } N_s = 3 + N_m^s \,, \:\: N_v = 3 + N_m^v  \,,
\label{eq:SPSOequiv}
\end{align}
with $N_m' = 2N_m^s + N_m^v + 6$.
Meanwhile, the low-energy dynamics of two such equivalent models
might look very different due to the presence of the light composite states
in the $SO(10)$ theory.


The full scalar potential of our generalized DCI models based on $SO(10)$ is given by,
\begin{align}
V \simeq C \, \Lambda^4
\left(\frac{\lambda_0\left|\phi\right|}{\Lambda}\right)^{14/11}
\prod_{s=1}^{N_m^s/2}
\left|\frac{M_s + \lambda_s\,\phi}{\Lambda}\right|^{8/11}
\prod_{v=1}^{N_m^v}
\left|\frac{M_v + \lambda_v\,\phi}{\Lambda}\right|^{2/11} \,,
\label{eq:VtotSO10}
\end{align}
where $C$ is an undetermined proportionality constant and with $\lambda_0$ corresponding
to the coupling $\lambda$ in Eq.~\eqref{eq:LambdaeffSO10}.
The $N_m^s/2$ mass parameters $M_s$ are the masses of the $N_m^s/2$
additional spinor-antispinor pairs, while the $N_m^v$ mass parameters
$M_v$ denote the masses of the $N_m^v$ extra vector fields.
In the large-field regime, this scalar potential reduces to
\begin{align}
V \simeq C \, \Lambda^4
\left(\frac{\lambda\left|\phi\right|}{\Lambda}\right)^p \,, \quad
\lambda = \prod_m\lambda_m^{(b_m-b_{m+1})/(b_{\textrm{eff}}-b)}
\,, \quad p = \frac{2\left(7 + 2\,N_m^s + N_m^v\right)}{11} \,,
\end{align}
where the product runs over all $4+N_m^s+N_m^v$ Yukawa coupling constants
$\lambda_m$ in the tree-level superpotential for the inflaton field,
$\lambda_0$, $\lambda_1$, $\lambda_2$, $\lambda_3$, $\lambda_s$, and $\lambda_v$.
In contrast to our $SP(N_c)$ models, we are now unable to reconstruct the vacuum energy
density during inflation simply based on $R$ symmetry arguments, cf.\ Sec.~\ref{subsec:SPNWeff}.
The simple reason for this is that the DSB model based on $SO(10)$ does not
admit an effective low-energy description in terms of a semiclassical K\"ahler potential
and superpotential---there simply is no superpotential to reconstruct.


\subsection[Models based on \texorpdfstring{$SU(5)$}{SU(5)} dynamics
(fractional power \texorpdfstring{$p\geq16/13$}{p >= 16/13})]
{Models based on \texorpdfstring{\boldmath{$SU(5)$}}{SU(5)} dynamics
(fractional power \texorpdfstring{\boldmath{$p\geq16/13$}}{p >= 16/13})}


The DSB model based on $SO(10)$ dynamics, which we
introduced at the beginning of Sec.~\ref{subsec:SO10massless} and which describes
the dynamics of our $SO(10)$ DCI models during inflation, is in fact a
simple generalization of another chiral DSB model based on the gauge
group $SU(5)$.
Shortly before the authors of Ref.~\cite{Affleck:1984mf} presented
their findings regarding the $SO(10)$ model, they had discussed
this alternative model in Ref.~\cite{Affleck:1983vc}.
The $SU(5)$ model contains one antifundamental
representation, $\bar{Q}^0 \sim \mathbf{5}^*$,
as well as one antisymmetric tensor, $A \sim \mathbf{10}$,
and it exhibits two nonanomalous global $U(1)$
symmetries, $U(1)_A$ and $U(1)_R$.
The $SO(10)$ model is obtained from the $SU(5)$ model
in three steps.
All we need to do is to (i) gauge the $U(1)_A$ symmetry,
(ii) add an additional $SU(5)$ singlet to cancel the
$\left[U(1)_A\right]^3$ anomaly and (iii) embed
the resulting $SU(5)\times U(1)_A$ theory into $SO(10)$.
The fact that both the $SO(10)$ as well as the $SU(5)$ theory dynamically
break supersymmetry in their respective ground states has ultimately been
proven in Ref.~\cite{Murayama:1995ng}.
As we shall now demonstrate in this section, the $SU(5)$ model
can be equally employed for the construction of viable DCI
scenarios.
The analysis proceeds exactly along the same lines as the
discussion in the previous sections, which is why we
only briefly summarize our results.


\begin{table}
\centering
\begin{tabular}{c||ccccc}
$SU(5)$ theory & \textnumero~1 & \textnumero~2 & \textnumero~3 &
\textnumero~4 & \textnumero~5 \\\hline\hline
$\left(\#\,\mathbf{5},\#\,\mathbf{5}^*,\#\,\mathbf{10},\#\,\overline{\mathbf{10}}\right)$ &
$(6,6,0,0)$ & $(4,5,1,0)$ & $(3,3,1,1)$ & $(0,3,3,0)$ & $(2,4,2,0)$
\end{tabular}
\caption{The complete set of s-confining $SU(5)$ theories according to
Ref.~\cite{Csaki:1996sm}.
As indicated, the different theories feature different numbers of matter fields
in the fundamental ($\mathbf{5}$), antifundamental ($\mathbf{5}^*$),
antisymmetric tensor ($\mathbf{10}$) and conjugated antisymmetric tensor
representation ($\overline{\mathbf{10}}$).}
\label{tab:SU5theories}
\end{table}


According to the general analysis presented in Ref.~\cite{Csaki:1996sm},
there are exactly five s-confining $SU(5)$ gauge theories,
differing from each other in terms of their matter content,
cf.\ Tab.~\ref{tab:SU5theories}.
Among these five theories, only theory \textnumero~2 can be mutated via mass
deformations into the DSB model based on $SU(5)$.
This theory is hence a unique candidate for the construction of a working
DCI model.%
\footnote{For more details on the dynamics of $SU(N_c)$ gauge theories with
one antisymmetric tensor, $N_f$ fundamentals and $N_f+N_c-4$
antifundamentals, cf.\ Ref.~\cite{Pouliot:1995me}.
Note that, within this class of theories, theory \textnumero~2
in Tab.~\ref{tab:SU5theories} corresponds to the special case of $N_c=5$ and $N_f=4$.}
Note that it cannot be obtained from the corresponding s-confining model
which we use in the case of our $SO(10)$ scenarios.
It is not possible to embed the matter content of theory \textnumero~2
in Tab.~\ref{tab:SU5theories} into the $SO(10)$ representations of theory
\textnumero~5 in Tab.~\ref{tab:SO10theories}.
In order to deform the s-confining theory into the DSB model,
we now introduce a tree-level superpotential, in which we couple four matter flavors
$\left(Q^i,\bar{Q}^i\right)$, where $i=1,2,3,4$, to the inflaton field,
\begin{align}
W_{\textrm{tree}} = \lambda_i \, \Phi \, Q^i \bar{Q}^i \,.
\label{eq:WtreeSU5}
\end{align}
The beta-function coefficients for the $SU(5)$ gauge coupling constant
above and below the heavy quark mass thresholds,
$b$ and $b_{\textrm{eff}}$, are then given as,
\begin{align}
b  = 3 \times 5 - \left(\frac{3}{2} + \frac{1}{2}\right) - 4 = 9 \,, \quad
b_{\textrm{eff}}  = 3 \times 5 - \left(\frac{3}{2} + \frac{1}{2}\right) = 13 \,.
\end{align}
Performing our standard analysis of the scalar potential in the large-field regime,
we then obtain the following inflaton potential,
\begin{align}
V \simeq C\,\Lambda^4 \left(\frac{\lambda\left|\phi\right|}{\Lambda}\right)^p \,,
\quad \lambda = \prod_{i=1}^4 \lambda_i^{1/4} \,, \quad
p = \frac{4\left(b_{\textrm{eff}}-b\right)}{b_{\textrm{eff}}} = \frac{16}{13} \,, 
\label{eq:VSU5}
\end{align}
where $C$ is some undetermined $\mathcal{O}(1)$ factor resulting from the strong
dynamics at low energies.
At small inflaton field values, our theory reaches the s-confinement phase,
where the dynamics are effectively described in terms of the composite fields
$Q\bar{Q}$, $A\bar{Q}^2$, $A^2Q$, $AQ^3$, and $\bar{Q}^5$.
Just as in the case of our $SO(10)$ scenario, these fields partly need to be
stabilized by means of higher-dimensional operators.
The exact form of the inflaton potential close to the true vacuum depends
on the explicit form of these operators.


Finally, we mention that the above model can also be generalized to a larger number
of matter fields coupling to the inflaton.
Adding $N_m$ further pairs of vector-like quarks $\left(P^a,\bar{P}^a\right)$,
where $P^a\sim\mathbf{5}$ and $\bar{P}^a \sim\mathbf{5}^*$, with explicit supersymmetric
masses $M_a$ and a superpotential
\begin{align}
W_{\textrm{tree}} \supset \left(M_a + \lambda_a\Phi\right)P^a\bar{P}^a \,,\quad
\Lambda \lesssim M_a \ll M_{\textrm{Pl}} \,,
\end{align}
the beta-function coefficient $b$ turns into $b = 9-N_m$.
This results in a fractional power $p$ of
\begin{align}
p = \frac{4\left(4 + N_m\right)}{13} \,.
\end{align}
Now, the bound in Eq.~\eqref{eq:pboundSPN}, $p\leq4$, is saturated for $N_m = 9$
additional flavors coupling to the inflaton field.
Comparing this result with our expression for $p$ in Eq.~\eqref{eq:pNm},
we observe that $SU(5)$ DCI models with $N_m$ extra quark fields predict
the same values for the power $p$ as $SP(12)$ DCI models with $N_m' \equiv 2 N_m + 7$
extra quark fields.
In analogy to Eq.~\eqref{eq:SPSOequiv}, we therefore have
\begin{align}
SP(12) \textrm{ with } N_f = N_c + 2 + N_m' \quad \leftrightsquigarrow \quad
SU(5) \textrm{ with } N_f = 4 + N_m \,.
\end{align}
The low-energy dynamics of two such equivalent DCI models are however
different from each other due to the presence of the light composite
states in the $SU(5)$ theory.


\section{Models based on \texorpdfstring{\boldmath{$SU(3)\times SU(2)$}}{SU(3)xSU(2)} dynamics}
\label{sec:32}


\subsection[Massless matter fields only (fractional power \texorpdfstring{$p=8/7$}{p = 8/7})]
{Massless matter fields only (fractional power \texorpdfstring{\boldmath{$p=8/7$}}{p = 8/7})}


So far, we have only considered DCI models based on simple Lie groups.
As we will demonstrate in this section, our general recipe can however also
be easily applied in the case of product groups.
To this end, we shall now construct a DCI scenario based on one of the
simplest chiral DSB models featuring a product group,
namely the so-called 3-2 model presented in Ref.~\cite{Affleck:1984xz}.%
\footnote{For a brief review of this model, cf.\ also Ref.~\cite{Intriligator:2007cp}.}
This model comes with an $SU(3)\times SU(2)$ gauge group and four chiral
superfields $q$, $\bar{u}$, $\bar{d}$ and $\ell$, transforming as follows in the
representations of this group,
\begin{align}
q \sim \left(\mathbf{3},\mathbf{2}\right) \,, \quad
\bar{u} \sim \left(\bar{\mathbf{3}},\mathbf{1}\right) \,, \quad
\bar{d} \sim \left(\bar{\mathbf{3}},\mathbf{1}\right) \,, \quad
\ell \sim \left(\mathbf{1},\mathbf{2}\right) \,.
\label{eq:fields32}
\end{align}
The model hence corresponds to supersymmetric QCD with three colors
and two flavors in combination with a gauged flavor symmetry,
where the additional $SU(2)$ doublet $\ell$ serves the purpose to cancel
the Witten anomaly of the $SU(2)$ theory~\cite{Witten:1982fp}.
In the absence of any tree-level superpotential, the classical moduli space
exhibits three flat directions corresponding to the following three gauge-invariant
composite operators,
\begin{align}
X = q \,\bar{d} \,\ell \,, \quad Y = q \,\bar{u}\,\ell \,, \quad Z = q\,q\,\bar{u}\,\bar{d}\,.
\end{align}
In order to accommodate dynamical supersymmetry breaking in this model,
one introduces a tree-level superpotential for the field $X$,
which lifts all of the classical $D$-flat directions,
\begin{align}
W_{\textrm{tree}} = \kappa \,X = \kappa\, q\,\bar{d}\,\ell \,.
\label{eq:W32kappa}
\end{align}
This superpotential is accompanied by a dynamically generated
superpotential originating from $SU(3)$ instanton effects,
\begin{align}
W = W_{\textrm{dyn}} + W_{\textrm{tree}} \,, \quad
W_{\textrm{dyn}} = \frac{\Lambda_3^b}{Z} \,, \quad b = 3\times3 -2 = 7 \,.
\label{eq:W32}
\end{align}
Here, $b$ denotes the coefficient of the beta function for the $SU(3)$
gauge coupling constant and $\Lambda_3$ is the associated dynamical scale.
The $SU(2)$ gauge dynamics, by contrast, do not yield a contribution to
$W_{\textrm{dyn}}$.
As our set of chiral superfields contains four $SU(2)$ doublets,
i.e.\ two flavors of matter fields in the fundamental representation of $SU(2)$,
the $SU(2)$ interactions rather result in a modified quantum moduli
constraint, similarly as in the case of an $SP(1)$ theory with $N_f = 2$ flavors,
cf.\ Eq.~\eqref{eq:dqmconstraint}.
In the following, we shall assume that the dynamical scale of the
$SU(2)$ theory, $\Lambda_2$, is significantly smaller than $\Lambda_3$.
This then allows us to neglect the moduli constraint induced by
the $SU(2)$ interactions.


The total superpotential in Eq.~\eqref{eq:W32} is now responsible for
the spontaneous breaking of supersymmetry, which can be seen as
follows~\cite{Affleck:1983vc,Affleck:1984xz,Intriligator:2007cp}:
The dynamically generated superpotential requires the field $Z$ to take
a large field value, which spontaneously breaks $R$ symmetry. 
This gives rise to a compact modulus space spanned by the Goldstone
boson of $R$ symmetry breaking.
In the case of unbroken supersymmetry, this $R$-axion would need to
be accompanied by a scalar partner without any potential.
That is, the scalar potential would need to be able to exhibit a flat
direction---which is however not possible due to the tree-level superpotential.
Thus, supersymmetry must be broken.
The vacuum energy density $V$ as well as the typical field values $v$
in the SUSY-breaking vacuum can be estimated
as follows~\cite{Affleck:1984xz},
\begin{align}
V \simeq \kappa^{10/7} \,\Lambda_3^4 \,, \quad v \simeq \frac{\Lambda_3}{\kappa^{1/7}} \,.
\label{eq:V32}
\end{align}
This expression relies on the assumption of a small value for the
coupling constant $\kappa$, such that the SUSY-breaking vacuum
is located at field values $v \gg \Lambda_3, \Lambda_2$.
Only in this situation, the theory is weakly coupled and the
K\"ahler potential can be approximated by the canonical classical
K\"ahler potential for the fundamental matter fields.%
\footnote{Note that requiring large field VEVs, $v \gg \Lambda_3, \Lambda_2$,
in combination with the expression for $v$ in Eq.~\eqref{eq:V32} implies that
$\Lambda_3 \gg \kappa^{1/7} \Lambda_2$.
This is consistent with our assumption that $\Lambda_2 \ll \Lambda_3$,
which led to Eq.~\eqref{eq:V32} in the first place.}


In order to construct a DCI scenario featuring the above outlined
dynamics at low energies, we need to find an s-confining theory which can
be mutated into the 3-2 model.
To do so, we shall exploit the fact that the $SU(2)$ theory with three flavors is
s-confining.
In addition to the four chiral superfields in Eq.~\eqref{eq:fields32},
let us therefore add six more chiral superfields transforming as follows
under the $SU(3) \times SU(2)$ gauge group,
\begin{align}
U \sim \left(\mathbf{3},\mathbf{1}\right) \,, \quad
D \sim \left(\mathbf{3},\mathbf{1}\right) \,, \quad
\bar{U} \sim \left(\bar{\mathbf{3}},\mathbf{1}\right) \,, \quad
\bar{D} \sim \left(\bar{\mathbf{3}},\mathbf{1}\right) \,, \quad
L \sim \left(\mathbf{1},\mathbf{2}\right) \,, \quad
\bar{L} \sim \left(\mathbf{1},\mathbf{2}\right) \,,
\label{eq:32extra}
\end{align}
so that now four flavors of matter fields participate in the
$SU(3)$ interactions.
Below the dynamical scale of the $SU(3)$ gauge interactions,
the model is described in terms of 24 color-singlet hadrons:
four baryons $B^I$, four antibaryons $\bar{B}^I$
and 16 mesons $M^{IJ}$, where $I = 1,..,4$,
\begin{align}
B = \begin{pmatrix} q_2\,UD \\ q_1\,UD \\ q_1q_2\,D \\ q_1q_2\, U \end{pmatrix} \,, \quad
\bar{B} = \begin{pmatrix} \bar{d}\,\bar{U}\bar{D} \\ \bar{u}\,\bar{U}\bar{D} \\
\bar{u}\bar{d}\,\bar{D} \\ \bar{u}\bar{d}\, \bar{U} \end{pmatrix} \,, \quad
M = \begin{pmatrix}
q_1\bar{u} & q_1\bar{d} & q_1\bar{U} & q_1\bar{D} \\
q_2\bar{u} & q_2\bar{d} & q_2\bar{U} & q_2\bar{D} \\
U\bar{u} & U\bar{d} & U\bar{U} & U\bar{D} \\
D\bar{u} & D\bar{d} & D\bar{U} & D\bar{D} \end{pmatrix} \,, \quad
q = \begin{pmatrix} q_1 \\ q_2 \end{pmatrix} \,,
\label{eq:32BBM}
\end{align}
as well as by the three $SU(2)$ doublets $\ell$, $L$ and $\bar{L}$.
Out of these 30 color-neutral fields, 16 fields combine to form eight $SU(2)$ doublets:
$qUD$, $q\bar{u}$, $q\bar{d}$, $q\bar{U}$, $q\bar{D}$,
$\ell$, $L$ and $\bar{L}$.
The doublets $q\bar{d}$ and $\ell$ receive masses due to the tree-level
superpotential in Eq.~\eqref{eq:W32}.
Here, we assume that the coupling constant $\kappa$ is sufficiently
large, so that $\ell$ and $q\bar{d}$ are always stabilized at zero
at energies below the dynamical scale $\Lambda_3$.
Note that this may require some tuning of $\kappa$, as the coupling $\kappa$
must also not be too large since otherwise corrections to the K\"ahler
potential may no longer be under control, cf.\ our comment below
Eq.~\eqref{eq:V32}.
After integrating out the fields $q\bar{d}$ and $\ell$, only
six $SU(2)$ doublets remain, such that at low energies our model eventually
turns into  an $SU(2)$ theory with three flavors.
If the separation between the dynamical scales of the $SU(3)$ and the $SU(2)$
interactions is large enough, this theory is s-confining.
Otherwise, the dynamical superpotential associated with the
$SU(3)$ dynamics may affect the $SU(2)$ gauge dynamics
and the low energy theory may not exhibit a phase of s-confinement.
However, as long as $\Lambda_3 \gg \Lambda_2$, which is what
we assume, the dynamical superpotential originating
from the $SU(3)$ interactions is guaranteed not to cause any problems.
In this case, our model then reaches an s-confined phase
at low energies.
In order to lift all flat directions present in the s-confined phase,
we finally need to add further singlet fields coupling to the
light $SU(2)$ flavors, similarly as we have done it in the case of
our DCI scenarios based on $SO(10)$.
We do not specify the concrete form of the superpotential responsible for
stabilizing all flat directions and leave a more detailed study of the
low-energy dynamics of our $SU(3) \times SU(2)$ model for future work.
For now, we shall merely focus on the high-energy dynamics during inflation.


The last step in constructing our DCI model based on $SU(3)\times SU(2)$
gauge dynamics is to identify the flow of the s-confining theory
to the DSB model with large inflaton field values.
Analogously to our procedure in Secs.~\ref{sec:SPN} and \ref{sec:SO10},
this is done by providing the extra flavors
introduced in Eq.~\eqref{eq:32extra} with inflaton-dependent supersymmetric
mass terms,
\begin{align}
W_{\textrm{tree}} \rightarrow W_{\textrm{tree}} = \kappa \, q \bar{d} \, \ell +
\Phi \left(\lambda_1 \,U\bar{U} + \lambda_2 \,D \bar{D} + \lambda_3 \,L \bar{L} \right) \,.
\label{eq:Wtree32}
\end{align}
Large inflaton field values therefore render the fields $U$, $\bar{U}$,
$D$, $\bar{D}$, $L$ and $\bar{L}$ heavy and thus mutate the s-confining theory
with in total elementary 30 fields into the DSB model
that we initially started out with in our discussion.
During inflation, the dynamical scale associated with the $SU(3)$
interactions, $\Lambda \equiv \Lambda_3$, consequently needs to be replaced
with an effective dynamical scale $\Lambda_{\textrm{eff}}$.
A calculation along exactly the same lines as in the previous two
sections yields
\begin{align}
\Lambda_{\textrm{eff}} \simeq \Lambda \left(\frac{\lambda\,\Phi}{\Lambda}
\right)^{(b_{\textrm{eff}} - b)/b_{\textrm{eff}}} \,, \quad
\lambda = \sqrt{\lambda_1\lambda_2} \,,
\label{eq:Lambdaeff32}
\end{align}
where we relied on our assumption of a large hierarchy between the scales $\Lambda_3$ and
$\Lambda_2$, ensuring that the running of the $SU(3)$ gauge coupling constant
is not significantly affected by the $SU(2)$ dynamics.
The beta-function coefficients above and below the heavy quark mass thresholds,
$b$ and $b_{\textrm{eff}}$, are given by the usual $SU(3)$ expressions for
four and two flavors, respectively,
\begin{align}
b = 3 \times 3 -4 = 5 \,, \quad 
b_{\textrm{eff}} =  3 \times 3 -  2  = 7 \,.
\label{eq:bbeff32}
\end{align}
Combining Eqs.~\eqref{eq:V32}, \eqref{eq:Lambdaeff32} and \eqref{eq:bbeff32},
we then finally obtain the following inflaton potential,
\begin{align}
V \simeq \kappa^{10/7} \, \Lambda_{\textrm{eff}}^4  \simeq \kappa^{10/7} \,
\Lambda^4 \left(\frac{\lambda\left|\phi\right|}{\Lambda}\right)^p \,, \quad
p = \frac{4\left(b_{\textrm{eff}}-b\right)}{b_{\textrm{eff}}} = \frac{8}{7} \,.
\label{eq:Vtot32}
\end{align}
The same fractional power as in this inflaton potential is also
generated in the DCI scenario based on $SP(6)$ with
in total $1+N_m = 4$ quark flavors coupling to the inflaton field,
cf.\ Eq.~\eqref{eq:pNm}.


\subsection[Massless and massive matter fields
(fractional powers \texorpdfstring{$p \geq 8/7$}{p >= 8/7})]
{Massless and massive matter fields
(fractional powers \texorpdfstring{\boldmath{$p \geq 8/7$}}{p >= 8/7})}


Just as in the case of our $SP(N_c)$ and $SO(10)$ scenarios,
also the DCI model based on the gauge group $SU(3)\times SU(2)$
can be generalized to a larger number of matter fields coupling
to the inflaton.
Let us supplement the field content of our s-confining theory
with $N_m$ pairs of quark and antiquark fields transforming in the
representations $\left(\mathbf{3},\mathbf{1}\right)$ and 
$\left(\bar{\mathbf{3}},\mathbf{1}\right)$, respectively.
Let us further equip all of these $N_m$ new quark flavors with
supersymmetric masses $M_a$ above the dynamical scale $\Lambda_3$ and
couple them to the inflaton field in the same way as the fields $U$, $\bar{U}$,
$D$ and $\bar{D}$ in Eq.~\eqref{eq:Wtree32}.
This eventually leads us to an inflaton potential of the very same form as
the potentials in Eqs.~\eqref{eq:Vtot} and \eqref{eq:VtotSO10},
\begin{align}
V \simeq \kappa^{10/7} \, \Lambda^4
\left(\frac{\lambda_0\left|\phi\right|}{\Lambda}\right)^{8/7}
\prod_{a=1}^{N_m} \left|\frac{M_a + \lambda_a'\phi}{\Lambda}\right|^{4/7} \,,
\label{eq:V32prod}
\end{align}
where $\lambda_0$ corresponds to the coupling $\lambda$
in Eq.~\eqref{eq:Vtot32}.
In the large-field regime, this scalar potential reduces to
a simple monomial featuring a fractional power $p$,
\begin{align}
V \simeq \kappa^{10/7} \, \Lambda^4 \left(\frac{\lambda\left|\phi\right|}{\Lambda}\right)^p
\,, \quad \lambda = \left(\lambda_1\lambda_2\prod_{a=1}^{N_m} \lambda_a'\right)^{1/(2+N_m)}
\,, \quad p = \frac{4\left(2+N_m\right)}{7} \,.
\label{eq:V32tot}
\end{align}
The bound on the power $p$ in Eq.~\eqref{eq:pboundSPN} is hence saturated for $N_m = 5$
additional $SU(3)$ flavors.
Besides that, the comparison between the expressions for $p$
in Eqs.~\eqref{eq:V32tot} and \eqref{eq:pNm} reveals
that models based on the product group $SU(3)\times SU(2)$
with $N_m$ extra quark fields yield the same prediction for the power $p$
as $SP(6)$ models with $N_m' \equiv 2 N_m + 3$ extra quark fields.
In addition to Eq.~\eqref{eq:SPSOequiv}, we therefore find a second
equivalence relation between different DCI models,
\begin{align}
SP(6) \textrm{ with } N_f = N_c + 2 + N_m' \quad \leftrightsquigarrow \quad
SU(3) \times SU(2) \textrm{ with } N_f = 4 + N_m \,.
\end{align}
Again, the low-energy dynamics of two such equivalent DCI models might look very
different due to the presence of the light composite states in the
$SU(3)\times SU(2)$ theory.


In the above analysis, we have consistently assumed that $\Lambda_3 \gg \Lambda_2$,
which implies that the dominant contribution to the dynamical breaking of
supersymmetry arises from the $SU(3)$ dynamics, while the effects
of the $SU(2)$ dynamics on the vacuum energy density are negligible.
In passing, we mention that also the opposite case, $\Lambda_2 \gg \Lambda_3$,
provides the basis for the construction of consistent DCI scenarios.
In this case, the dynamics of supersymmetry breaking are no longer governed by
the $SU(3)$ instanton effects encoded in $W_{\textrm{dyn}}$, cf.\ Eq.~\eqref{eq:W32},
but rather by the deformed moduli constraint on the $SU(2)$
moduli space.
The vacuum energy density then turns out to be $V \simeq \kappa^2 \Lambda_{\textrm{eff}}^4$,
with $\Lambda_{\textrm{eff}}$ being the effective dynamical scale of the 
$SU(2)$ interactions below all heavy quark mass thresholds.
For a total of $1 + N_m$ $SU(2)$ flavors coupling to the inflaton field,
this results in a monomial inflaton potential featuring a power
$p = 1 + N_m$.
This expression is consistent with our results for DCI models based on $SP(1)$,
cf.\ Eq.~\eqref{eq:pNm}, as well as our discussion of general $SU(N_c)$
DCI models in App.~\ref{app:SUN}, cf.\ Eq.~\eqref{eq:pSUN}.


\subsection[Effective superpotential from \texorpdfstring{$R$}{R} symmetry]
{Effective superpotential from \texorpdfstring{\boldmath{$R$}}{R} symmetry}


Last but not least, we point out that, just like before, our results
in Eqs.~\eqref{eq:V32prod} and \eqref{eq:V32tot} can also be derived
employing arguments based on $R$ symmetry. 
Taking into account the constraints on the $R$ charges of our matter fields
imposed by the tree-level superpotential, the only remaining independent charges
are the charge of the inflaton field, $x = R\left[\Phi\right]$, as well as the
charges of the fields $q$, $\bar{u}$ and $\bar{d}$,
$q_q = R[q]$, $q_{\bar{u}} = R[\bar{u}]$ and $q_{\bar{d}} = R[\bar{d}]$.
The $2+N_m$ flavors coupling to the inflaton as well as the $SU(2)$ doublets
$L$ and $\bar{L}$ all carry $R$ charge $1-x/2$, while the $R$ charge of the $SU(2)$
doublet $\ell$ is constrained by the tree-level term in Eq.~\eqref{eq:W32kappa},
$R[\ell] = 2 - q_q - q_{\bar{d}}$.
With this charge assignment, the coefficient $\mathcal{A}_R$ of the
$U(1)_R\left[SU(3)\right]^2$ anomaly is then computed to be
\begin{align}
\mathcal{A}_R = & \: 6 + 2(q_q - 1) + (q_{\bar{u}} - 1) + (q_{\bar{d}} - 1)
+ 2\left(-\frac{x}{2}\right)\left(2+N_m\right) \\
= & \: 2 + 2q_q + q_{\bar{u}} +
q_{\bar{d}} - x\,(2 + N_m) \,, \nonumber
\end{align}
which endows the scale $\Lambda \equiv \Lambda_3$ with an
$R$ charge $\ell = R[\Lambda] = \mathcal{A}_R / b$.
The only gauge-invariant term possibly appearing in the dynamical
superpotential is then of the following form,
\begin{align}
\frac{\Lambda^b \,\Phi^{2+N_m}}{q\,q\,\bar{u}\,\bar{d}} =
\frac{\Lambda^b \,\Phi^{2+N_m}}{Z} =
\frac{\Lambda^7}{Z} \left(\frac{\Phi}{\Lambda}\right)^{2+N_m}
\,, \quad b = 5 - N_m \,.
\end{align}
From this point on, the further calculation is analogous to the
analysis of the DSB model based on $SU(3)\times SU(2)$.
Eventually, it leads to the following superpotential,
\begin{align}
W = W_{\textrm{dyn}} + W_{\textrm{tree}} \,, \quad
W_{\textrm{dyn}} = \frac{\Lambda_{\textrm{eff}}^7}{Z} \,, \quad
\Lambda_{\textrm{eff}} = \Lambda \left(\frac{\lambda_0\Phi}{\Lambda}\right)^{2/7}
\prod_{a=1}^{N_m}\left(\frac{M_a + \lambda_a'\Phi}{\Lambda}\right)^{1/7} \,,
\end{align}
with $W_{\textrm{tree}}$ being given in Eq.~\eqref{eq:Wtree32},
as well as to the very same scalar potential as in Eq.~\eqref{eq:V32prod}.
In the large-field limit, we therefore find again the monomial potential
in Eq.~\eqref{eq:V32tot} featuring the fractional power
$p = 4/7(2+N_m)$.


\section{Dynamical chaotic inflation in supergravity}
\label{sec:SUGRA}


In the last three sections, we constructed various DCI models, all of which
eventually led to scalar inflaton potentials which, at large field values, 
were roughly of the following form,%
\begin{align}
V \simeq V_0 = \Lambda^4 \left(\frac{\lambda\left|\phi\right|}{\Lambda}\right)^p \,.
\label{eq:Vpropto}
\end{align}
In principle, the potential $V_0$ still needs to be multiplied by
a numerical $\mathcal{O}(1)$ factor $C$, which increases the number
of effective free parameters describing the potential from two ($\Lambda,\lambda$)
to three ($\Lambda,\lambda,C$), cf.\ our explicit results for the potential
in Eqs.~\eqref{eq:Vtot}, \eqref{eq:VtotSO10}, \eqref{eq:VSU5} and \eqref{eq:V32tot}.
In the following, we shall however omit the factor $C$ for simplicity.
A generalization of our subsequent analysis to the case of a scaled inflaton
potential, $V \simeq C \,V_0$, is straightforward.%
\footnote{Our choice of setting $C$ to $1$ can also be thought of as a
redefinition of the coupling  constant, $\lambda \rightarrow C^{1/p}\lambda$.
Note however that this redefinition affects the mass of the inflaton field in the
true vacuum, $m_\phi \sim \lambda\Lambda$, such that setting $C=1$ corresponds
indeed to a special choice and cannot be simply assumed w.l.o.g.}
In the last sections, we succeeded in developing a dynamical mechanism
for the generation of potentials such as the one in
Eq.~\eqref{eq:powerlawV}.
On the other hand, we have neglected any gravitational corrections to the
inflaton potential up to now.
A careful examination of these corrections is, however, mandatory
if we really intend to demonstrate that our DCI models are capable of
providing the basis for consistent scenarios of cosmic inflation.
Thus, we shall now discuss under which circumstances our DCI models
can be successively coupled to supergravity.


Inflation based on the potential $V_0$ is characterized by the slow-roll
motion of the inflaton field at very large field values.
This can be quantified in terms of the slow-roll conditions, which are only satisfied
for field values around or larger than $\left|\phi_0\right| \simeq p\, M_{\textrm{Pl}}$.
Unless $p$ is extremely small, the inflaton consequently exceeds the
Planck scale $M_{\textrm{Pl}}$ for almost the entire period of inflation.
In supergravity, the scalar potential $V_0$ picks up an exponential
prefactor, $\exp\left[K/M_{\textrm{Pl}}^2\right]$, where $K$ is the K\"ahler potential.
If we coupled our DCI models to supergravity merely assuming a canonical
K\"ahler potential for the inflaton field, $K = \left|\phi\right|^2$,
we would therefore quickly encounter a too steep potential,
\begin{align}
V \supset \exp\left[K/M_{\textrm{Pl}}^2\right] V_0 =
\exp\left[\left|\phi\right|^2/M_{\textrm{Pl}}^2\right] V_0 = 
V_0 + 3H_0^2\left|\phi\right|^2 + .. \,, \quad
H_0^2 = \frac{V_0}{3M_{\textrm{Pl}}^2} \,.
\label{eq:VHmass}
\end{align}
In the case of a canonical K\"ahler potential, the inflaton field acquires a mass
of the order of the Hubble scale $H_0$, which results in the slow-roll parameter
$\eta$ being too large at all times,%
\footnote{Note that, while this eta problem certainly becomes worse for super-Planckian
field values, the slow-roll parameter $\eta$ is also already too large for
sub-Planckian field values $\left|\phi\right|\lesssim M_{\textrm{Pl}}$.\smallskip}
\begin{align}
\eta = M_{\textrm{Pl}}^2 \frac{V''}{V} \simeq 1 +
\frac{\left|\phi\right|^2}{M_{\textrm{Pl}}^2} \,,
\end{align}
and hence spoils the inflationary dynamics.
To avoid running into the eta problem, we are thus led to assume a
shift symmetry in the direction of the inflaton field
$\Phi$~\cite{Kawasaki:2000yn,Kallosh:2010ug,Kallosh:2011qk,Harigaya:2014qza}.
That is to say, we assume that the K\"ahler potential is invariant
under the following transformation,
\begin{align}
\Phi \rightarrow \Phi + i\,c\, M_{\textrm{Pl}}  \,, \quad c \in \mathbb{R} \,.
\label{eq:shift}
\end{align}
While this transformation behavior is reminiscent of the behavior of
axions or more generally of Nambu-Goldstone bosons in the context
of spontaneously broken symmetries, we do not further specify
the origin of this shift symmetry.
In this sense, our general DCI mechanism can apparently not be regarded
as a complete description of the inflationary dynamics.
It rather needs to be embedded into a UV-complete
theory that is capable of explaining the origin of the shift symmetry.
A promising candidate for such a UV completion is certainly string
theory, in which, for instance, the low-energy description of brane
dynamics may feature shift symmetries that could be used for
inflation~\cite{Hsu:2004hi}.
In addition, the assignment of $R$ charges in the individual
DCI models may provide a hint as to why a shift symmetry is realized in
exactly the direction of the inflaton field.
For instance, if the inflaton was the only gauge singlet carrying
zero $R$ charge, it would be singled out as a unique direction in
field space along which a shift symmetry could potentially be realized.
In Sec.~\ref{subsec:SPNWeff}, we gave an example for such a situation
in the context of our DCI models based on $SP(N_c)$ gauge groups,
cf.\ the discussion related to Eq.~\eqref{eq:SPNR0}.


Given the shift symmetry in Eq.~\eqref{eq:shift}, the
K\"ahler potential now ends up being a function of
the linear combination $\phi+\phi^*$ only,%
\footnote{Of course, the K\"ahler potential is a function of the
chiral superfields $\Phi$ and $\Phi^\dagger$, which is
why $K$ is actually given as $K = \left(\Phi + \Phi^\dagger\right)^2/2$.
However, as we are only interested in the dynamics of the scalar components of 
the fields $\Phi$ and $\Phi^\dagger$, it is for our purposes sufficient to work
with the complex scalars $\phi$ and $\phi^*$ in Eq.~\eqref{eq:Kshif}.}
\begin{align}
K  = \frac{1}{2}\left( \phi + \phi^* \right)^2 = \sigma^2 \,, \quad
\phi = \frac{1}{\sqrt{2}}\left(\sigma + i \tau\right) \,.
\label{eq:Kshif}
\end{align}
Thanks to the shift symmetry, it only depends on the real part of the complex
inflaton field, $\sigma = \sqrt{2}\:\textrm{Re}\hspace{-0.05cm}\left\{\phi\right\}$, but
not longer on its imaginary part,
$\tau = \sqrt{2}\:\textrm{Im}\hspace{-0.05cm}\left\{\phi\right\}$.
Instead of the potential in Eq.~\eqref{eq:VHmass}, we now have for
the full scalar potential in supergravity,
\begin{align}
V(\sigma,\tau) \supset \exp\left[K/M_{\textrm{Pl}}^2\right] V_0(\sigma,\tau) =
V_0(\sigma,\tau) + 3H_0^2\sigma^2 + .. \,.
\end{align}
During inflation, the real field $\sigma$ is hence stabilized at
$\sigma = 0$ due to its Hubble-induced mass, while the field
$\tau$ slowly rolls in the potential $V_0(0,\tau)$.
The imaginary component of the complex field $\phi$ thus
exhibits a fractional power-law potential, even at super-Planckian field
values, which is why we can now identify the field $\tau$ as the actual
real degree of freedom driving inflation.


It is important to note that in all of our DCI models the shift symmetry 
is explicitly broken by the Yukawa terms in the tree-level superpotential
coupling the inflaton field to a subset of matter fields,
cf.\ Eqs.~\eqref{eq:Wtreefull}, \eqref{eq:SO10}, \eqref{eq:WtreeSU5}
and \eqref{eq:Wtree32}.
These Yukawa interactions are a crucial ingredient to our general
DCI recipe, without which our entire construction would collapse.
None of our DCI models can therefore ever be exactly shift-symmetric
in the direction of the inflaton field.
The shift symmetry is always explicitly broken in the superpotential at tree level
and, as a consequence, it is also always broken in the K\"ahler
potential at loop level.
In fact, the explicit symmetry-breaking terms in the superpotential
induce the following K\"ahler potential,
\begin{align}
\delta K(\mu) \sim c\,\frac{\lambda^2}{16\pi^2} \left|\phi\right|^2
\log\left(\frac{\mu^2}{M_{\textrm{Pl}}^2}\right) \,,
\label{eq:deltaK}
\end{align}
with $\mu$ being the energy scale at which $\delta K$ is supposed to be evaluated
and with $c$ denoting a model-dependent factor of $\mathcal{O}(1..10)$.
Here, we have assumed that the shift-symmetric K\"ahler potential
in Eq.~\eqref{eq:Kshif} is defined around the Planck scale,
so that for energies $\mu \simeq M_{\textrm{Pl}}$ no large logarithmic
correction appears.
A quick way to see why the term $\delta K \propto \lambda^2/\left(16\pi^2\right)
\left|\phi\right|^2$ is expected to be generated in the effective K\"ahler
potential is the following:
Formally, the symmetry-breaking terms in the superpotential can be rendered
invariant, once we promote the shift symmetry to a spurious symmetry under which
the coupling constant $\lambda$ transforms as
$\lambda \rightarrow \lambda\Phi / \left(\Phi + i\,c\, M_{\textrm{Pl}}\right)$.
This spurious shift symmetry still forbids the dangerous canonical term in
the tree-level K\"ahler potential, $K \not\supset \left|\phi\right|^2$, but
it allows for a loop-suppressed term
$\delta K \propto \lambda^2/\left(16\pi^2\right)\left|\phi\right|^2$
in the  effective K\"ahler potential.
The loop-induced breaking of the shift symmetry in the K\"ahler potential
then bears the risk of re-introducing the eta problem, as it results again
in a steep exponential potential at large inflaton field values, $\tau \gg M_{\textrm{Pl}}$.
In order to prevent the K\"ahler potential in Eq.~\eqref{eq:deltaK}
from spoiling the inflationary dynamics after all, we thus have to assume
that the value of $\lambda$ is rather suppressed, $\lambda \ll 1$.
In the sense of 't~Hooft~\cite{'tHooft:1979bh}, such an assumption however
does not confront us with a naturalness problem, since the limit
$\lambda \rightarrow 0$ corresponds to the restoration of the shift symmetry
in the superpotential as well as in the K\"ahler potential.
More precisely, the shift in the slow-roll parameters $\varepsilon$ and $\eta$
and hence also in the inflationary CMB observables induced by $\delta K$
in Eq.~\eqref{eq:deltaK} can be estimated as~\cite{Harigaya:2014qza},
\begin{align}
\Delta \varepsilon \sim  \Delta \eta \sim c\,\frac{\lambda^2}{16\pi^2}
\sim
10^{-3}\, \bigg(\frac{c}{10}\bigg)\left(\frac{\lambda}{0.1}\right)^2
\,, \label{eq:Deltaepseta}
\end{align}
where we have assumed rather large  $c$, so as to
be on the safe side and not to underestimate the upper bound on $\lambda$.
The Yukawa coupling $\lambda$ is therefore only allowed to
take values at most as large as $\lambda\simeq10^{-1}$; larger values of $\lambda$
result into too large corrections to the slow-roll parameters.


The fact that we are led to incorporate a shift symmetry into our general
DCI mechanism does not come as a surprise.
In Refs.~\cite{Kawasaki:2000yn,Kallosh:2010ug}, it has been shown that,
in order to successfully construct a model of chaotic inflation in the
context of supergravity, it is sufficient to impose a shift symmetry in
the direction of the inflaton and to require the superpotential to be
of the form $W = f(\Phi)\,\Xi$, where $f$ is an arbitrary holomorphic
function of $\Phi$ and the field $\Xi$ is a gauge singlet that can be
identified as the goldstino superfield responsible for the spontaneous
breaking of supersymmetry during inflation~\cite{Kallosh:2010xz}.
Now we recognize that several of our DCI scenarios actually belong
to this class of SUGRA models of chaotic inflation.
For example, our $SP(N_c)$ models feature an effective superpotential
$W_{\textrm{eff}} \propto \Lambda_{\textrm{eff}}^2\, X $ during inflation,
cf.\ Eq.~\eqref{eq:Weff}, such that the goldstino field $\Xi$ can be
straightforwardly identified as the linear combination $X \propto J^{ij}Z_{ij}$,
cf.\ Eq.~\eqref{eq:goldstino}.
A similar observation holds for our $SU(3)\times SU(2)$ models.
In the case of our $SO(10)$ scenarios, a description of the SUSY-breaking
dynamics in terms of a perturbative superpotential as well as a perturbative
K\"ahler potential is, by contrast, not feasible.


Finally, we mention that a second modification of the scalar potential related
to supergravity, besides the exponential factor involving the K\"ahler potential,
$\exp\left[K/M_{\textrm{Pl}}^2\right]$, arises due to the fact that
in supergravity also the superpotential itself, as a function of the scalar fields,
enters into the scalar potential.
In fact, for any chiral superfield $\Phi$ possessing a nonzero $F$-term,
$W \simeq -F_\Phi^* \phi $ with $F_\Phi \neq 0$ during inflation, the scalar potential
contains a term of the form,
\begin{align}
V \supset - \frac{3}{M_{\textrm{Pl}}^2}\left|W\right|^2 \simeq
-\frac{|\phi|^2}{M_{\textrm{Pl}}^2} |F_\Phi|^2 + .. \,.
\end{align}
This potential is negative and dominant over the positive
contribution, $|F_\Phi|^2$, for $|\phi| \gg M_{\textrm{Pl}}$.
Therefore, if in our DCI scenarios the inflaton had a large $F$-term,
significantly contributing to the vacuum energy density 
during inflation, the associated negative SUGRA contribution would ruin
the inflationary dynamics~\cite{Kawasaki:2000yn}.
Because of this, it is important that, as we already mentioned at the end of
Sec.~\ref{sec:idea}, it is the $F$-term of a different field other than the inflaton
which provides the dominant contribution to the vacuum energy density during inflation.
Fortunately, this requirement is always fulfilled in our DCI models by construction---according
to our general DCI algorithm, the dynamical breaking of supersymmetry and hence the
dynamical generation of the vacuum energy density is always taken care of by the
strongly interacting effective theory below all heavy mass thresholds, which no
longer involves the inflaton as a dynamical degree of freedom.
The breaking of supersymmetry is therefore attributed to the $F$-term of some other
field $\Xi \neq \Phi$ in this effective theory.
In the case of our $SP(N_c)$ models, this field corresponds for example to
the linear combination $X \propto J^{ij}Z_{ij}$, cf.\ Eq.~\eqref{eq:goldstino},
as we already pointed above.


\section{Phenomenological implications}
\label{sec:phenomenology}


In the previous sections, we have demonstrated how our general DCI mechanism
allows to construct monomial inflaton potentials such as the one in Eq.~\eqref{eq:powerlawV}
as well as how to consistently embed them into supergravity.
Now we shall study their phenomenological implications for the inflationary phase,
the reheating process after the end of inflation as well as for the
inflationary CMB observables in more detail.
Throughout our analysis, we will approximate the scalar potential for the real
inflaton degree of freedom $\tau$ as follows
\begin{align}
V \simeq V_0 = \Lambda_{\textrm{eff}}^4(\tau) \,, \quad
\Lambda_{\textrm{eff}}(\tau) = \Lambda \left(\frac{\lambda\,\tau}{\Lambda}\right)^{p/4} \,,
\label{eq:Vtau}
\end{align}
where, compared to Eq.~\eqref{eq:Vpropto}, we have absorbed a factor of
$1/\sqrt{2}$ in the coupling constant $\lambda$.
Effectively, for a given value of the fractional power $p$, every DCI model therefore
has two free parameters: the dynamical scale $\Lambda$ as well as the inflaton Yukawa coupling
constant $\lambda$.


\subsection{Constraints on parameter space}
\label{subsec:constraints}


The two parameters $\Lambda$ and $\lambda$ are subject to several constraints, which
follow from the requirement of internal theoretical consistency.
First of all, we note that the masses of the matter fields coupling to the inflaton,
which are all of $\mathcal{O}\left(\lambda\,\tau\right)$, should never exceed the
Planck scale.
Only as long as $\lambda\,\tau \lesssim M_{\textrm{Pl}}$ for all times during inflation,
we can rely on our ordinary field-theoretic analysis. 
Larger masses would, by contrast, require us to specify the UV completion
of our model and perform an explicit string-theoretic calculation.
As we are only interested in the last $N_e = 50..60$ $e$-folds of inflation, the
largest relevant field value of the inflaton is the one $N_e$ $e$-folds before
the end of inflation, $\tau_{N_e} \simeq \left(2\,p\,N_e\right)^{1/2} M_{\textrm{Pl}}$.
This field value typically exceeds the Planck scale by an order of magnitude,
$\tau_{N_e} \sim \mathcal{O}(10)M_{\textrm{Pl}}$,
which is why $\lambda$ must be rather small,
so as to guarantee that the product $\lambda\,\tau_{N_e}$ still
remains below the Planck scale,
\begin{align}
\lambda\,\tau_{N_e} \lesssim M_{\textrm{Pl}} \quad\Rightarrow\quad
\lambda \lesssim \sqrt{\frac{1}{2\,p\,N_e}} \simeq 7.1 \times 10^{-2}
\left(\frac{2}{p}\right)^{1/2}\left(\frac{50}{N_e}\right)^{1/2} \,.
\label{eq:lambdaupper}
\end{align}
This condition needs to be compared with the bound resulting from the requirement
of not too large shifts in the slow-roll parameters due to the explicit
breaking of shift symmetry in the K\"ahler potential,
$\lambda \lesssim 10^{-1}\left(\Delta \eta_{\textrm{max}}/10^{-3}\right)^{1/2}
\big(10/c\big)^{1/2}$, cf.\ Eq.~\eqref{eq:Deltaepseta}.
Thus, if we maximally allow for shifts $\Delta \eta_{\textrm{max}}$
of $\mathcal{O}\left(10^{-3}\right)$ and assume the coefficient $c$
to take some $\mathcal{O}(10)$ value, this latter bound is always significantly
weaker than the one in Eq.~\eqref{eq:lambdaupper}.
Once the condition in Eq.~\eqref{eq:lambdaupper} is satisfied, we therefore
do not have to worry about possibly too large corrections to the slow-roll
parameters---they are then automatically guaranteed to be negligibly small.


Furthermore, similarly as for the masses of the decoupling matter fields, we also must pay
attention that the total potential energy density $V$ never exceeds Planckian values.
Our field-theoretic analysis only remains valid as long as $V \lesssim M_{\textrm{Pl}}^4$
for all times during inflation.
Making use of the fact that the potential energy density is largest for field values
around $\tau \simeq \tau_{N_e}$, this condition then implies a second upper bound on the
coupling constant $\lambda$,
\begin{align}
V \lesssim M_{\textrm{Pl}}^4 \quad\Rightarrow\quad
\Lambda_{\textrm{eff}} \lesssim M_{\textrm{Pl}} \quad\Rightarrow\quad
\lambda \lesssim \sqrt{\frac{1}{2\,p\,N_e}}
\left(\frac{M_{\textrm{Pl}}}{\Lambda}\right)^{(4-p)/p} \,.
\label{eq:lambdaupper3}
\end{align}
Given that the dynamical scale $\Lambda$ is smaller than the Planck scale,
$\Lambda \lesssim M_{\textrm{Pl}}$, this bound is weaker than the one in
Eq.~\eqref{eq:lambdaupper} for fractional powers $p \leq 4$.
This means in turn that, for powers $p>4$, the masses of the matter fields
coupling to the inflaton field are forbidden to grow to values as large as the
Planck mass, because otherwise super-Planckian energy densities would occur
during the last $N_e$ $e$-folds of inflation.
For this reason, as we wish to retain the possibility of reaching
$\mathcal{O}\left(M_{\textrm{Pl}}\right)$ values with the masses
$\lambda\,\tau_{N_e}$, we will restrict ourselves to powers $p \leq 4$ in the
following.
In more formal terms, denoting the two maximally possible values of $\lambda$
according to Eqs.~\eqref{eq:lambdaupper} and \eqref{eq:lambdaupper3} by 
$\lambda_{\textrm{max},1}$ and $\lambda_{\textrm{max},2}$,
we then have
\begin{align}
\lambda_{\textrm{max},1} \,\tau_{N_e} \simeq M_{\textrm{Pl}} \,, \quad
\Lambda \left(\frac{\lambda_{\textrm{max},2}\,\tau_{N_e}}{\Lambda}\right)^{p/4}
\simeq M_{\textrm{Pl}} \,, \quad \lambda_{\textrm{max},1} \leq \lambda_{\textrm{max},2}
\quad\Rightarrow\quad p \leq 4 \,.
\label{eq:pbound}
\end{align}
In addition to the bound on the power $p$ in Eq.~\eqref{eq:pboundSPN}, this
condition provides another, slightly more physical reason why it is sensible
to restrict our analysis to $p$ values not larger than $4$.
Besides that, as discussed in Sec.~\ref{subsec:observables} and 
illustrated in Fig.~\ref{fig:nsr}, powers larger than $4$ are,
of course, almost ruled out observationally in any case.


Third, we must ensure that all matter fields coupling to the inflaton decouple
\textit{perturbatively} at energy scales above the dynamical scale $\Lambda$.
This places a lower bound on the masses of the decoupling
matter fields, $\lambda\,\tau \gtrsim \Lambda$.
For smaller masses, the dynamics of the matter fields are dominated
by strong-coupling effects and we can no longer integrate
fields out by naively employing the techniques of perturbation theory.
The smallest relevant value of the inflaton during inflation
is the one reached at the end of inflation, when the slow-roll
conditions become violated, $\tau_0 \simeq p \,M_{\textrm{Pl}}$.
We consequently find the following lower bound on the coupling $\lambda$,
\begin{align}
\lambda\,\tau_0 \gtrsim \Lambda \quad\Rightarrow\quad
\lambda \gtrsim \frac{1}{p}\frac{\Lambda}{M_{\textrm{Pl}}} \simeq 2.1 \times 10^{-3}
\left(\frac{2}{p}\right)\left(\frac{\Lambda}{10^{16}\,\textrm{GeV}}\right) \,.
\label{eq:lambdalower}
\end{align}
Eqs.~\eqref{eq:lambdaupper} and \eqref{eq:lambdalower} hence tell us
that, for typical values of $p$ and $\Lambda$, the coupling $\lambda$ should
take values of $\mathcal{O}\left(10^{-3}..\,10^{-1}\right)$.
Such small values of $\lambda$ are also sufficient to adequately suppress
the shift symmetry-breaking corrections in the effective K\"ahler potential,
cf.\ Eq.~\eqref{eq:Deltaepseta}.


Fourth, it is important that during inflaton the inflaton field value
does not vary too rapidly.
The typical timescale of physical interactions in the strongly coupled sector
is of the order of $\Lambda_{\textrm{eff}}^{-1}$.
During this time, the inflaton rolls a distance
$\Delta\tau \simeq \dot{\tau}/\Lambda_{\textrm{eff}}$ in field space,
which should be small compared to its actual field value,
\begin{align}
\left|\Delta\tau\right| \lesssim \left|\tau\right|  \quad\Rightarrow\quad
\left|\frac{\dot{\tau}}{\tau}\right| \lesssim \Lambda_{\textrm{eff}} \,.
\label{eq:slowtau}
\end{align}
Otherwise backreactions from the strongly coupled sector could potentially perturb
the inflationary dynamics in an uncontrollable manner.
The inflaton velocity $\dot{\tau}$ follows from the slow-roll equation of motion,
$3H_0\dot{\tau} \approx - V_{,\tau}$, such that the condition in
Eq.~\eqref{eq:slowtau} can be rewritten as
\begin{align}
\left|\frac{\dot{\tau}}{\tau\,\Lambda_{\textrm{eff}}}\right| \simeq 
\frac{p\,\lambda^2}{\sqrt{3}} \left(\frac{M_{\textrm{Pl}}}{\Lambda}\right)
\left(\frac{\lambda\,\tau}{\Lambda}\right)^{p/4-2} \lesssim 1 \,.
\end{align}
Together with the condition in Eq.~\eqref{eq:lambdaupper},
$\lambda\,\tau \lesssim M_{\textrm{Pl}}$, we then find a second upper bound on $\lambda$,
\begin{align}
\frac{p\,\lambda^2}{\sqrt{3}}
\left(\frac{M_{\textrm{Pl}}}{\Lambda}\right)^{p/4-1} \lesssim 1
\quad\Rightarrow\quad
\lambda \lesssim \frac{3^{1/4}}{p^{1/2}}
\left(\frac{\Lambda}{M_{\textrm{Pl}}}\right)^{(p/4-1)/2}
\label{eq:lambdaupper2}
\end{align}
This constraint on $\lambda$ only becomes tighter than the one in
Eq.~\eqref{eq:lambdaupper} for large fractional powers, $p \gtrsim 4$,
and small values of $\Lambda$.
However, as we are only interested in powers $p \leq 4$,
this case is irrelevant for us.
For $p \leq 4$, Eq.~\eqref{eq:lambdaupper2} is always satisfied for
$\lambda \lesssim 6.6 \times 10^{-1}$, which is a significantly weaker bound
than the one in Eq.~\eqref{eq:lambdaupper}.


\begin{figure}
\centering
\includegraphics[width=0.81\textwidth]{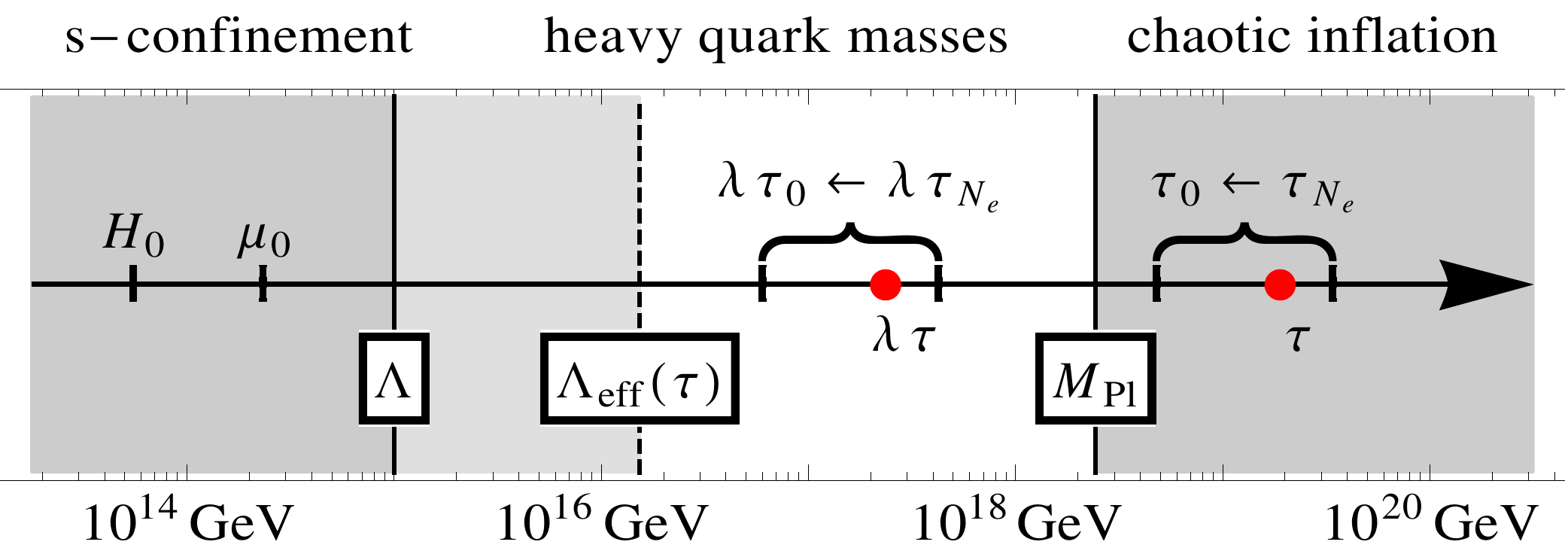}
\caption{Hierarchy among all the relevant energy scales involved in dynamical chaotic inflation
($H_0$: Hubble rate; $\mu_0$: Wilsonian cut-off scale; $\Lambda$ and $\Lambda_{\textrm{eff}}$:
fundamental and effective dynamical scale, respectively; $\lambda\,\tau$: inflaton-dependent
heavy quark mass; $M_{\textrm{Pl}}$: Planck scale; $\tau$: real inflaton field).
Here, we have chosen the following exemplary parameter values:
$p=2$, $\Lambda = 1.0\times 10^{15}\,\textrm{GeV}$, $\lambda = 1.2 \times 10^{-2}$, $N_e =50$,
which also result in the correct value for the scalar spectral amplitude
$A_s$, cf.\ the black dot in Fig.~\ref{fig:norm}.
The red dots indicate the inflaton field value $\tau$ as well as the inflaton-dependent
heavy quark mass $\lambda\tau$ at the time 15 $e$-folds before the end of inflation.}
\label{fig:scales}
\end{figure}


Finally, we mention as a fifth condition that the inflationary Hubble
rate $H_0$ must always be smaller than the dynamical scale $\Lambda_{\textrm{eff}}$,
\begin{align}
H_0 \lesssim \Lambda_{\textrm{eff}} \,.
\end{align}
The Hubble scale $H_0$ determines the size of the scalar fluctuations of
the inflaton field.
If the Hubble rate was larger than $\Lambda_{\textrm{eff}}$, we would not be able to
base the calculation of our predictions for the inflationary observables
on the effective scalar potential in Eq.~\eqref{eq:Vtau}.
This last condition is, however, always trivially fulfilled,
since the effective dynamical scale $\Lambda_{\textrm{eff}}$ is
always smaller than the Planck scale $M_{\textrm{Pl}}$ according
to our second condition in Eq.~\eqref{eq:lambdaupper3},
\begin{align}
H_0 = \frac{V_0^{1/2}}{\sqrt{3}M_{\textrm{Pl}}} =
\frac{\Lambda_{\textrm{eff}}^2}{\sqrt{3}M_{\textrm{Pl}}} \lesssim
\Lambda_{\textrm{eff}} \quad\Rightarrow\quad
\Lambda_{\textrm{eff}} \lesssim \sqrt{3}M_{\textrm{Pl}} \,.
\end{align}


A summary of all the relations listed above is given in Fig.~\ref{fig:scales}, in which
we illustrate the hierarchy between the scales $H_0$, $\Lambda$, $\Lambda_{\textrm{eff}}$,
$\lambda\,\tau_0$, $\lambda\,\tau_{N_e}$, $M_{\textrm{Pl}}$, $\tau_0$ and $\tau_{N_e}$
for a representative parameter example.
In addition to all relevant physical scales, we also indicate the approximate location
of the Wilsonian cut-off scale $\mu_0$, which defines the validity range of the effective 
superpotentials derived in the previous sections.
As for the coupling constant $\lambda$, we conclude that the constraints in
Eqs.~\eqref{eq:lambdaupper} and \eqref{eq:lambdalower} are the strongest ones.
We will come back to these bounds when discussing our prediction for the
amplitude $A_s$ of the scalar power spectrum, cf.\ Fig.~\ref{fig:norm}.


\subsection{Initial conditions for inflation and inflationary phase}


The scalar potential in Eq.~\eqref{eq:Vtau} gives rise to a phase of a chaotic
inflation~\cite{Linde:1983gd}.
In contrast to many other inflationary scenarios, standard chaotic inflation based on
a potential such as $m^2\left|\phi\right|^2$ or $\lambda/4\left|\phi\right|^4$
has the virtue that it does not require any particular fine-tuning of the initial
position and velocity of the inflaton field.
The reason for this is the following:
As is well known, a necessary condition for inflation to occur is
that at early times the potential energy of the inflaton field as well as
its kinetic and gradient energies must be sufficiently homogeneous
across an entire Hubble volume $H^{-3}$~\cite{Goldwirth:1991rj}.
If inflation was preceded by a phase of, for instance, radiation domination and the
actual inflationary phase only began at sub-Planckian expansion rates,
$H \ll M_{\textrm{Pl}}$, the initial inflation field value and the initial
inflaton velocity would need to be adjusted to homogeneous values across
a number of roughly $\left(M_{\textrm{Pl}}/H\right)^{3/2}$ nonequilibrated Planck domains.
Ordinary chaotic inflation avoids this initial horizon problem, as it starts out with arbitrary
or \textit{chaotic} initial conditions in the vicinity of the Planck scale, such that
\begin{align}
t_{\textrm{ini}} \sim M_{\textrm{Pl}}^{-1} \,, \quad
\tau_{\textrm{ini}} \gg M_{\textrm{Pl}} \,, \quad
V(\tau_{\textrm{ini}})\sim \frac{1}{2}{\dot \tau_{\textrm{ini}}}^2 \sim
\frac{1}{2}(\vec{\nabla} \tau)_{\textrm{ini}}^2
\sim H_{\textrm{ini}}^4  \lesssim M_{\rm Pl}^4 \,.
\end{align}
In this picture, the entire observable universe then originates from a single
homogeneous Planck domain, which begins to inflate at times $t_{\textrm{ini}}$
close to the Planck time $M_{\textrm{Pl}}^{-1}$.


In the context of dynamical chaotic inflation, the situation is similar,
but only to a certain extent.
The crucial difference here is that at times around $M_{\textrm{Pl}}^{-1}$ the
scalar potential for the inflaton typically does not yet exist.
It rather takes the interactions in the strongly coupled sector a time
of $\mathcal{O}\left(\Lambda_{\textrm{eff}}^{-1}\right)$ to dynamically generate
a potential for the inflaton field.
Inflaton can therefore only set in once the Hubble rate has dropped to
a value $H \sim \Lambda_{\textrm{eff}}$.
As we require the mass of the heavy quarks coupling to the inflaton
not to exceed the Planck scale for any time during inflation and as we focus
on DCI models with a power $p\leq4$, the effective dynamical scale can become
at most as large as $M_{\textrm{Pl}}$, cf.\ Eqs.~\eqref{eq:lambdaupper},
\eqref{eq:lambdaupper3} and \eqref{eq:pbound}.
Similarly, the Hubble rate can maximally become as large as
$H \simeq \left(\Lambda/M_{\textrm{Pl}}\right)^{2-p/2}M_{\textrm{Pl}}/3^{1/2}$.
For powers much smaller than $p=4$, the delay between the Planck time,
$t\sim M_{\textrm{Pl}}^{-1}$, and the onset of inflation at
$t \sim \Lambda_{\textrm{eff}}^{-1}$ then seems to re-introduce the
initial horizon problem.
However, as we will argue in the following, this is actually not the case.
In doing so, we will first present a rough order of magnitude estimate for the degree
of inhomogeneity at the time when  $H \sim \Lambda_{\textrm{eff}}$;
then we will follow and describe the time evolution of the field fluctuations
between  $t \sim M_{\textrm{Pl}}$ and $t \sim \Lambda_{\textrm{eff}}^{-1}$
in more detail.%
\footnote{Despite the absence of any initial
horizon problem in dynamical chaotic inflation, the delay between the Planck time
and the onset of inflation forces us to assume that the spatial curvature of the
universe has negative sign or is at most zero.
That is, we have to assume an open or at most flat universe.
In the case of a closed universe, the chance of having the universe survive
until the onset of inflation at $t \sim \Lambda_{\textrm{eff}}^{-1}$ is
vanishingly small.\smallskip}


At the time when the inflaton potential is dynamically generated,
$t \sim H^{-1} \sim \Lambda_{\textrm{eff}}^{-1}$, the energy
contained in field gradients ranging over a single Hubble patch can be
estimated to be,
\begin{align}
\big(\vec{\nabla} \delta\tau\big)^2 \sim \left(\frac{\delta\tau}{1/H}\right)^2
= H^2 \,\delta\tau^2 \,.
\end{align}
This energy needs to be compared with the total energy density $\rho$ at this
time, $\rho = 3 H^2 M_{\textrm{Pl}}^2$.
We thus see that the fluctuations in the inflaton field configuration,
$\delta\tau$, can at most be of the order of the Planck scale,
$\delta\tau \lesssim \sqrt{3}M_{\textrm{Pl}}$.
On the other hand, we know that the initial inflaton field value
in chaotic inflation is at least one, if not more, orders of magnitude
larger than the Planck scale, $\tau \sim \mathcal{O}\left(10..\,10^3\right)M_{\textrm{Pl}}$.
That is, thanks to the large initial field value in chaotic inflation---which ultimately
only becomes possible due to the shift symmetry discussed in Sec.~\ref{sec:SUGRA}---the
universe is automatically homogeneous to a very large precision 
at the time when the strong interactions become effective,
$\delta\tau/\tau \ll 1$.
Thus, owing to the large initial field value and the underlying shift symmetry,
we do not encounter any initial horizon problem in dynamical chaotic inflation.


Now, after this sketch of an argument, let us follow the time evolution of inflaton field perturbations
between the Planck time and the onset of inflation more carefully.
We will do this in three steps, which will eventually provide us with three 
explicit reasons why the horizon problem is also resolved in dynamical
chaotic inflation.
Here, our discussion mainly proceeds along the lines of Ref.~\cite{Linde:2005ht}.%
\footnote{The first two out of our three points actually hold in general
for many large-field models, in which inflation only begins at a time
$t\sim H^{-1} \gg M_{\textrm{Pl}}^{-1}$.
Only the third point specifically applies to dynamical chaotic inflation.}
To start with, let us consider the situation around $t\sim M_{\textrm{Pl}}^{-1}$.
We assume that at this time the initial field
values fall into the range
$-M_{\textrm{Pl}}/\lambda \lesssim \tau \lesssim M_{\textrm{Pl}}/\lambda$.
This is mainly because, at larger field values, we no longer have any control
over the dynamics of the inflaton field.
For one thing, some of the quark flavors in the inflaton sector
acquire masses above $M_{\textrm{Pl}}$ for $\tau \gtrsim M_{\textrm{Pl}}/\lambda$;
for another thing, we loose control over the
effective K\"ahler potential at very large inflaton field values.
Now, given such an initial inflaton field configuration, the inflaton
value typically fluctuates over distances $l \gg M_{\textrm{Pl}}^{-1}$
with an amplitude of $\mathcal{O}\left(M_{\textrm{Pl}}/\lambda\right)$.
These long-wavelength modes thus stretch over a large number
of individually causally connected Hubble patches, the spatial
extent of which is determined
by the Hubble radius, $H^{-1}\sim M_{\textrm{Pl}}^{-1}$.
In each Hubble patch, the perturbations with comoving
wavenumber $k \ll M_{\textrm{Pl}}$
therefore appear as contributions to a homogeneous background value of
$\mathcal{O}\left(M_{\textrm{Pl}}/\lambda\right)$.
The only noticeable perturbations within a Hubble patch correspond
in turn to the inflaton modes with wavenumber $k \gtrsim M_{\textrm{Pl}}$
and an amplitude not larger than $M_{\textrm{Pl}}$.
Thanks to the large value of the homogeneous background, these
fluctuations are hence already comparatively small from the outset.


Second, as time goes on and the universe expands, no perturbation
modes with wavenumbers $k \lesssim M_{\textrm{Pl}}$ and hence dangerously
large amplitude can enter into the Hubble horizon.
This is due to the fact that, in consequence of the large gradients in
the inflaton field configuration at early times, the expansion is initially
mostly driven by the gradient energy of the inflaton field, which implies that $H$
decreases as the inverse of the scale factor $a$.%
\footnote{The same holds true for a curvature dominated universe.
The following discussion also applies in that case.\smallskip}
The product $aH$ is then a constant, indicating that no large-wavelength modes
can cross inside the horizon.
Because of that, the already quite homogeneous field values within each Hubble
patch fortunately avoid being perturbed by incoming modes with large amplitude.
Instead, the modes with $k \lesssim M_{\textrm{Pl}}$ stay outside the horizon,
such that their amplitude remains preserved, whereas the modes with
$k \gtrsim M_{\textrm{Pl}}$ stay inside the horizon, where they decay like $1/a$.
In addition to that, the gradient as well as the kinetic energy of the inflaton
field are redshifted due to the expansion.%
\footnote{It is easy to see that the kinetic energy decreases much more rapidly than
the gradient energy.}
Therefore, at the time when the Hubble rate has dropped to the
effective dynamical scale, $H\sim\Lambda_{\textrm{eff}}$, the
kinetic energy of the inflaton field has already become very
small, while the inflaton field values end up being homogeneous within
each Hubble patch to very good precision.


Finally, at times around $t \sim \Lambda_{\textrm{eff}}^{-1}$, the inflaton
potential is eventually generated dynamically.
The exact time at which the inflaton potential emerges in a given region
of space solely depends on the local value of the Hubble rate.
As we argued above, the energy density and hence also the Hubble rate is,
however, homogeneous within each Hubble patch to very good approximation.
The inflaton potential is therefore generated
simultaneously in all parts of a given Hubble patch.
That is why, within each Hubble patch, the inflaton field has a homogeneous
field value, a small velocity as well as a homogeneous potential energy,
once the expansion rate has reached a value of the order of the effective
dynamical scale, $\Lambda_{\textrm{eff}}$, of the strongly coupled sector.
These are ideal initial conditions for the onset of inflation!
In summary, we conclude that dynamical chaotic inflation does not
suffer from an initial horizon problem for three reasons:
(i) the large initial field values and gradients at the Planck time,
(ii) the gradient-dominated expansion at early times as well as
(iii) the fact that the dynamical generation of the inflaton potential
is triggered by nothing else than the reaching of a specific value for the Hubble rate.
All in all, we can hence be confident that our general DCI mechanism
in combination with the scalar potential in Eq.~\eqref{eq:Vtau} allows
for a successful realization of chaotic inflation.
Let us now discuss the exact shape of the potential and the ensuing
inflationary phase in more detail.

\begin{figure}
\centering
\includegraphics[width=0.58\textwidth]{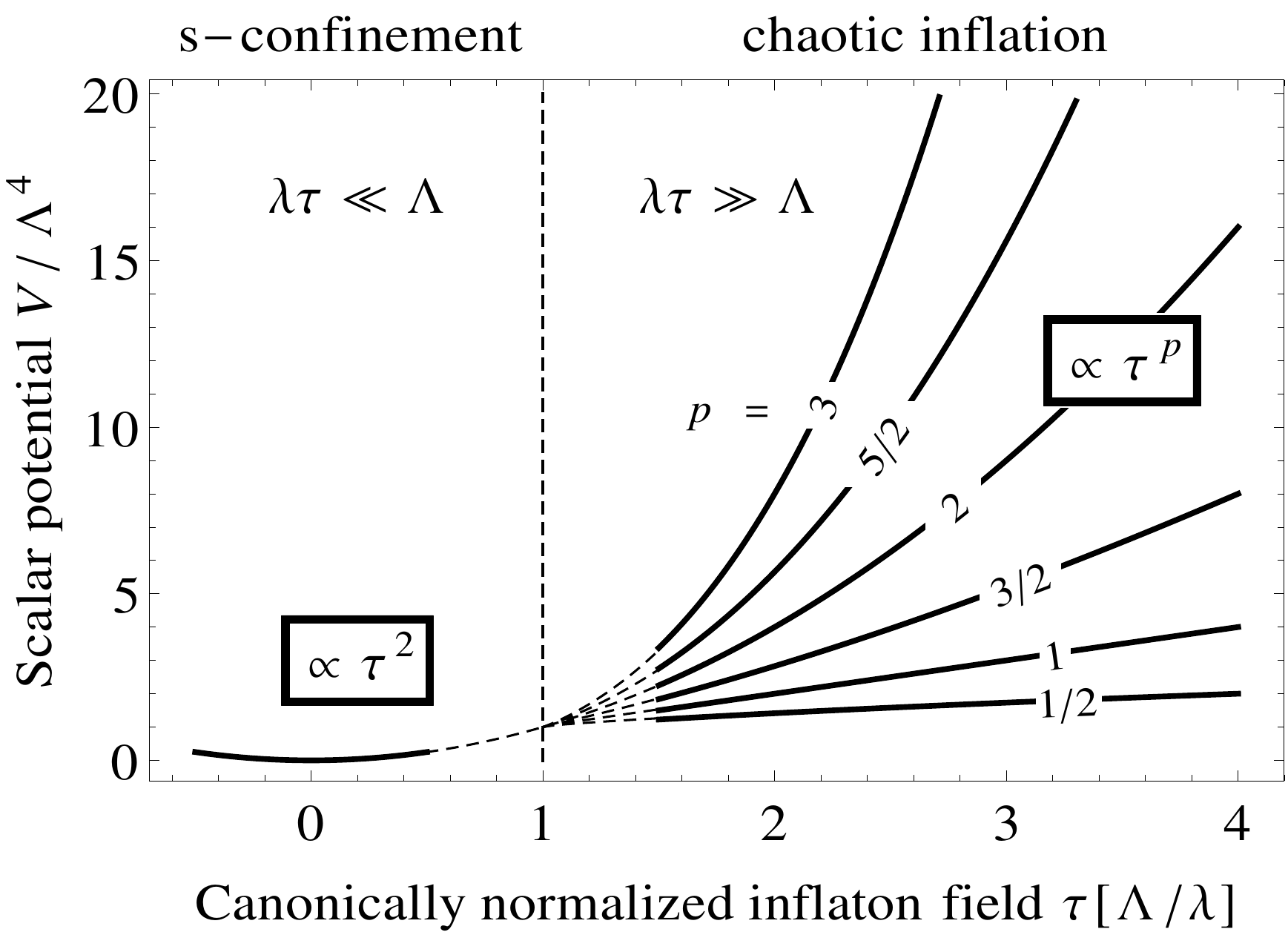}
\caption{Schematic shape of the scalar potential for the canonically
normalized inflaton field $\tau$.
At large field values, $\lambda\,\tau \gg \Lambda$, the inflaton slowly rolls
in a power-law potential, cf.\ Eq.~\eqref{eq:Vtau}, thereby giving rise to a
stage of chaotic inflation.
As the inflaton field decreases to very small values, $\lambda\,\tau \ll \Lambda$
the strongly interacting sector reaches the phase of s-confinement and the inflaton
begins the oscillate in a quadratic potential, cf.\ Eq.~\eqref{eq:Vmphi}.}
\label{fig:potential}
\end{figure}

In all of our DCI models, large inflaton field values,
$\lambda\,\tau \gg \Lambda$, result in the perturbative decoupling of all
matter fields coupling to the inflaton via Yukawa interactions, such that the
strongly coupled sector gives rise to dynamical supersymmetry breaking.
The vacuum energy density associated with this spontaneous breaking of
supersymmetry, cf.\ Eq.~\eqref{eq:Vtau}, then acts as the scalar potential
for the inflaton field during inflation.
Similarly, we know that at small inflaton field values, $\lambda\,\tau \ll \Lambda$,
the strongly coupled sector is in the s-confinement phase, which is well-behaved and
free of singularities at the origin in field space.
In this s-confined phase, the inflaton as well as all the gauge-invariant
composite fields are massive.
In particular, we expect the inflaton to receive a mass
$m_{\phi} \sim \lambda\Lambda$ from its coupling to the meson
fields in the tree-level superpotential as well as in the effective K\"ahler potential,
\begin{align}
\lambda\,\tau \ll \Lambda \,, \quad
V \sim \lambda^2 \Lambda^2 \left|\phi\right|^2 \,.
\label{eq:Vmphi}
\end{align}
In between these two regimes, i.e.\ for $\lambda\,\tau \simeq \Lambda$,
the scalar potential is dominated by strong-coupling effects and hence
unfortunately not calculable.
If we are unlucky, the potential might even not be monotonic
at these intermediate field values and exhibit a local minimum,
in which the inflation might become trapped before reaching the
true vacuum at $\tau = 0$.
In the following, we shall however assume that no such peculiar
feature exists in the scalar potential for field values
$\lambda\,\tau \simeq \Lambda$.
Instead, we suspect that the potential in the large-field regime,
$V \propto \tau^p$, is monotonically connected to the potential
in the small-field regime, $V \propto \tau^2$, so that the
inflaton does not encounter any hindrance on its way to the true vacuum,
cf.\ Fig.~\ref{fig:potential}, in which we give a schematic overview of
the shape of the inflaton potential in the different field regimes.


The actual inflationary stage is entirely realized in the large-field regime,
where $\lambda\,\tau \gg \Lambda$.
It is characterized by the slow-roll motion of the inflaton from very large
field values, $\tau \gtrsim \tau_{N_e}$, to smaller
field values, $\tau\simeq \tau_0$, shortly above the dynamical scale,
cf.\ Eqs.~\eqref{eq:lambdaupper} and \eqref{eq:lambdalower}.
Here, the concrete initial value of the inflaton field at the very onset
of inflation may be some arbitrary value high above the Planck scale,
$\tau_{\textrm{ini}} \sim M_{\textrm{Pl}}/\lambda$.
Once the inflaton passes below the value $\tau\simeq \tau_0$, the
slow-roll conditions become violated and the exponential expansion terminates.
Unfortunately, we cannot make any exact statements about the subsequent
transition from the large-field to the small-field regime, as it is dominated
by strong-coupling effects.
Assuming that nothing peculiar happens at field values around
$\tau \simeq \Lambda/\lambda$, we can only say that, after traversing the
intermediate field regime, the inflaton eventually
ends up oscillating in its quadratic potential around the origin.


\subsection{Preheating and reheating after inflation}


The violation of the slow-roll conditions at $\tau\simeq \tau_0$
as well as the subsequent oscillations around $\tau=0$ mark the
onset of the reheating process, i.e.\ the conversion of the energy
stored in the vacuum to a thermal plasma of SM particles.
In fact, reheating proceeds in two steps: nonperturbative preheating
and perturbative reheating due to inflaton decay.
Preheating is a model-dependent nonlinear process, which needs to be
treated numerically on a case-by-case basis in order to obtain reliable
results.
Existing studies of preheating in the literature either describe preheating
in the case of small-field and hybrid models of inflation, featuring 
negatively curved scalar potentials, or preheating in the case of large-field
models of inflation, featuring positively curved scalar potentials.
As for the former models, preheating was found to proceed via
\textit{tachyonic oscillations}\,\cite{Brax:2010ai} of the inflaton field
or alternatively via \textit{tachyonic preheating}\,\cite{Felder:2000hj},
whereas for the latter models preheating was identified to occur
via \textit{parametric resonance}\,\cite{Kofman:1994rk}.
Except for very small powers $p$, all of our DCI models are large-field
models of inflaton.
Therefore, if the inflaton potential has positive curvature
around $\tau \simeq \tau_0$, i.e.\ for $p > 1$, we expect
preheating to take place via ordinary parametric resonance.
On the other hand, if the inflaton potential has negative curvature
around $\tau \simeq \tau_0$, i.e.\ for $p < 1$, we are confronted with
the unusual situation of large inflaton field values in combination with
a negatively curved potential---a scenario which has not been studied
so far.
We conjecture that in this unusual case preheating ends up being a
combination of tachyonic inflaton oscillations as well as parametric resonance.
A proof of this conjecture however requires a detailed numerical analysis
and is beyond the scope of this paper.
Because of this, we leave a further study of preheating in the context of dynamical
chaotic inflation for future work---expecting that it might bear some interesting
phenomenological aspects such as a characteristic spectrum of gravitational waves
or an influence on the reheating temperature.
For now we content ourselves with the fact that, once preheating is over,
the actual reheating process based on the perturbative
decay of the inflaton field is guaranteed to set in.
At this stage, the inflaton oscillates around the minimum of its quadratic
potential, dissipating energy due to the expansion and its decay into
SM particles, until it eventually reaches the true supersymmetric vacuum
at the origin of moduli space.


The perturbative decays of the inflaton field into SM particles derive
from higher-dimensional operators in the K\"ahler potential coupling
the inflaton to gauge-invariant products of SM fields,
\begin{align}
K \supset  \frac{C_5}{M_{\textrm{Pl}}} H_u H_d \left(\Phi + \Phi^\dag\right)
+ .. + \textrm{h.c.} \,.
\label{eq:Kinfdec}
\end{align}
Here, $H_u$ and $H_d$ denote the Higgs fields of the supersymmetric standard model
and $C_5$ is a dimensionless numerical coefficient, which we
assume to be of $\mathcal{O}(1)$.
The $H_u H_d\,\Phi^\dag$ coupling in the K\"ahler potential induces the
following dimension-five term in the scalar Lagrangian,
\begin{align}
\mathcal{L} \sim C_5 \frac{m_{\phi}^2}{M_{\textrm{Pl}}} H_u H_d \,\phi^* +
.. + \textrm{h.c.} \,, \quad m_\phi \sim \lambda \Lambda. \label{eq:Linfdec1}
\end{align}
A similar term is also contained in the $F$-term contributions to the scalar
potential from the meson fields $M^n = P^n\bar{P}^n/\Lambda$ coupling to the inflaton.
If the symmetries of a given DCI model, in particular $R$ symmetry, allow for
the operator $H_u H_d\,\Phi$ in the K\"ahler potential, then the superpotential
also always contains a term $H_u H_d \,P^n \bar{P}^n$, with the $\left(P^n,\bar{P}^n\right)$
being some strong-sector quark flavors.
Schematically, we then have,%
\footnote{The presence of the term $H_u H_d\,\Phi$ in the K\"ahler potential indicates
that $H_u H_d$ carries $R$ charge $0$, which is well motivated from, for instance,
the \textit{pure gravity} mechanism for the mediation of supersymmetry
breaking~\cite{Ibe:2011aa}.
In this case, however, $H_u H_d$ also couples to the singlet fields $Z_{ij}$,
which stabilize the meson directions in moduli space,
in the superpotential, $W \supset Z_{ij}\left(Q^iQ^j +H_uH_d\right)$, cf.\
Eq.~\eqref{eq:SPNR0}.
This is undesirable, as the $F$-term conditions induced by the singlets $Z_{ij}$ then cause
the Higgs fields to acquire very large VEVs, $H_uH_d \sim \Lambda^2$.
These VEVs cancel the meson VEVs, $M^{ij}\sim \Lambda^2$, in the superpotential,
such that singlet-field $F$-terms vanish, supersymmetry is restored and
the inflaton potential disappears.
One possible way out of this problem is to forbid the $Z\,H_uH_d$ coupling
by means of a discrete $\mathbb{Z}_2$ symmetry, under which
$\Phi$ and $H_uH_d$ carry even charge and the fields $Z_{ij}$ and $Q^iQ^j$
carry odd charge, cf.\ Ref.~\cite{Kawasaki:2000ws} for a
more detailed discussion
in the case of quadratic chaotic inflation.}
\begin{align}
W \supset \lambda_0 \,\Phi\, P^n \bar{P}^n + \frac{\lambda_1}{M_{\textrm{Pl}}}
H_u H_d\, P^n \bar{P}^n \,.
\end{align}
Below the dynamical scale, where the meson fields $M^n$ correspond to
the dynamical degrees of freedom, this superpotential turns into
\begin{align}
W \supset \lambda_0 \,\Lambda \,\Phi\, M^n + \lambda_1\,\frac{\Lambda}{M_{\textrm{Pl}}}
H_u H_d\, M^n \,,
\end{align}
so that the $F$-term contributions to the potential from the fields $M^n$ include
the following term,
\begin{align}
\mathcal{L} \sim C_5' \frac{m_{\phi}^2}{M_{\textrm{Pl}}} H_u H_d \,\phi^*
+ \textrm{h.c.} \,, \quad C_5' = \frac{m_\phi^2}{\lambda_0\lambda_1 \Lambda^2} \,.
\label{eq:Linfdec2}
\end{align}


Based on naive dimensional analysis,
the perturbative inflaton decay rate resulting from the interactions in
Eqs.~\eqref{eq:Linfdec1} and \eqref{eq:Linfdec2} can then be estimated as,
\begin{align}
\Gamma_\phi \sim \frac{m_\phi}{8\pi}\left(C_5 + C_5'\right)^2
\left(\frac{m_\phi}{M_{\textrm{Pl}}}\right)^2 \,.
\label{eq:Gammaphi}
\end{align}
At the same time, the couplings in Eqs.~\eqref{eq:Linfdec1} and \eqref{eq:Linfdec2}
are also responsible for the non-adiabatic production of radiation during
preheating.
As they are strongly suppressed by inverse powers of the Planck mass, we can take
it for granted that during preheating most of the initial vacuum energy is transferred into
nonrelativistic inflaton particles and only a small
fraction into radiation.
This implies in particular that the dynamics of preheating can be safely neglected
when estimating the reheating temperature.
Instead, we can simply employ the standard expression for the reheating
temperature $T_{\textrm{RH}}$ in dependence of the inflaton decay
rate, $T_{\textrm{RH}} \simeq 0.5 \left(\Gamma_\phi M_{\textrm{Pl}}\right)^{1/2}$.
Assuming that  $C_5$ and $C_5'$ are not required to vanish
due to particular symmetry reasons, the dimension-five operators in the
Lagrangian then yield the dominant contribution to $T_{\textrm{RH}}$,
\begin{align}
T_{\textrm{RH}} \sim 10^{10}\,\textrm{GeV}
\left(C_5 + C_5'\right)
\left(\frac{m_\phi}{3\times10^{13}\,\textrm{GeV}}\right)^{3/2} \,,
\label{eq:TRH1}
\end{align}
and reheating predominantly occurs via the decay of the inflaton field into
the SM Higgs fields $H_u$ and $H_d$.
Note that here we have worked
with an inflaton mass $m_\phi$ of $\mathcal{O}\left(10^{13}..\,10^{14}\right)\,\textrm{GeV}$.
As we will see in the next section, this actually turns out
to be a typical value, if one also takes account that the power spectrum
of the scalar CMB perturbations must be correctly normalized.%
\footnote{If the mass of the inflaton is as large as
$\mathcal{O}\left(10^{14}\right)\,\textrm{GeV}$, its decay
products have extremely large momenta.
Nonetheless, one can show that also in this case the inflaton decay products
thermalize soon after their production~\cite{Harigaya:2013vwa}.
We can therefore safely use the standard expression for the reheating temperature.}
As a result, we find that high reheating temperatures of at least
$\mathcal{O}\left(10^9\right)\,\textrm{GeV}$ can be achieved rather easily,
which is favorable for the successful realization of thermal leptogenesis~\cite{Fukugita:1986hr}.


The above estimate of the reheating temperature is solely based on
perturbative considerations; and indeed we are rather confident that to good
approximation it is not necessary to include the effect of nonperturbative
particle production during preheating into these estimates.
On the other hand, there might be additional nonperturbative effects
that could potentially have a sizable impact on the reheating temperature, i.e.\ the
formation and evaporation of so-called oscillons or $I$-balls~\cite{Kolb:1993hw}.
For fractional powers $p < 2$, the inflaton potential is shallower than a
quadratic one for large field values, but quadratic around the origin.
In this case, it is known that, during the oscillating phase of the
inflaton field, instabilities in the spatial field configuration
are able to grow, until they eventually form quasi-stable lumps
of inflaton particles in a coherent state.
These solitonic field configurations are commonly referred to
as oscillons or $I$-balls.
In a sense, they represent the real analogue of $Q$-balls~\cite{Coleman:1985ki},
which might be formed at the end of inflation if the inflaton was able to freely
move in the complex plane.
However, since in our case the real inflaton component $\sigma$
is stabilized at $\sigma=0$ due its Hubble-induced mass term for
all times during inflation, $Q$-balls have no chance of being formed
and we rather have to contemplate the possibility of $I$-balls
emerging at the end of inflation.
If the inflaton field should indeed fragment into $I$-balls, the spatial
distribution of inflaton particles would end up being segregated into
small, sharply localized density peaks with a spatial extent $L$ of
the order of the inverse inflaton mass, $L \sim m_\phi^{-1}$.
Reheating then occurs via $I$-ball evaporation, which is why it is
in general subject to quantum effects such as Pauli
blocking or Bose enhancement~\cite{Cohen:1986ct}.
Fortunately, we however do not have to care about the impact of
$I$-balls on the reheating temperature in the case of our DCI models.
Even if $I$-balls should be formed, we expect our above estimate
of the reheating temperature to remain valid.
Since the inflaton decay width is suppressed by at least two
inverse powers of the Planck mass, cf.\ Eq.~\eqref{eq:Gammaphi}, and
hence very small, the number density of inflaton decay products around
individual $I$-balls is also always very small, so that neither Pauli blocking
nor Bose enhancement ever become effective.


Finally, let us mention one more nonperturbative effect at the end of inflation, which
might have interesting phenomenological consequences, namely the formation and decay of
primordial black holes (PBHs)~\cite{GarciaBellido:1996qt}.
As we mentioned above, we are not able to calculate the inflaton
potential at intermediate field values around $\tau \sim \Lambda/\lambda$,
as our gauge theory is always strongly coupled in this regime.
However, under certain circumstances an interesting phenomenon might
occur in this region.
Provided that the inflaton potential is very flat around $\tau \sim \Lambda/\lambda$,
the motion of the inflaton field is largely governed by quantum fluctuations,
since the classical driving force due to the potential gradient is weak.
The metric fluctuations leaving the Hubble horizon at this point therefore
have a very large amplitude.
Such large fluctuations then lead to the formation of PBHs, which give rise
to a rich phenomenology.
PBHs may account for the relic density of dark matter~\cite{Hawking:1971ei},
result in nonthermal baryogenesis~\cite{Toussaint:1978br},
act as an alternative origin of the primordial density perturbations~\cite{Fujita:2013bka},
and/or seed super-massive black holes~\cite{Carr:1984id}.
Furthermore, if the inflaton potential is indeed very flat around
$\tau \sim \Lambda/\lambda$, a smaller number of $e$-folds is required
in the large-field regime, where $V\propto \tau^p$.
This effectively results in a smaller value of $N_e$ and hence
in a larger tensor-to-scalar ratio and a smaller scalar spectral index,
which improves in particular the consistency between our predictions and
the Planck constraints on these observables for DCI models with $p \lesssim 1$,
cf.\ Sec.~\ref{subsec:observables}.
In summary, we conclude that it would be very interesting to know whether
and, if so, to which extent PBHs are indeed formed in scenarios of dynamical chaotic inflation.
Due to our poor knowledge of the inflaton potential at intermediate
field values, we are however at present unable to make any further
quantitative statements regarding PBHs.
Further investigations into this direction are therefore very desirable.


\subsection{Predictions for the inflationary CMB observables}
\label{subsec:observables}


\begin{figure}[t]
\centering
\includegraphics[width=0.54\textwidth]{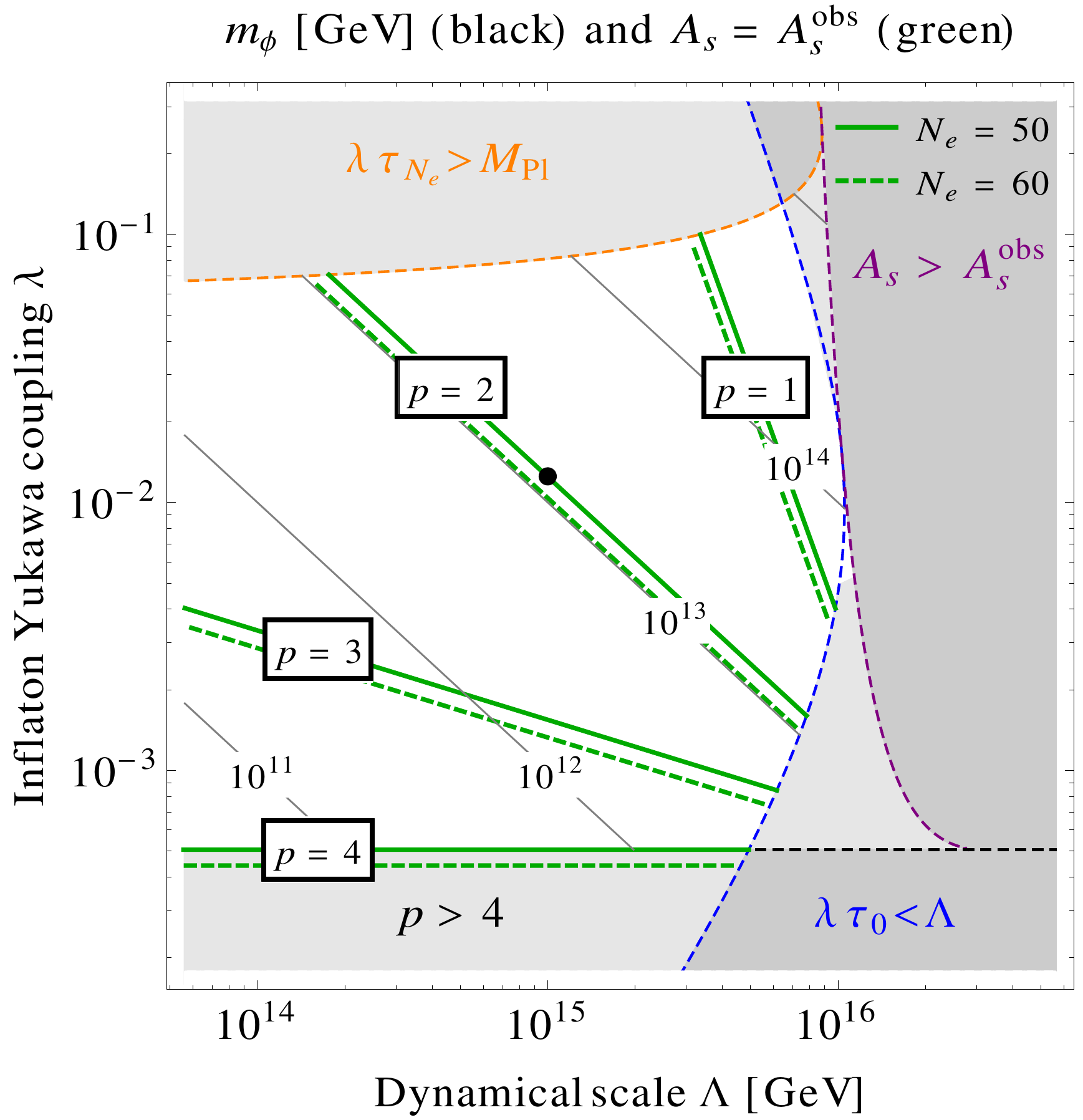}
\caption{Viable region in the $(\lambda,\Lambda)$ plane.
The green lines indicate where the amplitude of the scalar power
spectrum can be successfully reproduced for different values of
the power $p$, while the gray contours correspond to different values
of the inflaton mass $m_\phi \sim \lambda\Lambda$.
The regions of parameter space where one or more of our theoretical
requirements listed in Eqs.~\eqref{eq:lambdaupper}, \eqref{eq:pbound}
and \eqref{eq:lambdalower} are violated are also correspondingly marked.
The black dot marks the position of the parameter pointed used in Fig.~\ref{fig:scales}.}
\label{fig:norm}
\end{figure}


In the last step of our phenomenological discussion, we shall now finally turn to
the predictions for the inflationary CMB observables.%
\footnote{For a comprehensive comparison between the predictions of chaotic inflation
and the predictions of other small-field and large-field models of inflation,
cf.\ also Ref.~\cite{Alabidi:2010sf}.}
All observables that we are interested in can be readily calculated in terms of the
inflationary slow-roll parameters $\varepsilon$, $\eta$ and $\xi$,
which are defined as~\cite{Lyth:1998xn} 
\begin{align}
\varepsilon = \frac{M_{\textrm{Pl}}}{2}\left(\frac{V'}{V}\right)^2\,, \quad
\eta = M_{\textrm{Pl}}^2 \frac{V''}{V} \,, \quad
\xi^2 = M_{\textrm{Pl}}^4 \frac{V'V'''}{V^2} \,, \quad
(\cdot)' \equiv \frac{d}{d\tau} \,.
\end{align}
According to our expression for the scalar potential in Eq.~\eqref{eq:Vtau}
and making use of the fact that $\tau_{N_e} \simeq \left(2pN_e\right)^{1/2}M_{\textrm{Pl}}$,
these three parameters are found to take the following form,
\begin{align}
\varepsilon \simeq \frac{p}{4N_e} \,, \quad
\eta \simeq \frac{p-1}{2N_e} \,,\quad
\xi^2 \simeq \frac{(p-1)(p-2)}{4N_e^2} \,.
\label{eq:srpara}
\end{align}
Meanwhile, the scalar spectral amplitude $A_s$, the scalar spectral index $n_s$, the running
of the scalar spectral index $\alpha_s \equiv dn_s/d\ln k$
as well as the tensor-to-scalar ratio $r$ are given by
\begin{align}
A_s = \frac{V/M_{\textrm{Pl}}^4}{24\pi^2\varepsilon} \,, \quad
n_s = 1 +2 \eta - 6 \varepsilon \,, \quad
\alpha_s = 16\varepsilon\eta - 24 \varepsilon^2 - 2 \xi^2 \,, \quad
r = 16\varepsilon \,.
\end{align}
Inserting our results for $\varepsilon$, $\eta$ and $\xi$ in Eq.~\eqref{eq:srpara}
into these expressions, we then obtain
\begin{align}
A_s = \frac{\lambda^p}{12\pi^2p^2}
\left(\frac{\Lambda}{M_{\textrm{Pl}}}\right)^{4-p}\left(2 pN_e\right)^{1+p/2} \,, \quad
n_s - 1 = -\frac{p+2}{2N_e}, \quad
\alpha_s = -\frac{2+p}{2N_e^2} \,, \quad
r = \frac{4p}{N_e} \,.
\label{eq:obspre}
\end{align}
In consequence of the simple form of the potential, all observables except for $A_s$
turn out to be independent of the actual model parameters, $\Lambda$ and $\lambda$.
Instead, $n_s$, $\alpha_s$ and $r$ are fully determined by the power $p$ as well
as the number of $e$-folds $N_e$ which elapse between the time when the CMB scales cross
outside the Hubble horizon and the end of inflation.
The amplitude $A_s$ does, by contrast, depend on $\Lambda$ and $\lambda$,
which allows us to eliminate one free parameter, i.e.\ $\Lambda$ or $\lambda$,
by demanding that our prediction
for $A_s$ must reproduce the observed value,
$A_s^{\textrm{obs}} \simeq 2.21\times 10^{-9}$~\cite{Ade:2013zuv},
\begin{align}
\lambda = C_p^{1/p} \left(\frac{\Lambda}{M_{\textrm{Pl}}}\right)^{1-4/p} \,, \quad
\Lambda = C_p^{1/(4-p)}\lambda^{p/(p-4)}M_{\textrm{Pl}} \,, \quad
C_p = \frac{12\pi^2p^2A_s^{\textrm{obs}}}{\left(2pN_e\right)^{1+p/2}} \,.
\label{eq:LLrel}
\end{align}


\begin{figure}
\centering
\includegraphics[width=0.485\textwidth]{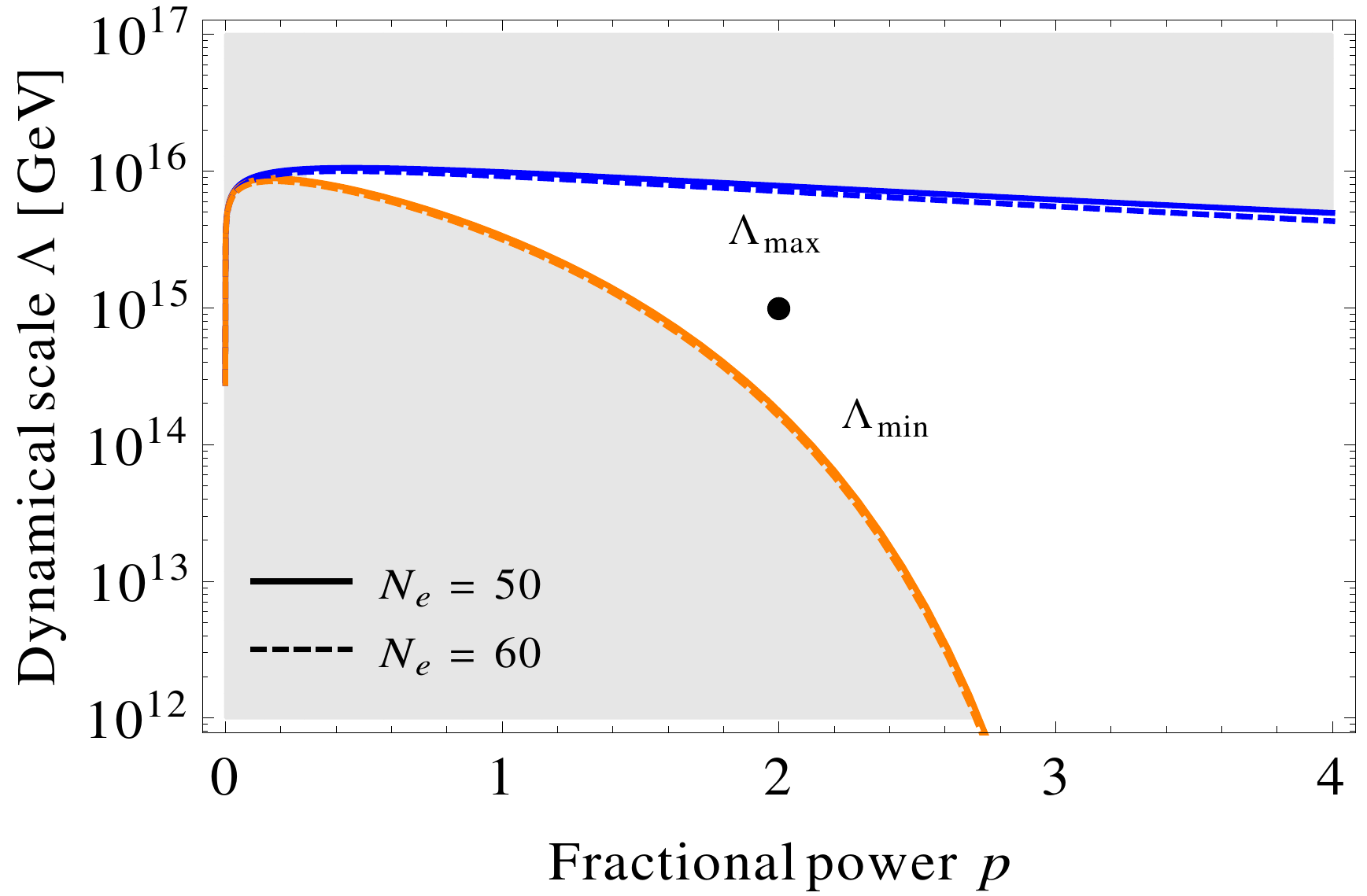}\hspace{0.02\textwidth}
\includegraphics[width=0.485\textwidth]{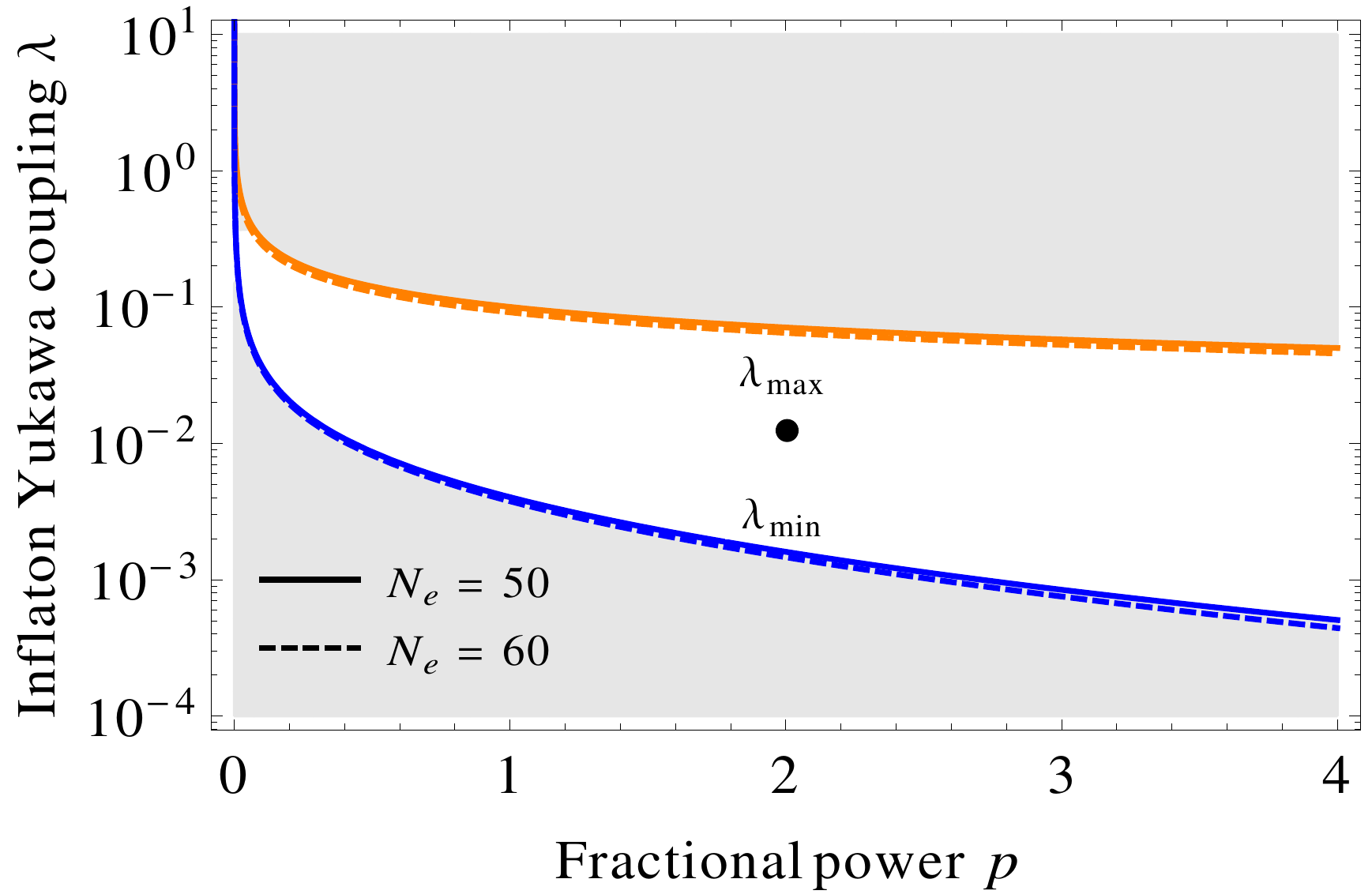}
\caption{Viable ranges for the dynamical scale $\Lambda$ \textbf{(left panel)}
and the inflaton Yukawa coupling $\lambda$ \textbf{(right panel)}
according to the bounds in Eqs.~\eqref{eq:lambdaupper} and \eqref{eq:lambdalower}
and after imposing the condition that the scalar power spectrum be correctly normalized,
$A_s = A_s^{\textrm{obs}}\simeq 2.21\times 10^{-9}$~\cite{Ade:2013zuv},
for fixed $N_e$ and as functions of the power $p$, cf.\ Eq.~\eqref{eq:Llminmax}.
The black dots mark again the position of the parameter pointed used in Fig.~\ref{fig:scales}.}
\label{fig:ranges}
\end{figure}


In Fig.~\ref{fig:norm}, we present the combinations of $\Lambda$ and $\lambda$
values that yield the correct amplitude $A_s$ for fixed values of $p$ and with $N_e$ being
varied between $50$ and $60$.
Besides that, we also indicate the value of the inflaton
mass, $m_\phi \sim \lambda\Lambda$, in the respective parts of
the $\left(\Lambda,\lambda\right)$ plane and illustrate which
regions of parameter space are theoretically inaccessible due
to the upper and lower bounds on the Yukawa coupling $\lambda$ in
Eqs.~\eqref{eq:lambdaupper} and \eqref{eq:lambdalower}.
In fact, the relation between $\Lambda$ and $\lambda$ in
Eq.~\eqref{eq:LLrel} in combination with
the bounds in Eqs.~\eqref{eq:lambdaupper} and \eqref{eq:lambdalower}
results in explicit intervals, $\left[\Lambda_{\textrm{min}},
\Lambda_{\textrm{max}}\right]$ and $\left[\lambda_{\textrm{min}},
\lambda_{\textrm{max}}\right]$, in which $\Lambda$ and $\lambda$
are allowed to take values, cf.\ Fig.~\ref{fig:ranges},
\begin{align}
\lambda_{\textrm{min}} \simeq & \: p^{p/4-1} C_p^{1/4} \,, \quad
\lambda_{\textrm{max}} \simeq  \left(2pN_e\right)^{-1/2} \,,
\label{eq:Llminmax}\\ \nonumber
\Lambda_{\textrm{min}} \simeq & \: 
\left(2pN_e\right)^{p/(8-2p)} C_p^{1/(4-p)} M_{\textrm{Pl}}\,, \quad
\Lambda_{\textrm{max}} \simeq p^{p/4} C_p^{1/4} M_{\textrm{Pl}}
\end{align}
As we can see from Figs.~\ref{fig:norm} and \ref{fig:ranges}, the coupling constant $\lambda$
typically falls into the range $\mathcal{O}\left(10^{-3}..\,10^{-1}\right)$,
as already anticipated below Eq.~\eqref{eq:lambdalower}, while
the dynamical scale $\Lambda$ is required to take a value remarkably close to
the GUT scale, $\Lambda \sim 10^{15}..\,10^{16}\,\textrm{GeV}$.
This leads us to the conclusion that our DCI scenarios featuring a dynamical
scale $\Lambda \sim \Lambda_{\textrm{GUT}}$ seem to provide a possible and in fact
very appealing explanation for why the energy scale of inflation is so close
to the GUT scale.
At the same time, we find that the inflaton mass is of
$\mathcal{O}\left(10^{14}..\,10^{15}\right)\,\textrm{GeV}$ for most of the
interesting $p$ values.
However, for very large powers, $p\gtrsim 2$, the inflaton mass can also become
significantly smaller and take values in the range
$10^{12}..\,10^{13}\,\textrm{GeV}$.


After this discussion of the normalization of the scalar power spectrum
and its implications, we shall now comment on our predictions for $n_s$ and $r$.
For a fixed number of $e$-folds $N_e$, our expressions in Eq.~\eqref{eq:obspre}
imply a linear relation
between $n_s$ and $r$ that applies independently of the value chosen for the power $p$,
\begin{align}
r = 8 \left[1 - n_s - \frac{1}{N_e}\right]
= 0.16 + 8 \left[(0.96-n_s) + \left(\frac{1}{50}-\frac{1}{N_e}\right)\right] \,,
\label{eq:nsrrel}
\end{align}
and which may be regarded as a consistency relation that needs to be satisfied
in every DCI model.
If we combine this result with the standard relation between $r$ and the
tensor spectral tilt, $n_t = -r/8$, which generically holds in every single-field
slow-roll model of inflation~\cite{Liddle:1992wi}, we obtain a second prediction
that is characteristic for all DCI models and which relates the two spectral tilts,
$n_s$ and $n_t$, to each other,
\begin{align}
n_t = -1 + n_s + \frac{1}{N_e}
= -0.02 + (n_s-0.96) + \left(\frac{1}{N_e}-\frac{1}{50}\right)\,. 
\end{align}
Remarkably enough, we find that for a spectral index $n_s$ close to $0.96$,
the tensor-to-scalar ratio predicted by dynamical chaotic inflation is in
very good agreement with the value reported by the BICEP2 collaboration.
On the other hand, the current experimental sensitivity to $n_t$ is not yet
sufficient to allow for a meaningful comparison between our prediction and the data.
It is interesting to note, though, that at the present stage it seems as if
the BICEP2 data slightly favored a positive value of $n_t$, i.e.\ a blue-tilted
tensor spectrum~\cite{Gerbino:2014eqa}, which, if confirmed, would challenge
the entire inflationary paradigm~\cite{Brandenberger:2014faa}.
But for the time being, it is definitely too early to jump to any conclusions
and further observations are needed in order to determine the sign and eventually
the magnitude of $n_t$.


\begin{figure}
\centering
\includegraphics[width=0.485\textwidth]{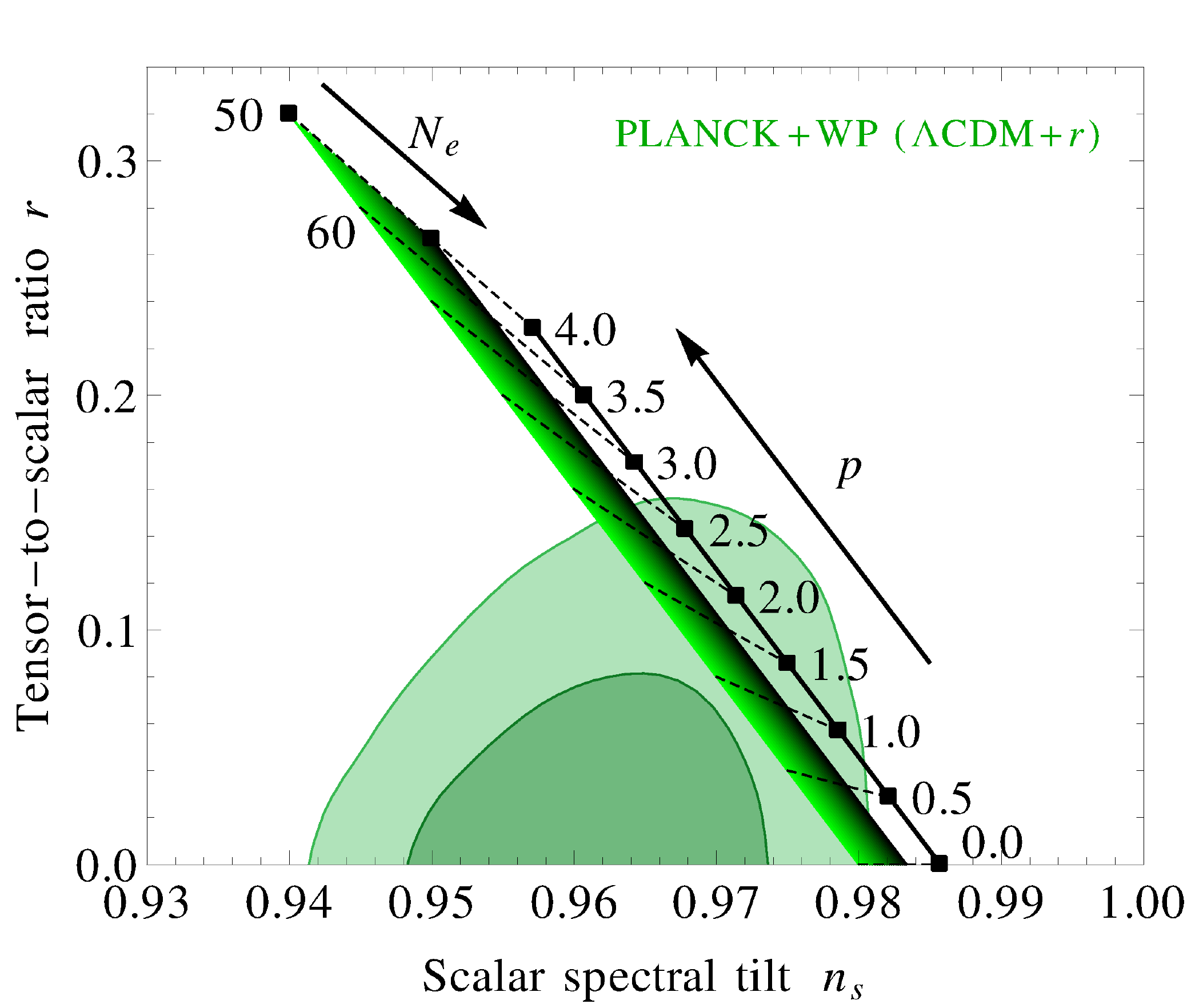}\hspace{0.02\textwidth}
\includegraphics[width=0.485\textwidth]{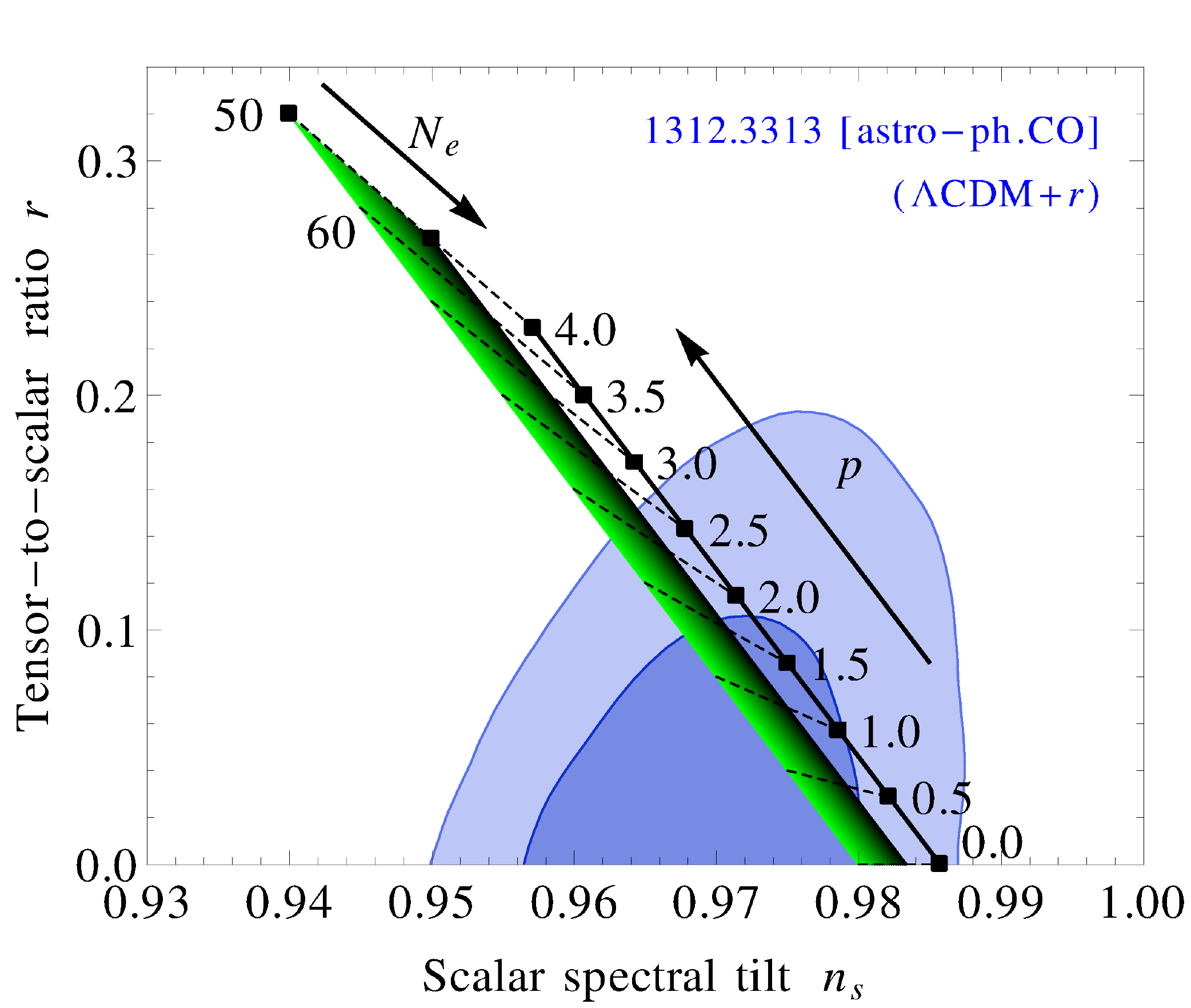}
\caption{Comparison between our predictions for the scalar spectral tilt $n_s$
and the tensor-to-scalar ratio $r$, cf.\ Eq.~\eqref{eq:obspre}, and the constraints
on these two observables according to the PLANCK collaboration~\cite{Ade:2013zuv} \textbf{(left panel)}
and the authors of Ref.~\cite{Spergel:2013rxa} \textbf{(right panel)}, respectively.
Note the linear relation given in Eq.~\eqref{eq:nsrrel}.}
\label{fig:nsr}
\end{figure}


The $r$ value measured by the BICEP2 experiment, $r = 0.20_{-0.05}^{+0.07}$,
conflicts with the upper limits on this observable, which had previously been
deduced from the WMAP, $r<0.13$ (at 95\,\%\,CL)~\cite{Hinshaw:2012aka},
as well as from the PLANCK data, $r<0.11$ (at 95\,\%\,CL)~\cite{Ade:2013zuv}.
And indeed, particularly for this reason, the BICEP2 analysis is presently
under intense scrutiny~\cite{Mortonson:2014bja}.
At the moment, it is not entirely clear which fraction of the BICEP2 signal could 
also come about simply due to polarized dust emission in our own galaxy and it might be that
the true value of $r$ is in fact much smaller than $r\simeq0.2$.
Ultimately, only additional experimental data will help us settle the question whether
$r$ is really of $\mathcal{O}(0.1)$ or whether the BICEP2 signal is eventually nothing
but a foreground effect and $r$ actually lies orders of magnitude below the
current experimental sensitivity.
Fortunately, a multitude of dedicated experiments is currently underway, so that we will
soon know much better how to correctly interpret the BICEP2 measurement.
For the moment, we will take the attitude that the BICEP2 collaboration has indeed
detected a signal of primordial gravitational waves in the CMB and we shall argue
that a tensor-to-scalar ratio of $\mathcal{O}(0.1)$ may  very well be explained
in the context of dynamical chaotic inflation.


\begin{figure}[t]
\centering
\includegraphics[width=0.68\textwidth]{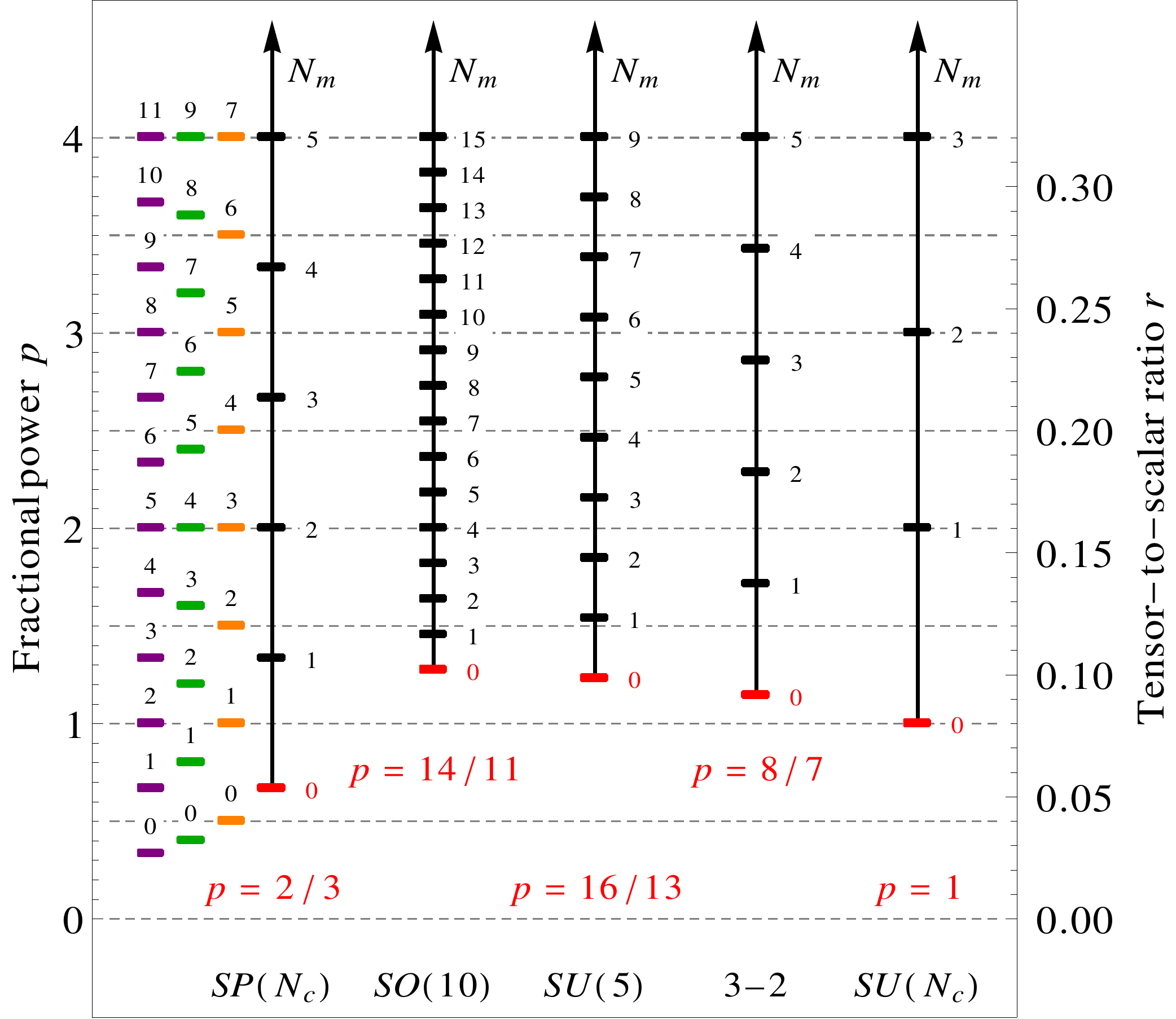}
\caption{Predictions of all DCI models constructed in this paper for the power $p$
and the tensor-to-scalar ratio $r$, respectively.
For definiteness, we have set $N_e$ to $50$.
In the case of the $SP(N_c)$ theories, the four different columns respectively refer to
(from right to left): $N_c = 2$ (black), $N_c = 3$ (orange), $N_c = 4$ (green), $N_c =5$
(purple).
As for the $SP(1)$ case, the predictions for $p$ and $r$ are identical to those obtained
in the general $SU(N_c)$ case, cf.\ Eq.~\eqref{eq:SPSUequi}.
The small numbers next to each vertical bar denote the respective numbers of additional
massive quark flavors $N_m$ coupling to the inflaton field.
Hence, the bars labeled with a red $0$ correspond to the models without any additional
mass input scales.
For these models, we state the predicted power explicitly in each column.}
\label{fig:powers}
\end{figure}


The predictions for the scalar spectral index as well as the tensor-to-scalar ratio
resulting from the fractional power-law potential in Eq.~\eqref{eq:Vtau} are shown in
Fig.~\ref{fig:nsr}.%
\footnote{For other recent studies of polynomial
chaotic inflation in the light of BICEP2, cf.\ for instance
Ref.~\cite{Kobayashi:2014jga}.}
As pointed out above, for fixed $N_e$, the predicted value for $n_s$ and $r$ exhibit
a linear relation controlled by the power $p$, cf.\ Eq.~\eqref{eq:nsrrel}.
In the two panels of Fig.~\ref{fig:nsr}, we compare our predictions with the
constraints on $n_s$ and $r$ derived from the PLANCK data as well as with the
constraints obtained by the authors of Ref.~\cite{Spergel:2013rxa}.
In this latter work, the PLANCK data has been re-analyzed taking particular care of
possible systematics in the 217 GHz temperature map.
The results of Ref.~\cite{Spergel:2013rxa} differ from the PLANCK results mainly
by the fact that they imply a spectral index roughly larger by $1\,\sigma$
than the PLANCK best-fit value.
From Fig.~\ref{fig:nsr}, we see that all powers $p$ in the range
$0\lesssim p\lesssim 2$ are still consistent with the PLANCK data at the 
$2\,\sigma$ level.
This observation also justifies that we neglected powers larger than $4$
in our analysis, cf.\ Eq.~\eqref{eq:pbound}.
From the perspective of dynamical chaotic inflation, we expect that
future updates of the constraints on $n_s$ and $r$
will eventually single out a particular region or even a particular point within
the green band in Fig.~\ref{fig:nsr}.
If that should really be the case, it would allow us to infer
highly non-trivial information about the strongly interacting gauge dynamics 
in the inflaton sector.
In Fig.~\ref{fig:powers}, we present a summary of the predictions for the
tensor-to-scalar ratio for all DCI models that we have constructed in this paper.
If it should turn out feasible to determine $r$ with large precision, this
figure would enable us to read off the corresponding
gauge dynamics underlying the inflationary phase.


Finally, we remark that dynamical chaotic inflation always predicts a negative
and rather small running of the scalar spectral index.
According to Eq.~\eqref{eq:obspre}, $\left|\alpha_s\right|$
never exceeds values of $\mathcal{O}(10^{-3})$.
Since the release of the BICEP2 results, many authors have pointed out
that a sizable negative running of the scalar spectral index
$\left|\alpha_s\right| \sim 10^{-2}$ would allow to relieve the
tension between the BICEP2 best-fit value for $r$ and the upper bounds
on $r$ according to the PLANCK and WMAP data~\cite{Ma:2014vua}.
The BICEP2 collaboration itself noticed, for instance, that the
upper bound on $r$ deduced from the PLANCK data significantly increases
as soon as the scalar spectral index is allowed to run.
Including $\alpha_s$ as an additional free parameter in the analysis of the PLANCK
data, one obtains a best-fit value for $\alpha_s$ of roughly $-0.02$ and
the maximally allowed $r$ value (at 95\,\%\,CL) becomes as large as $0.26$.
In the context of dynamical chaotic inflation, the clear result however is
that $\left|\alpha_s\right|$ is too small; unfortunately, it cannot be of any
help in reconciling the BICEP2 best-fit value with the PLANCK and WMAP bounds on $r$.
As already mentioned several times, the true value for $r$ may be
smaller than $r\simeq 0.2$ by some $\mathcal{O}(1)$ factor in any case
due to unaccounted contributions to the B-mode polarization spectrum
from polarized dust.
Besides that, a variety of alternative explanations for the discrepancy between
BICEP2 and PLANCK/WMAP have recently been presented in the literature.
These reach from non-standard features in the primordial scalar
power spectrum that result in a suppression
of adiabatic perturbations on large scales~\cite{Contaldi:2014zua}
over additional relativistic particle species such as
sterile neutrinos~\cite{Giusarma:2014zza} to
the presence of isocurvature perturbations as they
occur, for instance, in the curvaton model~\cite{Kawasaki:2014lqa}.
While all of these approaches look very interesting,
it remains to be seen which of them might be brought
into accordance with dynamical chaotic inflation.
A further discussion of this question is beyond the
scope of the paper and left for future work.


\section{Conclusions and outlook}
\label{sec:conclusions}


In view of the most recent CMB observations,
chaotic inflation based on a simple monomial
potential such as the one in Eq.~\eqref{eq:powerlawV}
appears to be a viable, particularly attractive and theoretically well
motivated scenario of cosmic inflation.
So far, it has however been unclear what kind of dynamical mechanism might underlie
the emergence of such a potential in the low-energy effective description of
supergravity.
In conventional SUGRA models of chaotic inflation, the inflaton potential
is usually merely derived from \textit{ad hoc}
input functions, i.e.\ specifically chosen superpotentials
and K\"ahler potentials, which may or may not be motivated or inspired by
more fundamental concepts.
In these models, the energy scale of inflation then typically ends up
being determined by a combination of dimensionful input parameters,
so that a successful description of inflation around the GUT scale
is only achieved as long as certain input mass scales are adjusted to
particular values by hand.
These shortcomings are remedied in the framework of dynamical chaotic
inflation (DCI), which had originally been proposed in Ref.~\cite{Harigaya:2012pg}.
In dynamical chaotic inflation, the inflaton field couples to the
quark flavors of a strongly interacting supersymmetric gauge theory in
such a way that it acquires a fractional power-law potential
via the purely quantum mechanical effect of dimensional transmutation.
The virtue of dynamical chaotic inflation is therefore twofold:
On the one hand, the general DCI recipe outlined in Sec.~\ref{sec:idea}
allows to construct monomial inflaton potentials
featuring some fractional power $p \in \mathbb{Q}^+$ in combination with a canonical
kinetic term for the inflaton \textit{in field theory}.
Up to now, such constructions had only been achieved in string theory in the context
of axion monodromy.
On the other hand, dynamical chaotic inflation in its simplest realizations
is conformally invariant at the classical level.
As long as the quark flavors coupling to the inflaton field do not possess
any explicit mass terms, no input mass scales are involved in the 
generation of the inflaton potential $V$. 
Instead, the scale of inflation, $V^{1/4}$, is directly related to the
effective dynamical scale of the strong interactions in the inflaton sector,
$V^{1/4}\sim\Lambda_{\textrm{eff}}$.
If we assume that the gauge dynamics in the hidden sector are
perturbative and eventually asymptotically free at energies around and
above $M_{\textrm{Pl}}$, the RGE running of the gauge coupling constant 
automatically implies that the effective dynamical scale $\Lambda_{\textrm{eff}}$
must be located a few orders of magnitude below the Planck scale,
$\Lambda_{\textrm{eff}} \sim 10^{-3}..\,10^{-1} M_{\textrm{Pl}}$.
Dynamical chaotic inflation thus explains, purely based on the behavior of
dimensionless coupling constants, why $V^{1/4}$ has to be close to the GUT scale.


In this paper, we have undertaken a first step towards the formulation of
a more complete theory of dynamical chaotic inflation.
While we had exclusively studied the simplest case of $SP(N_c)$ gauge dynamics
in our earlier work~\cite{Harigaya:2012pg}, we have now shown how to construct
consistent DCI models for a variety of different supersymmetric gauge
theories.
The idea behind the construction of these models is always the same:
All one needs to do is to couple the inflaton field to an s-confining supersymmetric
gauge theory in such a way that it mutates into a model of dynamical supersymmetry
breaking in the limit of large inflaton field values.
The nonzero vacuum energy density associated with dynamical supersymmetry breaking
at large inflaton values then acts as a scalar potential for the inflaton
field during inflation.
At the same time, the fact that all DCI models reach a phase of s-confinement
after the end of inflation ensures good control over the particle spectrum in the
infrared as well as over the properties of the true ground state.
In particular, s-confinement allows us to stabilize all flat directions in moduli
space in an easy manner, i.e.\ by simply coupling the fundamental degrees of freedom
at high energies to an appropriate number of singlet fields. 
Moreover, s-confinement implies that all global flavor symmetries are preserved
at the origin of field space.
We therefore do not have to worry about the presence of massless Goldstone bosons
after the end of inflation.
All these advantages of s-confinement, however, do not mean that an s-confining
true vacuum is the only option
for the further evolution of the inflaton sector at small field values.
As discussed in App.~\ref{app:dynSUSY}, it is also conceivable that the DSB model
governing the inflationary dynamics flows to another, second DSB model at low energies.
This scenario offers the possibility that dynamical chaotic inflation and low-energy
supersymmetry breaking could potentially be embedded into a common theory.
Alternatively, the inflaton sector may also reach a non-Abelian Coulomb phase
in the conformal window.
In this case, the inflaton would turn into an unparticle after the end of
inflation, so that the reheating process would occur in a very exotic fashion.
The explicit construction of consistent DCI models featuring one of these two alternatives
to an s-confining vacuum at low energies is beyond the scope of this paper, but certainly desirable.


The various DCI models we constructed in this paper are either based on 
$SP(N_c)$, $SO(10)$, $SU(5)$, $SU(3)\times SU(2)$ or $SU(N_c)$ gauge dynamics.
In their simplest, conformally invariant formulation, these models predict a power $p$
in the inflaton potential of $2/3$, $14/11$, $16/13$, $8/7$, and $1$, respectively.%
\footnote{Here, $p=2/3$ refers to $SP(2)$;
the $SP(1)$ model yields the same prediction as the $SU(N_c)$ models, $p=1$.}
If we extend these models by $N_m$ additional massive flavors coupling to the inflaton
field, we are able to respectively increment our predictions for the power $p$ in
discrete steps by $N_m$ times a specific model-dependent rational number,
cf.\ Fig.~\ref{fig:powers}.
While some values of the power $p$ can be generated in different DCI models,
such as, for instance, $p=2$, most of our predictions are unique and directly
point to the underlying gauge dynamics.
Under fortunate circumstances, a precise determination of the inflationary observables
would hence allow to pinpoint the gauge dynamics of the inflaton sector.
A further possible implication of additional massive flavors is the fact
that the power $p$ in the potential may actually be an effective quantity, $p_{\textrm{eff}}$,
that smoothly varies during inflation, cf.\ Eq.~\eqref{eq:Vtot}.
While we briefly commented on this possibility in Sec.~\ref{subsec:SPNcmassive},
this scenario certainly also deserves a more detailed investigation.
Furthermore, we note that all DSB models that we considered as possible sources
for the vacuum energy density during inflation actually realize supersymmetry
breaking in a stable vacuum.
As an alternative, one might also attempt to construct DCI scenarios based
on models of dynamical SUSY breaking in meta-stable vacua such as the 
ISS model~\cite{Intriligator:2006dd}.
However, such models typically introduce explicit mass parameters for the
strongly interacting quark flavors, which may conflict with our idea
that every model of dynamical chaotic inflation should also admit a
conformally invariant limit.
The same holds true for DSB models in which supersymmetry breaking
takes place in the conformal window of some conformal field theory at
high energies~\cite{Izawa:2009nz}.
These DSB models represent yet another potential alternative to the DSB scenarios
discussed in this paper; but just like the ISS model, they depend on the
introduction of explicit mass parameters.
A dedicated analysis of the prospects for dynamical chaotic inflation in the
context of ISS-like models and DSB models with SUSY breaking
``in the conformal window'' is therefore needed.


Finally, we emphasize that the mechanism of dynamical chaotic inflation
relies on the assumption of a shift symmetry in the direction of the inflaton
field, which helps us avoid the eta problem in supergravity.
While we do not consider any effects of shift symmetry breaking
in the tree-level K\"ahler potential, we are led to introduce explicit
shift symmetry-breaking terms in the superpotential, i.e.\ the Yukawa
interactions between the inflaton field and the strongly interacting
quark fields.
These terms induce a shift symmetry-breaking contribution to the
effective K\"ahler potential at the one-loop level,
$\delta K \propto \lambda^2/\left(16\pi^2\right)\left|\phi\right|^2$,
which needs to be adequately suppressed, so as not to obtain too large
corrections to the inflationary slow-roll parameters.
The Yukawa coupling $\lambda$ is thus required to take
a rather small value, $\lambda \lesssim 0.1$, although one might probably expect
that $\lambda$ should actually be of $\mathcal{O}(1)$.
On the other hand, the limit $\lambda\rightarrow0$ just corresponds to the
restoration of the shift symmetry in the Lagrangian, so that small $\lambda$
values are still natural in the sense of 't Hooft.
Perhaps more importantly, one might worry that the shift symmetry is not present
in the Lagrangian in the first place, not even an approximate one.
As is well known, all global symmetries are believed to be eventually
violated by quantum gravity effects~\cite{Giddings:1988cx}, so that
the shift symmetry in the direction of the inflaton field cannot be 
a fundamental symmetry of the UV completion.
It is therefore important to understand how an approximate shift
symmetry may survive or emerge in the low-energy effective description.
Or put differently, dynamical chaotic inflation needs to be embedded into a
concrete string construction that explains the origin of the shift symmetry.
Such an endeavor would certainly be worth the effort.
After all, if we assume that the shift symmetry is indeed realized
at low energies, dynamical chaotic inflation appears to be one of the
most promising frameworks for the description of large-field inflation
based on a simple monomial potential.
If the idea behind the concept of dynamical chaotic inflation should
indeed turn out to be correct, we would learn that inflation is nothing but a
natural consequence of strongly interacting supersymmetric gauge dynamics shortly below
the Planck scale, a fascinating notion that calls for further exploration.


\subsubsection*{Acknowledgements}


The authors wish to thank N.\ Takeda for fruitful discussions and comments.
This work has been supported by Grant-in-Aid for Scientific Research
No.~26287039 (T.\,T.\,Y.),  No.~24740151 (M.\,I.) and No.~25105011 (M.\,I.)
from the Ministry of Education, Science, Sports, and Culture (MEXT), Japan,
Grant-in-Aid No.~26287039 (M.I.) from the Japan Society for the Promotion
of Science (JSPS) as well as by the World Premier
International Research Center Initiative (WPI), MEXT, Japan.
K.\,H.\ is supported in part by a JSPS Research Fellowship
for Young Scientists.

\newpage


\appendix


\section{Models based on \texorpdfstring{\boldmath{$SU(N_{\lowercase{c}})$}}{SU(Nc)} dynamics}
\label{app:SUN}


In Sec.~\ref{sec:SPN}, we construct and discuss models of dynamical chaotic
inflation based on strong $SP(N_c)$ gauge dynamics.
These DCI scenarios are particularly attractive due to their rather minimal
field content and the simple structure of their superpotential, which
dispenses with any higher-dimensional operators.
In this appendix, we now show how these scenarios can be generalized to the
case of $SU(N_c)$ gauge dynamics.
As we will see, the field content as well as the superpotential
of these $SU(N_c)$ models will end up being more involved than in their $SP(N_c)$
counterparts, whereby it becomes clear why we focused our attention
to the $SP(N_c)$ case in the main text.


\subsection[Massless matter fields only (fractional power \texorpdfstring{$p = 1$}{p = 1})]
{Massless matter fields only (fractional power \texorpdfstring{\boldmath{$p = 1$}}{p = 1})}
\label{subsec:SUNmassless}


Let us consider supersymmetric QCD (SQCD), i.e.\ an $SU(N_c)$ gauge theory,
with $N_f = N_c+1$ quark flavors $\big(Q^I, \bar{Q}^{\bar{I}}\big)$,
where the $Q^I$ and $\bar{Q}^{\bar{I}}$ transform in the fundamental and
antifundamental representation of $SU(N_c)$, respectively,
$Q^I \sim \Yfund\hskip2pt$ and $\bar{Q}^{\bar{I}} \sim \overline{\Yfund}\hskip2pt$.
This theory is the best-known example for an s-confining theory.
In fact, SQCD with $N_f = N_c+1$ is the first supersymmetric gauge theory
that was shown to exhibit an s-confined phase at low
energies, namely by Seiberg in Ref.~\cite{Seiberg:1994pq}, and it is the
discovery of s-confinement in SQCD that triggered the revelation of
many exact properties concerning the vacuum structure and infrared spectrum
of other gauge theories.
SQCD with $N_c+1$ flavors is therefore an excellent candidate for an s-confining theory
that is possibly suited for the construction of a working DCI model.
The low-energy dynamics of this theory are well described by a smooth effective
theory in terms of $N_f(N_f+2)$ gauge-invariant composite fields:
$N_f$ pairs of baryons $B^I$ and antibaryons $\bar{B}^{\bar{I}}$
as well as $N_f^2$ meson fields $M^{I\bar{J}}$,
\begin{align}
B_I = \epsilon_{I I_1..\,I_{N_c}} \frac{Q^{I_1} ..\,Q^{I_{N_c}}}{\Lambda^{N_c-1}} \,, \quad
\bar{B}_{\bar{I}} = \epsilon_{\bar{I} \bar{I}_1 ..\, \bar{I}_{N_c}}
\frac{\bar{Q}^{\bar {I}_1} ..\, \bar{Q}^{\bar{I}_{N_c}}}{\Lambda^{N_c-1}} \,, \quad
M^{I\bar{J}} = \frac{Q^I \bar{Q}^{\bar{J}}}{\Lambda} \,.
\end{align}


Similar as in the case of our s-confining $SP(N_c)$ theories, SQCD with $N_c+1$
flavors can be mutated into a DSB model by reducing the number
of flavors by one and stabilizing all other flat directions of the moduli space
via couplings to an appropriate number of singlet fields.
Analogously to Eq.~\eqref{eq:tree1}, we therefore introduce the
tree-level superpotential
\begin{align}
\label{eq:WtreeSUN}
W_{\textrm{tree}}  = \lambda_{I\bar{J}} \,Z_{I \bar{J}} \,Q^I \bar{Q}^{\bar{J}}
+ \frac{\kappa_I}{M_{\textrm{Pl}}^{N_c-2}} \,\epsilon_{I I_1 ..\, I_{N_c}} Y^I\,
Q^{I_1}..\, Q^{I_{N_c}}
+ \frac{\bar{\kappa}_I}{M_{\textrm{Pl}}^{N_c-2}}\,
\epsilon_{\bar{I} \bar{I}_1..\, \bar{I}_{N_c}}\bar Y^{\bar I} \,
\bar{Q}^{\bar{I}_1}..\, \bar{Q}^{\bar{I}_{N_c}} \,.
\end{align}
with the fields $Z_{I\bar{J}}$, $Y^I$ and $\bar{Y}^{\bar{I}}$, where
$I,\bar{I},\bar{J} = 1,2,..,N_f$, representing
gauge singlet fields and with $\lambda_{I\bar{J}}$, $\kappa_I$ and $\bar{\kappa}_{\bar{I}}$
denoting dimensionless coupling constants defined at the dynamical scale $\Lambda$.
Note that now, in contrast to our $SP(N_c)$ models, we not only have to introduce
singlet fields $Z_{I\bar{J}}$, which couple to quark bilinears, but also
two further sets of singlet fields, $Y^I$ and $\bar{Y}^{\bar{I}}$, which couple
to the higher-dimensional quark operators $Q^{N_c}$ and $\bar{Q}^{N_c}$.
These latter couplings are in particular Planck-suppressed---which is not the
case for any operator in the tree-level superpotential of our $SP(N_c)$
scenarios and which, as we will see further below, has important phenomenological
implications.
Again, we identify the chiral inflaton superfield $\Phi$ w.l.o.g.\
as the $(N_f,N_f)$-component of the singlet-field tensor $Z_{I\bar{J}}$.
For large inflaton field values, $\lambda\,\tau \gg \Lambda$, where 
$\lambda \equiv \lambda_{N_f N_f}$, the $N_f$-th quark flavor
$\big(Q^{N_f}, \bar{Q}^{N_f}\big)$ then decouples perturbatively,
such that the low-energy effective theory
at energies below $\lambda\,\tau$ corresponds to the
$SU(N_c)$ gauge theory with $N_f^{\textrm{eff}} = N_c$.
Let us now show that supersymmetry is dynamically broken in this theory.


At low energies, the dynamical degrees of freedom of the effective SQCD
theory with $N_c$ flavors are the baryon-antibaryon pair
$\left(B,\bar{B}\right) \equiv \left(B_{N_f},\bar{B}_{N_f}\right)$, which does
not contain the $N_f$-th quark flavor, as well as the $N_c^2$ mesons fields
$M^{i\bar{\jmath}}$ composed out of the quark and antiquark fields
$Q^i$ and $\bar{Q}^{\bar{\jmath}}$, where $i,\bar{\jmath} =1,2,..\,,N_c$.
Eventually, this theory flows to a confined theory with a
deformed moduli constraint,
\begin{align}
\bar{B} B + \Lambda_{\textrm{eff}}^{2-N_c} \,
\textrm{det}^{(N_c)}\left(M^{i\bar{\jmath}}\right) = \Lambda_{\textrm{eff}}^2 \,,
\label{eq:constraintSUN}
\end{align}
where $\Lambda_{\textrm{eff}}$ again denotes the effective dynamical scale
below the heavy quark mass threshold.
The beta-function coefficients $b$ and $b_{\textrm{eff}}$
for the $SU(N_c)$ gauge coupling constant are given as $b = 3N_c - N_f = 2N_c-1$
and $b_{\textrm{eff}} = 3N_c - N_f^{\textrm{eff}} = 2N_c$, such that 
$\Lambda_{\textrm{eff}}$ turns out to be, cf.\ Eq.~\eqref{eq:Lambdaeff},
\begin{align}
\Lambda_{\textrm{eff}} \simeq \Lambda \left(\frac{\lambda\,\Phi}{\Lambda}
\right)^{\left(b_{\rm eff} - b\right)/b_{\rm eff}} =
\Lambda \left(\frac{\lambda\,\Phi}{\Lambda}\right)^{1/(2N_c)} \,.
\label{eq:LambdaeffSUN}
\end{align}
At the same time, the tree-level superpotential in Eq.~\eqref{eq:WtreeSUN}
can now be approximated as,
\begin{align}
\label{eq:SUNeff}
W_{\textrm{tree}} \simeq \lambda_{i\bar{\jmath}}\,\Lambda_{\textrm{eff}}\,
Z_{i \bar{\jmath}}\,M^{i\bar{\jmath}}
+ \kappa \left(\frac{\Lambda_{\textrm{eff}}}{M_{\textrm{Pl}}}\right)^{N_c-2}
\Lambda_{\textrm{eff}}\,YB
+ \bar{\kappa} \left(\frac{\Lambda_{\textrm{eff}}}{M_{\textrm{Pl}}}\right)^{N_c-2}
\Lambda_{\textrm{eff}}\,\bar{Y}\bar{B} \,,
\end{align}
where we have defined $Y\equiv Y^{N_f}$, $\bar{Y} \equiv \bar{Y}^{N_f}$,
$\kappa \equiv \kappa_{N_f}$, and $\bar{\kappa} \equiv \bar{\kappa}_{N_f}$.
The $F$-term conditions deriving from this superpotential and the deformed moduli
constraint in Eq.~\eqref{eq:constraintSUN} conflict with each other and cannot be
fulfilled simultaneously.
Hence, supersymmetry is dynamically broken in the confined phase of SQCD with
$N_f^{\textrm{eff}} = N_c$.


In the low-energy vacuum, the composite fields $B$, $\bar{B}$ and $M^{i\bar{\jmath}}$
acquire the following VEVs,
\begin{align}
B \simeq \left(\frac{\bar{\kappa}}{\kappa}\right)^{1/2}
\Lambda_{\textrm{eff}} \,, \quad
\bar{B} \simeq \left(\frac{\kappa}{\bar{\kappa}}\right)^{1/2}
\Lambda_{\textrm{eff}} \,, \quad
M^{i\bar{\jmath}} \simeq 0 \,.
\end{align}
We thus observe that the deformed moduli constraint in Eq.~\eqref{eq:constraintSUN}
is satisfied thanks to the nonzero VEVs of the baryon and antibaryon fields,
$\bar{B} B \simeq \Lambda_{\textrm{eff}}^2$, rather than due to nonzero meson VEVs.
Replacing the composite fields in Eq.~\eqref{eq:WtreeSUN} with their VEVs
provides us with the effective superpotential describing the low-energy
dynamics around the vacuum,
\begin{align}
W_{\textrm{eff}} \simeq  \left(\kappa\,\bar{\kappa}\right)^{1/2}
\left(\frac{\Lambda_{\textrm{eff}}}{M_{\textrm{Pl}}}\right)^{N_c-2}
\Lambda_{\textrm{eff}}^2 \left(Y + \bar{Y}\right) =
\left(2\kappa\,\bar{\kappa}\right)^{1/2}
\left(\frac{\Lambda_{\textrm{eff}}}{M_{\textrm{Pl}}}\right)^{N_c-2}
\Lambda_{\textrm{eff}}^2 \:\,\Xi\,,
\label{eq:WeffSUN0}
\end{align}
where we have identified the linear combination $\Xi = \left(Y+\bar{Y}\right)/\sqrt{2}$
as the goldstino superfield.
This superpotential can be written more compactly in the following way,
\begin{align}
W_{\textrm{eff}} \simeq C_{\textrm{eff}} \: \Lambda^2 \: \Xi
\left(\frac{\Lambda_{\textrm{eff}}}{\Lambda}\right)^{N_c} \,, \quad
C_{\textrm{eff}} = \left(2\kappa\,\bar{\kappa}\right)^{1/2}
\left(\frac{\Lambda}{M_{\textrm{Pl}}}\right)^{N_c-2} \,.
\label{eq:WeffSUN}
\end{align}
The corresponding $F$-term scalar potential is
nothing but the inflaton potential during inflation,
\begin{align}
V \simeq  \left|C_{\textrm{eff}}\right|^2 \Lambda^4
\left|\frac{\Lambda_{\textrm{eff}}}{\Lambda}\right|^{2N_c} \simeq
\left|C_{\textrm{eff}}\right|^2 \Lambda^4 
\left(\frac{\lambda\left|\phi\right|}{\Lambda}\right) \,.
\label{eq:VSUN}
\end{align}


Interestingly enough, the inflaton field hence always has a linear potential,
irrespectively of the number of colors $N_c$.
This is in stark contrast to our DCI models based on $SP(N_c)$ gauge dynamics,
for which the explicit dependence of the fractional power $p$ in the inflaton
potential on the size of the gauge group represents one of our main results,
cf.\ Eq.~\eqref{eq:pSPNc}.
Now we arrive at the remarkable conclusion that $p$ is a constant,
$p=1$, for all possible $SU(N_c)$ gauge groups.
Besides that, another interesting difference to all other DCI models considered in
the main text is the rather small constant factor $\left|C_{\textrm{eff}}\right|^2$ in
the scalar potential.
This factor enters linearly into the amplitude $A_s$ of the scalar 
power spectrum, cf.\ Eq.~\eqref{eq:obspre}.
Hence, for too small values of the dynamical scale $\Lambda$,
it suppresses our prediction for $A_s$ too severely, so that we
are no longer able to reproduce the observed value.
For this reason, our DCI models based on $SU(N_c)$ dynamics
require particularly large values of $\Lambda$. 
Due to our requirement that the mass of the decoupling quark flavor
should exceed the dynamical scale for all times during inflation,
$\lambda\,\tau \gg \Lambda$, large $\Lambda$ values necessitate in turn
large values of the inflaton coupling constant $\lambda$,
cf.\ Eq.~\eqref{eq:lambdalower}.
Depending on the exact numerical value of $C_{\textrm{eff}}$, such large
$\lambda$ values may then be in conflict with the upper bound on $\lambda$
in Eq.~\eqref{eq:lambdaupper} as well as the requirement that the
shift symmetry-breaking terms in the effective K\"ahler potential,
cf.\ Eq.~\eqref{eq:Deltaepseta}, should not be too large.
Compared to the DCI models discussed in the main text,
the phenomenological viability of dynamical chaotic inflation
based on $SU(N_c)$ gauge dynamics is therefore not as safely guaranteed.
A more comprehensive analysis of this class of DCI models, while beyond
the scope of this paper, is certainly imperative.


Finally, before concluding this section, let us check that the effective
superpotential in the small-field regime, $\lambda\,\tau\ll \Lambda$,
coincides with the effective superpotential in the large-field regime,
cf.\ Eq.~\eqref{eq:WeffSUN}.
In the s-confined phase, the full superpotential of the $SU(N_c)$ theory
with $N_c+1$ flavors is given by
\begin{align}
W \simeq W_{\textrm{dyn}} +
\lambda_{I\bar{J}} \, \Lambda \,Z_{I \bar{J}} \,M^{I\bar{J}}
+ \kappa_I\left(\frac{\Lambda}{M_{\textrm{Pl}}}\right)^{N_c-2} \, Y^I B_I
+ \bar{\kappa}_I\left(\frac{\Lambda}{M_{\textrm{Pl}}}\right)^{N_c-2} \, \bar{Y}^I \bar{B}_I \,,
\label{eq:WtreeSUNcon}
\end{align}
where the dynamically generated superpotential $W_{\textrm{dyn}}$ has the following form,
\begin{align}
W_{\textrm{dyn}} = -B_I \,M^{I\bar{J}} \bar{B}_{\bar{J}}  -
\Lambda^{2-N_c} \,\textrm{det}^{(N_c+1)}\big(M^{I\bar{J}}\big) \,.
\end{align}
We can easily separate all terms involving the composite fields $B$, $\bar{B}$
and $M \equiv M^{N_fN_f}$,
\begin{align}
W = & \: - BM\bar{B} - \Lambda^{2-N_c} M \,\textrm{det}^{(N_c)}
\big(M^{i\bar{\jmath}}\big) \label{eq:WBBM} \\ \nonumber
+ & \: \lambda \,\Lambda\, \Phi \,M
+ \kappa\left(\frac{\Lambda}{M_{\textrm{Pl}}}\right)^{N_c-2} \, \Lambda \,Y B
+ \bar{\kappa}\left(\frac{\Lambda}{M_{\textrm{Pl}}}\right)^{N_c-2} \, \Lambda\, \bar{Y} \bar{B}
+ .. \,,
\end{align}
where the ellipses stands for further terms that involve none of the three fields
$B$, $\bar{B}$ and $M$ and which are hence irrelevant for our discussion.
From Eq.~\eqref{eq:WBBM}, we then see that the $F$-term condition deriving from
the meson field $M$ is identical to the deformed moduli constraint of SQCD with
$N_c$ flavors and a field-dependent dynamical scale,
\begin{align}
-\bar{B} B - \Lambda_{\textrm{eff}}^{2-N_c} \,
\textrm{det}^{(N_c)}\left(M^{i\bar{\jmath}}\right) + \lambda \,\Lambda\, \Phi = 0\,,
\end{align}
In combination with the other $F$-term conditions deriving from 
Eq.~\eqref{eq:WBBM} this constraint is solved for the following baryon,
antibaryon and meson VEVs,
\begin{align}
B \simeq \left(\frac{\bar{\kappa}}{\kappa}\right)^{1/2}
\left(\lambda \,\Lambda\, \Phi\right)^{1/2} \,, \quad
\bar{B} \simeq \left(\frac{\kappa}{\bar{\kappa}}\right)^{1/2}
\left(\lambda \,\Lambda\, \Phi\right)^{1/2} \,, \quad
M^{i\bar{\jmath}} \simeq 0 \,.
\end{align}
Inserting these VEVs back into Eq.~\eqref{eq:WBBM}, 
we obtain the effective superpotential for $Y$ and $\bar{Y}$,
\begin{eqnarray}
W_{\textrm{eff}} \simeq \left(\kappa\,\bar{\kappa}\right)^{1/2}
\left(\frac{\Lambda}{M_{\textrm{Pl}}}\right)^{N_c-2}
\Lambda \left(\lambda \,\Lambda\, \Phi\right)^{1/2} \left(Y + \bar{Y}\right)
\simeq  C_{\textrm{eff}} \: \Lambda^2 \: \Xi
\left(\frac{\lambda \Phi}{\Lambda}\right)^{1/2}  \,,
\label{eq:WeffYY}
\end{eqnarray}
which indeed coincides with the effective superpotential in
the large-field regime, cf.\ Eq.~\eqref{eq:WeffSUN}.


As discussed in Sec.~\ref{subsec:SPNWeff}, this result is expected
from the holomorphicity of the superpotential.
In this sense, the fact that we are able to show that the effective superpotential
takes the same form in the small-field regime as in the large-field regime
is a nontrivial and useful consistency check.
At the same time, it is important to notice that the K\"ahler potential
does change its form as soon as strong-coupling effects become important.
For this reason, the scalar potential for the inflaton does no longer
have the same form as in Eq.~\eqref{eq:VSUN} at small field values.
In fact, due to the non-calculable K\"ahler potential at strong coupling
we are unable to compute the exact form of the inflaton potential in
the small-field regime.
Based on the fact that SQCD with $N_c+1$ flavors is s-confining and
given the tree-level superpotential in Eq.~\eqref{eq:WtreeSUNcon},
we can merely say that the inflaton acquires a mass of
$\mathcal{O}\left(\lambda\Lambda\right)$ at low energies.
Close to the true vacuum in field space, the inflaton potential
can therefore be approximated by Eq.~\eqref{eq:Vmphi},
cf.\ also Fig.~\ref{fig:potential}.


\subsection[Massless and massive matter fields (fractional powers \texorpdfstring{$p \geq 1$}{p >= 1})]
{Massless and massive matter fields (fractional powers \texorpdfstring{\boldmath{$p \geq 1$}}{p >= 1})}


Similarly as for the DCI scenarios discussed in the main text,
we can also generalize the model described in the previous section
to a larger number of matter fields coupling to the inflaton.
Let us extend the tree-level superpotential in Eq.~\eqref{eq:WtreeSUN}
as follows,
\begin{align}
W_{\textrm{tree}} \rightarrow W_{\textrm{tree}} +
\left(M_a + \lambda_a \Phi\right) P^a\bar{P}^a \,, \quad
P^a \sim \Yfund\textrm{\hskip2pt} \,, \quad
\bar{P}^{\bar{a}} \sim \overline{\Yfund}\textrm{\hskip2pt} \,, \quad
a = 1,2,..,N_m \,,
\label{eq:WtreeSUNfull}
\end{align}
where we assume that all of the supersymmetric masses $M_a$ have
a value in between the dynamical scale and the Planck scale,
$\Lambda \lesssim M_a \lesssim M_{\textrm{Pl}}$.
The beta-function coefficient for the $SU(N_c)$ gauge
coupling constant at high energies, $b$, is then given
as $b = 2N_c - 1 - N_m$, while the corresponding coefficient
at low energies, $b_{\textrm{eff}}$, remains unchanged,
$b_{\textrm{eff}} = 2N_c$.
In analogy to the $SP(N_c)$ case, this results in the following
effective dynamical scale $\Lambda_{\textrm{eff}}$
below all heavy quark mass thresholds, cf.\ Eq.~\eqref{eq:Lambdaefffull},
\begin{align}
\Lambda_{\textrm{eff}} = \Lambda \prod_{n=0}^{N_m}
\left(\frac{M_n + \lambda_n\Phi}{\Lambda}\right)^{1/(2N_c)} \,, \quad
M_0 \equiv 0 \,,
\end{align}
with $\lambda_0$ denoting the Yukawa coupling constant belonging to the
$(N_c+1)$-th quark flavor coupling to the inflaton.
Inserting this expression for $\Lambda_{\textrm{eff}}$ into Eq.~\eqref{eq:WeffSUN},
we obtain the effective superpotential and consequently also the scalar potential
of our extended $SU(N_c)$ model,
\begin{align}
\label{eq:WeffSUNfull}
W_{\textrm{eff}} \simeq C_{\textrm{eff}} \: \Lambda^2 \: \Xi
\prod_{n=0}^{N_m} \left(\frac{M_a + \lambda_a\Phi}{\Lambda}\right)^{1/2} \,, \quad
V \simeq \left|C_{\textrm{eff}}\right|^2 \Lambda^4 \prod_{n=0}^{N_m}
\left|\frac{M_n + \lambda_n\phi}{\Lambda}\right| \,.
\end{align}
In the large-field regime, where $\lambda_n\left|\phi\right| \gg M_n$, the
inflaton potential hence turns out to be a simple monomial with an integer 
power $p$,
\begin{align}
p = 1 + N_m \,.
\label{eq:pSUN}
\end{align}
We thus make the remarkable observation that, in our DCI models based on $SU(N_c)$
dynamics, the number of quark fields coupling to the inflaton, $1+N_m$, directly
determines the power $p$ appearing in the inflaton potential.
In this sense, determining the power $p$ based on CMB observations could
be regarded as a means to actually count the number of strong-sector quarks
coupling to the inflaton.
According to the bound on the power $p$ in Eq.~\eqref{eq:pboundSPN}, $p\leq4$, this number
could be as large as four, i.e.\ $N_m$ could either be $0$, $1$, $2$ or $3$.
Furthermore, we observe that, for a fixed number of flavors $N_f$, all DCI models
based on $SU(N_c)$ yield the same prediction for the power $p$ as the DCI model
based on $SP(N_c)$ featuring one flavor more, i.e., in analogy to Eq.~\eqref{eq:SPSOequiv},
we now have the following equivalence relation,
\begin{align}
SP(1) \textrm{ with } N_f = N_c + 2 + N_m \quad \leftrightsquigarrow \quad
SU(N_c) \textrm{ with } N_f = N_c + 1+ N_m \,.
\label{eq:SPSUequi}
\end{align}
The crucial difference between two such scenarios is that
the superpotential as well as the infrared spectrum of the $SP(1)$ model
is always simpler and more minimal than in the case of its $SU(N_c)$ counterpart.
Moreover, as pointed out above, the $SU(N_c)$ models all feature the additional
suppression factor $\left|C_{\textrm{eff}}\right|^2$ in their scalar potential,
which leads to further complications.
It is for these reasons that we devote our attention to the DCI models based on $SP(N_c)$
dynamics in the main text and discuss the $SU(N_c)$ DCI models only here in this appendix.


\subsection[Effective superpotential from \texorpdfstring{$R$}{R} symmetry]
{Effective superpotential from \texorpdfstring{\boldmath{$R$}}{R} symmetry}


For most of the DCI models discussed in the main text, we managed to 
re-derive the effective superpotential solely based on $R$ symmetry
arguments.
As we shall now demonstrate, the $SU(N_c)$ models are no exceptions in this
respect.
According to the tree-level superpotential in Eq.~\eqref{eq:WtreeSUNfull},
the $R$ charges of our matter and singlet fields have to be related as follows,
\begin{align}
R\left[Q^I\right] = & \: R\big[\bar{Q}^{\bar{J}}\big] =
R\big[P^a\big] = R\big[\bar{P}^{\bar{a}}\big] = 1 - \frac{x}{2} \,, \quad
R\big[Z_{I\bar{J}}\big] = R\big[\Phi\big] = x \,, \\ \nonumber
R\big[Y^I\big] = & \: R\big[\bar{Y}^{\bar{J}}\big] = 2 - N_c \left(1-\frac{x}{2}\right) \,.
\end{align}
With this charge assignment, we obtain for the coefficient $\mathcal{A}_R$ of the
$U(1)_R\left[SU(N_c)\right]^2$ anomaly,
\begin{align}
\mathcal{A}_R = & \: 2N_c + 2\left(-\frac{x}{2}\right)\left(N_c+1+N_m\right) \,,
\end{align}
such that the scale $\Lambda$ acquires a spurious charge
$R[\Lambda] = \mathcal{A}_R / b$, where $b = 2N_c - 1 -N_m$.
Hence, only the following gauge-invariant terms can possibly appear
in the effective superpotential,
\begin{align}
\kappa \, M_{\textrm{Pl}}^{2-N_c}
\Lambda^{b/2} \, \Phi^{(1+N_m)/2} \,Y \,, \quad
\bar{\kappa}\,\Lambda^{b/2} \, \Phi^{(1+N_m)/2} \,\bar{Y}
\,, \quad b = 2N_c - 1 - N_m \,, 
\label{eq:SUNcand}
\end{align}
where $\kappa$ and $\bar{\kappa}$ are undetermined proportionality
constants.
Similar terms featuring any of the singlet fields $Y^i$ or
$\bar{Y}^{\bar{\imath}}$ instead of $Y$ and $\bar{Y}$, respectively,
are forbidden by the $SU(N_f-1)$ flavor symmetry under which the superpotential
must be invariant in the case of nonzero inflaton VEV.
Due to their similar structure, the two terms in Eq.~\eqref{eq:SUNcand}
can be easily combined into one term.
In particular for $\kappa \neq \bar{\kappa}$, it is convenient to replace both constants
by their geometric mean, $\kappa,\bar{\kappa} \rightarrow \left(\kappa\bar{\kappa}\right)^{1/2}$.
In addition to that, we can also shift the $1+N_m$ factors of $\Phi$
in the power $\Phi^{(1+N_m)/2}$ by the mass parameters $M_n$.
In the end, the effective superpotential therefore takes the following form,
\begin{align}
W_{\textrm{eff}} \simeq \left(\kappa\bar{\kappa}\right)^{1/2}
\left(\frac{\Lambda}{M_{\textrm{Pl}}}\right)^{N_c-2}
\Lambda^2  \prod_{n=0}^{N_m} \left(\frac{M_a + \lambda_a\Phi}{\Lambda}\right)^{1/2}
\left(Y + \bar{Y}\right) \,,
\end{align}
which is identical to the superpotential in Eq.~\eqref{eq:WeffSUNfull}.
Hence, also our analysis based on $R$ symmetry leads us to the conclusion
that, at large inflaton field values, the inflaton potential corresponds to
a monomial potential with an integer power $p = 1+N_m$.


\section{Supersymmetry breaking in the true vacuum after inflation}
\label{app:dynSUSY}


Our general DCI recipe as outlined in Sec.~\ref{sec:idea} relies on the idea that the
strong dynamics in the inflaton sector approach a phase of s-confinement at small inflaton
field values.
This guarantees the existence of a smooth effective theory describing the strong
gauge dynamics at low energies including the true ground state at the origin of field
space.
On the other hand, this construction automatically implies that all fields of the
inflaton sector always settle in a supersymmetric vacuum.
Within the class of DCI models discussed in the main text, we can therefore never realize
supersymmetry breaking in the true vacuum after the end of inflation.%
\footnote{As a loophole to this statement, we point out that supersymmetry may be spontaneously
broken in the final ground state of our DCI models after all, if this ground state does not
coincide with the origin of field space for some reason.
To accomplish such a situation, one would need to modify the inflaton potential at small
field values, $V \simeq m_\phi^2\left|\phi\right|^2$---for instance, by means of
strong-coupling effects---so that $\phi \neq 0$ in the true vacuum.}
In order to obtain a phenomenologically complete model, soft supersymmetry breaking
in our current vacuum then needs to be attributed to the dynamics of yet another
hidden sector different from the sector responsible for inflation.
This is certainly a viable and not very unlikely possibility.
As we shall now demonstrate, it is however also possible to actually retain
a small amount of supersymmetry breaking in the true vacuum after inflation, if
we just slightly modify our mechanism for the generation of the inflaton potential.
In the following, we shall only sketch our main idea.
The construction of an explicit and complete model is left for future work.


Every DCI model discussed in the main text involves an effective model of dynamical
supersymmetry breaking that is mutated into an s-confining theory at small inflaton
field values.
Let us now abandon this paradigm and try to construct a model in which the effective
SUSY-breaking model valid during inflation flows to a second alternative
model of dynamical supersymmetry breaking rather than to an s-confining theory.
Instead of one source of supersymmetry breaking, we then have to deal with two independent
sources of symmetry breaking: one active during inflation and providing the vacuum energy
density driving inflation and a second one, which only becomes effective after the end of
inflation.
This second source of supersymmetry breaking may for instance be the strong
gauge dynamics of an $SU(N_c)$ theory with $N_c$ flavors.
In our DCI models based on $SU(N_c)$ dynamics, the deformed moduli constraint
of an $SU(N_c)$ theory with $N_c$ flavors in combination with the tree-level
superpotential in Eq.~\eqref{eq:SUNeff} is in fact responsible for supersymmetry
breaking \textit{during} inflation.
Now we assume that such dynamics are actually responsible for supersymmetry breaking
\textit{after} inflation.


Consider an $SU(N_c)$ theory with $N_c$ quarks flavors $\left(Q^i,\bar{Q}^{\bar{\imath}}\right)$,
where $Q^i \sim \Yfund\hskip2pt$ and $\bar{Q}^{\bar{\imath}} \sim \overline{\Yfund}\hskip2pt$,
and $N_c^2+2$ singlet fields $Z_{ij}$, $Y$ and $\bar{Y}$, which interact via the 
following tree-level superpotential,
\begin{align}
\label{eq:WeffB}
W_{\textrm{tree}}  = \lambda_{i\bar{\jmath}} \,Z_{i \bar{\jmath}} \,Q^i \bar{Q}^{\bar{\jmath}}
+ \frac{\kappa}{M_{\textrm{Pl}}^{N_c-2}} \,\epsilon_{i_1 ..\, i_{N_c}} Y\,
Q^{i_1}..\, Q^{i_{N_c}}
+ \frac{\bar{\kappa}}{M_{\textrm{Pl}}^{N_c-2}}\,
\epsilon_{\bar{\imath}_1..\, \bar{\imath}_{N_c}}\bar Y \,
\bar{Q}^{\bar{\imath}_1}..\, \bar{Q}^{\bar{\imath}_{N_c}} \,.
\end{align}
From our analysis in Sec.~\ref{subsec:SUNmassless} we know that the effective
superpotential describing the low-energy dynamics of this theory can eventually
be brought into the following form, cf.\ Eq.~\eqref{eq:WeffSUN0},
\begin{align}
W_{\textrm{eff}} \simeq \left(2\kappa\,\bar{\kappa}\right)^{1/2}
\left(\frac{\Lambda}{M_{\textrm{Pl}}}\right)^{N_c-2} \Lambda^2 \:\,\Xi \,.
\end{align}
In Sec.~\ref{subsec:SUNmassless}, we noted that, if we intended to use this effective
superpotential to describe the inflationary phase, the suppression factor
$C_{\textrm{eff}} \propto \left(\Lambda/M_{\textrm{Pl}}\right)^{N_c-2}$
might turn out to be problematic, as it may prevent us from successively
reproducing the measured value of the scalar spectral amplitude $A_s$.
Now we find ourselves in quite the opposite situation.
We benefit from the suppression factor in the superpotential, since
it allows us to generate realistic SUSY-breaking scales or, equivalently,
obtain realistic gravitino masses $m_{3/2}$,
\begin{align}
m_{3/2} \sim \left(\frac{\Lambda}{M_{\textrm{Pl}}}\right)^{N_c}
\frac{M_{\textrm{Pl}}}{\sqrt{3}} \,.
\end{align}
While we have to assume a large dynamical scale, $\Lambda \gg 10^{16}\,\textrm{GeV}$,
in our models in Sec.~\ref{subsec:SUNmassless}, we can now easily set $\Lambda$ to a value
around the GUT scale.
For $N_c=5$, for instance, this then results in phenomenologically perfectly reasonable
values for the gravitino mass,
\begin{align}
N_c = 5 \,, \quad
m_{3/2} \sim \left(\frac{10^{15}..\,10^{16}\,\textrm{GeV}}
{M_{\textrm{Pl}}}\right)^5 \frac{M_{\textrm{Pl}}}{\sqrt{3}} \sim
10\,\textrm{GeV} ..\,1000\,\textrm{TeV}\,.
\end{align}


Next, we have to embed this SUSY-breaking model into a larger model that is also
capable of accommodating inflation.
In addition to the singlet fields in Eq.~\eqref{eq:WeffB}, let us introduce two further
singlet chiral superfields, $\Phi$ and $X$, where $\Phi$ will play the role of the inflaton
and $X$ is the Polonyi field responsible for spontaneous supersymmetry breaking during inflation.
We endow $\Phi$ with Yukawa couplings to all quark flavors and couple $X$
to the field strength field $\mathcal{W}_\alpha^{\,a}$,
\begin{align}
W \supset \lambda_{i\bar{\jmath}}' \,\Phi \,Q^i \bar{Q}^{\bar{\jmath}} +
\left(\frac{1}{4g^2} + \alpha\,\frac{X}{M_{\textrm{Pl}}}\right)
\mathcal{W}^{\alpha a} \mathcal{W}_\alpha^{\,a} + .. \,.
\label{eq:WX}
\end{align}
The ellipses stands for additional terms responsible for stabilizing the field $X$
during and after inflation, which we do not specify any further.
Given this superpotential, all flavors perturbatively decouple
at large inflaton field values, such that our initial $SU(N_c)$ theory
with $N_c$ flavors turns into a pure super
Yang-Mills (SYM) theory with zero flavors.
The low-energy dynamics of this pure SYM theory then lead to the formation
of a gaugino condensate, cf.\ Eq.~\eqref{eq:LambdaeffSUN},
\begin{align}
\left<\lambda^a \lambda^a\right> \sim \Lambda_{\textrm{eff}}^3 \,, \quad
\Lambda_{\textrm{eff}} \simeq \Lambda \left(\frac{\lambda\,\Phi}{\Lambda}\right)^{1/3}
\,,
\end{align}
as well as to the spontaneous breaking of supersymmetry,
\begin{align}
W_{\textrm{eff}} \sim \alpha\frac{X}{M_{\textrm{Pl}}}
\Lambda_{\textrm{eff}}^3 + \mathcal{O}\left(\frac{X^2}{M_{\textrm{Pl}}^2}\right)
= C_{\textrm{eff}}' \: \Lambda^2 \: X
\left(\frac{\Lambda_{\textrm{eff}}}{\Lambda}\right)^3
 + \mathcal{O}\left(\frac{X^2}{M_{\textrm{Pl}}^2}\right) \,, \quad
C_{\textrm{eff}}' = \alpha \frac{\Lambda}{M_{\textrm{Pl}}} \,,
\end{align}
where we have assumed that the goldstino field $X$ is appropriately
stabilized at some sub-Planckian value during inflation, $X/M_{\textrm{Pl}} \ll 1$, and
where we have absorbed all numerical prefactors in the coefficient $\alpha$.
Inflation is therefore driven by a quadratic inflaton potential,
\begin{align}
V \sim \left|C_{\textrm{eff}}'\right|^2 \Lambda^4
\left(\frac{\lambda\left|\phi\right|}{\Lambda}\right)^2 \,.
\end{align}
Luckily, this potential is not as severely suppressed as the one
in Eq.~\eqref{eq:WeffSUN}, $\left|C_{\textrm{eff}}'\right| \gtrsim
\left|C_{\textrm{eff}}\right|$, which is why it is surely capable
of providing appropriate conditions for successful inflation.


In summary, we conclude that the above toy model may very likely serve
as a basis for a complete model that is able to account for inflation
\textit{and} low-energy supersymmetry breaking:
At large inflaton field values, the strong dynamics are described by
a pure SYM theory and the inflaton slowly rolls in a scalar potential
that is generated by means of gaugino condensation.
As the inflaton field value becomes smaller, more and more quark flavors
become dynamical, the gaugino condensate dissolves and the strong interactions
enter into a regime where they are described by an $SU(N_c)$ theory with $N_c$ flavors
and a deformed moduli constraint.
Of course, the above outlined model is far from complete.
For instance, we did not specify how the Polonyi
field $X$ may be stabilized during and after inflation---which
is, however, crucial since the field $X$ would otherwise run away
to infinity simply given the superpotential in Eq.~\eqref{eq:WX}.
Here, it is important to note that the stabilization mechanism during inflation
may, in particular, also affect the final shape of the scalar potential.
Furthermore, the model is also potentially endangered by the possibility that the
inflaton produces too many gravitinos in its decays.
Therefore, a more careful analysis as well as some sort of extension
of our toy model are definitely needed.
But we are confident that our main idea has a good chance of surviving further
refinements and that dynamical chaotic inflation
and low-energy supersymmetry breaking can eventually be embedded
into a common theory.




\begin{thebibliography}{99}

\bibitem{Ade:2014xna} 
  P.~A.~R.~Ade {\it et al.}  [BICEP2 Collaboration],
  Phys.\ Rev.\ Lett.\  {\bf 112}, 241101 (2014),\newline
  1403.3985 [astro-ph.CO].

\bibitem{Linde:1983gd} 
  A.~D.~Linde,
  Phys.\ Lett.\ B {\bf 129}, 177 (1983);
%
  JETP Lett.\  {\bf 38}, 176 (1983)
  [Pisma Zh.\ Eksp.\ Teor.\ Fiz.\  {\bf 38}, 149 (1983)].

\bibitem{Silverstein:2008sg} 
  E.~Silverstein and A.~Westphal,
  Phys.\ Rev.\ D {\bf 78}, 106003 (2008),
  0803.3085 [hep-th].

\bibitem{McAllister:2008hb} 
  L.~McAllister, E.~Silverstein and A.~Westphal,
  Phys.\ Rev.\ D {\bf 82}, 046003 (2010),\newline
  0808.0706 [hep-th].

\bibitem{Kaloper:2011jz} 
  N.~Kaloper, A.~Lawrence and L.~Sorbo,
  JCAP {\bf 1103}, 023 (2011),
  1101.0026 [hep-th];
%
  E.~Palti and T.~Weigand,
  JHEP {\bf 1404}, 155 (2014),
  1403.7507 [hep-th].

\bibitem{Marchesano:2014mla} 
  F.~Marchesano, G.~Shiu and A.~M.~Uranga,
  1404.3040 [hep-th].

\bibitem{Kawasaki:2000yn} 
  M.~Kawasaki, M.~Yamaguchi and T.~Yanagida,
  Phys.\ Rev.\ Lett.\  {\bf 85}, 3572 (2000),\newline
  hep-ph/0004243.

\bibitem{Kallosh:2010ug} 
  R.~Kallosh and A.~Linde,
  JCAP {\bf 1011}, 011 (2010),
  1008.3375 [hep-th].

\bibitem{Kallosh:2010xz} 
  R.~Kallosh, A.~Linde and T.~Rube,
  Phys.\ Rev.\ D {\bf 83}, 043507 (2011),
  1011.5945 [hep-th].

\bibitem{Kallosh:2011qk} 
  R.~Kallosh, A.~Linde, K.~A.~Olive and T.~Rube,
  Phys.\ Rev.\ D {\bf 84}, 083519 (2011),\newline
  1106.6025 [hep-th].

\bibitem{Linde:2014nna} 
  A.~Linde,
  1402.0526 [hep-th].
  
\bibitem{Takahashi:2010ky} 
  F.~Takahashi,
  Phys.\ Lett.\ B {\bf 693}, 140 (2010),
  1006.2801 [hep-ph].
  
\bibitem{Ellis:2014rxa} 
  J.~Ellis, M.~A.~G.~García, D.~V.~Nanopoulos and K.~A.~Olive,
  JCAP {\bf 1405}, 037 (2014),
  1403.7518 [hep-ph].

\bibitem{Harigaya:2012pg} 
  K.~Harigaya, M.~Ibe, K.~Schmitz and T.~T.~Yanagida,
  Phys.\ Lett.\ B {\bf 720}, 125 (2013),
  1211.6241 [hep-ph];
%
  Phys.\ Lett.\ B {\bf 733}, 283 (2014),
  arXiv:1403.4536 [hep-ph].

\bibitem{Harigaya:2014eta} 
  K.~Harigaya and M.~Ibe,
  1404.3511 [hep-ph];
%
  M.~Dine, P.~Draper and A.~Monteux,\newline
  1405.0068 [hep-th].
  
\bibitem{Yonekura:2014oja} 
  K.~Yonekura,
  1405.0734 [hep-th].

\bibitem{Dimopoulos:1997fv} 
  S.~Dimopoulos, G.~R.~Dvali and R.~Rattazzi,
  Phys.\ Lett.\ B {\bf 410}, 119 (1997),
  hep-ph/9705348;
%
  K.~I.~Izawa, M.~Kawasaki and T.~Yanagida,
  Phys.\ Lett.\ B {\bf 411}, 249 (1997),
  hep-ph/9707201;
%
  K.~I.~Izawa,
  Prog.\ Theor.\ Phys.\  {\bf 99}, 157 (1998),
  hep-ph/9708315.
  
\bibitem{Brodsky:2009zd} 
  S.~J.~Brodsky and R.~Shrock,
  Proc.\ Nat.\ Acad.\ Sci.\  {\bf 108}, 45 (2011),
  0905.1151 [hep-th].
  
\bibitem{Appelquist:1996dq} 
  T.~Appelquist, J.~Terning and L.~C.~R.~Wijewardhana,
  Phys.\ Rev.\ Lett.\  {\bf 77}, 1214 (1996),
  hep-ph/9602385.
  
\bibitem{Seiberg:1994bz} 
  N.~Seiberg,
  Phys.\ Rev.\ D {\bf 49}, 6857 (1994),
  hep-th/9402044;

\bibitem{Intriligator:1995ne} 
  K.~A.~Intriligator and P.~Pouliot,
  Phys.\ Lett.\ B {\bf 353}, 471 (1995),
  hep-th/9505006;
%
\bibitem{Csaki:1996sm} 
  C.~Csaki, M.~Schmaltz and W.~Skiba,
  Phys.\ Rev.\ Lett.\  {\bf 78}, 799 (1997),
  hep-th/9610139;
%
  Phys.\ Rev.\ D {\bf 55}, 7840 (1997),
  hep-th/9612207.

\bibitem{Izawa:1996pk} 
  K.~-I.~Izawa and T.~Yanagida,
  Prog.\ Theor.\ Phys.\  {\bf 95}, 829 (1996),
  hep-th/9602180;
%
  K.~A.~Intriligator and S.~D.~Thomas,
  Nucl.\ Phys.\ B {\bf 473}, 121 (1996),
  hep-th/9603158.
  
\bibitem{Seiberg:1994pq} 
  N.~Seiberg,
  Nucl.\ Phys.\ B {\bf 435}, 129 (1995),
  hep-th/9411149.

\bibitem{Banks:1981nn} 
  T.~Banks and A.~Zaks,
  Nucl.\ Phys.\ B {\bf 196}, 189 (1982);
%
  H.~Georgi,
  Phys.\ Rev.\ Lett.\  {\bf 98}, 221601 (2007),
  hep-ph/0703260.
  
\bibitem{Seiberg:1993vc} 
  N.~Seiberg,
  Phys.\ Lett.\ B {\bf 318}, 469 (1993),
  hep-ph/9309335.
  
\bibitem{Affleck:1984mf} 
  I.~Affleck, M.~Dine and N.~Seiberg,
  Phys.\ Lett.\ B {\bf 140}, 59 (1984).

\bibitem{Affleck:1983vc} 
  I.~Affleck, M.~Dine and N.~Seiberg,
  Phys.\ Lett.\ B {\bf 137}, 187 (1984).

\bibitem{Murayama:1995ng} 
  H.~Murayama,
  Phys.\ Lett.\ B {\bf 355}, 187 (1995),
  hep-th/9505082.
  
\bibitem{Pouliot:1995me} 
  P.~Pouliot,
  Phys.\ Lett.\ B {\bf 367}, 151 (1996),
  hep-th/9510148.
  
\bibitem{Affleck:1984xz} 
  I.~Affleck, M.~Dine and N.~Seiberg,
  Nucl.\ Phys.\ B {\bf 256}, 557 (1985).
  
\bibitem{Intriligator:2007cp} 
  K.~A.~Intriligator and N.~Seiberg,
  Class.\ Quant.\ Grav.\  {\bf 24}, S741 (2007),
  hep-ph/0702069.
  
\bibitem{Witten:1982fp} 
  E.~Witten,
  Phys.\ Lett.\ B {\bf 117}, 324 (1982).

\bibitem{Harigaya:2014qza} 
  K.~Harigaya and T.~T.~Yanagida,
  1403.4729 [hep-ph].
  
\bibitem{Hsu:2004hi} 
  J.~P.~Hsu and R.~Kallosh,
  JHEP {\bf 0404}, 042 (2004),
  hep-th/0402047.
  
\bibitem{'tHooft:1979bh} 
  G.~'t Hooft,
  NATO Adv.\ Study Inst.\ Ser.\ B Phys.\  {\bf 59}, 135 (1980).

\bibitem{Goldwirth:1991rj}
  D.~S.~Goldwirth and T.~Piran,
  Phys.\ Rept.\  {\bf 214}, 223 (1992).
  
\bibitem{Linde:2005ht} 
  A.~D.~Linde,
  Contemp.\ Concepts Phys.\  {\bf 5}, 1 (1990),
  hep-th/0503203.
  
\bibitem{Brax:2010ai} 
  P.~Brax, J.~-F.~Dufaux and S.~Mariadassou,
  Phys.\ Rev.\ D {\bf 83}, 103510 (2011),\newline
  1012.4656 [hep-th].
  
\bibitem{Felder:2000hj} 
  G.~N.~Felder, J.~Garcia-Bellido, P.~B.~Greene, L.~Kofman, A.~D.~Linde and I.~Tkachev,
  Phys.\ Rev.\ Lett.\  {\bf 87}, 011601 (2001),
  hep-ph/0012142;
%
  G.~N.~Felder, L.~Kofman and A.~D.~Linde,
  Phys.\ Rev.\ D {\bf 64}, 123517 (2001),
  hep-th/0106179.

\bibitem{Kofman:1994rk} 
  L.~Kofman, A.~D.~Linde and A.~A.~Starobinsky,
  Phys.\ Rev.\ Lett.\  {\bf 73}, 3195 (1994),\newline
  hep-th/9405187;
  
\bibitem{Ibe:2011aa} 
  M.~Ibe and T.~T.~Yanagida,
  Phys.\ Lett.\ B {\bf 709}, 374 (2012),
  1112.2462 [hep-ph];
%
  M.~Ibe, S.~Matsumoto and T.~T.~Yanagida,
  Phys.\ Rev.\ D {\bf 85}, 095011 (2012),
  1202.2253 [hep-ph];
%
  B.~Bhattacherjee, B.~Feldstein, M.~Ibe, S.~Matsumoto and T.~T.~Yanagida,
  Phys.\ Rev.\ D {\bf 87}, 015028 (2013),
  1207.5453 [hep-ph].

\bibitem{Kawasaki:2000ws} 
  M.~Kawasaki, M.~Yamaguchi and T.~Yanagida,
  Phys.\ Rev.\ D {\bf 63}, 103514 (2001),
  hep-ph/0011104;
%
  K.~Harigaya and T.~T.~Yanagida,
  arXiv:1407.1580 [hep-ph].
  
\bibitem{Harigaya:2013vwa} 
  K.~Harigaya and K.~Mukaida,
  JHEP {\bf 1405}, 006 (2014),
  1312.3097 [hep-ph].
  
\bibitem{Fukugita:1986hr}
  M.~Fukugita and T.~Yanagida,
  Phys.\ Lett.\ B {\bf 174} (1986) 45;
%
  W.~Buchmuller, P.~Di Bari and M.~Plumacher,
  Annals Phys.\  {\bf 315}, 305 (2005),
  hep-ph/0401240;
%
  for reviews of leptogenesis, cf.\ for example:
  W.~Buchmuller, R.~D.~Peccei and T.~Yanagida,
  Ann.\ Rev.\ Nucl.\ Part.\ Sci.\  {\bf 55}, 311 (2005),
  hep-ph/0502169;
%
  S.~Davidson, E.~Nardi and Y.~Nir,
  Phys.\ Rept.\  {\bf 466}, 105 (2008),
  0802.2962 [hep-ph].

\bibitem{Kolb:1993hw} 
  E.~W.~Kolb and I.~I.~Tkachev,
  Phys.\ Rev.\ D {\bf 49}, 5040 (1994),
  astro-ph/9311037;
%
  E.~J.~Copeland, M.~Gleiser and H.~-R.~Muller,
  Phys.\ Rev.\ D {\bf 52}, 1920 (1995),
  hep-ph/9503217;
%
  S.~Kasuya, M.~Kawasaki and F.~Takahashi,
  Phys.\ Lett.\ B {\bf 559}, 99 (2003),
  hep-ph/0209358;
%
  K.~Mukaida and M.~Takimoto,
  1405.3233 [hep-ph].
  
\bibitem{Coleman:1985ki} 
  S.~R.~Coleman,
  Nucl.\ Phys.\ B {\bf 262}, 263 (1985)
  [Erratum-ibid.\ B {\bf 269}, 744 (1986)];
%
  A.~G.~Cohen, S.~R.~Coleman, H.~Georgi and A.~Manohar,
  Nucl.\ Phys.\ B {\bf 272}, 301 (1986);
%
  S.~Kasuya and M.~Kawasaki,
  Phys.\ Rev.\ D {\bf 61}, 041301 (2000),
  hep-ph/9909509;
%
  K.~Enqvist, S.~Kasuya and A.~Mazumdar,
  Phys.\ Rev.\ D {\bf 66}, 043505 (2002),
  hep-ph/0206272.
  
\bibitem{Cohen:1986ct} 
  A.~G.~Cohen, S.~R.~Coleman, H.~Georgi and A.~Manohar,
  Nucl.\ Phys.\ B {\bf 272}, 301 (1986);
  %
  M.~P.~Hertzberg,
  Phys.\ Rev.\ D {\bf 82}, 045022 (2010),
  1003.3459 [hep-th];
  %
  M.~Kawasaki and M.~Yamada,
  1311.0985 [hep-ph].
  
\bibitem{GarciaBellido:1996qt} 
  J.~Garcia-Bellido, A.~D.~Linde and D.~Wands,
  Phys.\ Rev.\ D {\bf 54}, 6040 (1996),
  astro-ph/9605094;
%
  M.~Kawasaki, N.~Sugiyama and T.~Yanagida,
  Phys.\ Rev.\ D {\bf 57}, 6050 (1998),
  hep-ph/9710259;
%
  J.~'i.~Yokoyama,
  Phys.\ Rev.\ D {\bf 58}, 083510 (1998),
  astro-ph/9802357.

\bibitem{Hawking:1971ei} 
  S.~Hawking,
  Mon.\ Not.\ Roy.\ Astron.\ Soc.\  {\bf 152}, 75 (1971).

\bibitem{Toussaint:1978br} 
  D.~Toussaint, S.~B.~Treiman, F.~Wilczek and A.~Zee,
  Phys.\ Rev.\ D {\bf 19}, 1036 (1979);
%
  M.~S.~Turner,
  Phys.\ Lett.\ B {\bf 89}, 155 (1979);
%
  J.~D.~Barrow, E.~J.~Copeland, E.~W.~Kolb and A.~R.~Liddle,
  Phys.\ Rev.\ D {\bf 43}, 984 (1991);
%
  for a recent analysis, cf.:
  T.~Fujita, K.~Harigaya, M.~Kawasaki and R.~Matsuda,
  Phys.\ Rev.\ D {\bf 89}, 103501 (2014),
  1401.1909 [astro-ph.CO].

\bibitem{Fujita:2013bka} 
  T.~Fujita, K.~Harigaya and M.~Kawasaki,
  Phys.\ Rev.\ D {\bf 88}, 123519 (2013),\newline
  1306.6437 [astro-ph.CO].

\bibitem{Carr:1984id} 
  B.~J.~Carr and M.~J.~Rees,
  Trieste Int. Sch. Advanc. Stud. - 83-18 A. (83,REC.FEB.84) 21p;
%
  R.~Bean and J.~Magueijo,
  Phys.\ Rev.\ D {\bf 66}, 063505 (2002),
  astro-ph/0204486;
%
  N.~Duechting,
  Phys.\ Rev.\ D {\bf 70}, 064015 (2004),
  astro-ph/0406260;
%
  M.~Kawasaki, A.~Kusenko and T.~T.~Yanagida,
  Phys.\ Lett.\ B {\bf 711}, 1 (2012),
  1202.3848 [astro-ph.CO].
     
\bibitem{Alabidi:2010sf} 
  L.~Alabidi and I.~Huston,
  JCAP {\bf 1008}, 037 (2010),
  1004.4794 [astro-ph.CO];
%
  J.~Martin, C.~Ringeval and R.~Trotta,
  Phys.\ Rev.\ D {\bf 83}, 063524 (2011),
  1009.4157 [astro-ph.CO].

\bibitem{Lyth:1998xn} 
  D.~H.~Lyth and A.~Riotto,
  Phys.\ Rept.\  {\bf 314}, 1 (1999),
  hep-ph/9807278.

\bibitem{Ade:2013zuv} 
  P.~A.~R.~Ade {\it et al.}  [Planck Collaboration],
  1303.5076 [astro-ph.CO].
  
\bibitem{Spergel:2013rxa} 
  D.~Spergel, R.~Flauger and R.~Hlozek,
  1312.3313 [astro-ph.CO].

\bibitem{Liddle:1992wi} 
  A.~R.~Liddle and D.~H.~Lyth,
  Phys.\ Lett.\ B {\bf 291}, 391 (1992),
  astro-ph/9208007.

\bibitem{Gerbino:2014eqa} 
  M.~Gerbino, A.~Marchini, L.~Pagano, L.~Salvati, E.~Di Valentino and A.~Melchiorri,
  1403.5732 [astro-ph.CO];
%
  A.~Ashoorioon, K.~Dimopoulos, M.~M.~Sheikh-Jabbari and G.~Shiu,
  1403.6099 [hep-th];
%
  F.~Wu, Y.~Li, Y.~Lu and X.~Chen,
  1403.6462 [astro-ph.CO];
%
  C.~Cheng and Q.~-G.~Huang,
  1403.7173 [astro-ph.CO];
%
  B.~Chang and L.~Xu,
  1404.1558 [astro-ph.CO].

\bibitem{Brandenberger:2014faa} 
  R.~H.~Brandenberger, A.~Nayeri and S.~P.~Patil,
  1403.4927 [astro-ph.CO];
%
  Y.~Wang and W.~Xue,
  1403.5817 [astro-ph.CO];
%
  T.~Biswas, T.~Koivisto and A.~Mazumdar,
  1403.7163 [hep-th].

\bibitem{Hinshaw:2012aka} 
  G.~Hinshaw {\it et al.}  [WMAP Collaboration],
  Astrophys.\ J.\ Suppl.\  {\bf 208}, 19 (2013),\newline
  1212.5226 [astro-ph.CO].
  
\bibitem{Mortonson:2014bja} 
  M.~J.~Mortonson and U.~Seljak,
  1405.5857 [astro-ph.CO];
%
  R.~Flauger, J.~C.~Hill and D.~N.~Spergel,
  1405.7351 [astro-ph.CO].
  
\bibitem{Kobayashi:2014jga} 
  T.~Kobayashi and O.~Seto,
  Phys.\ Rev.\ D {\bf 89}, 103524 (2014),
  1403.5055 [astro-ph.CO];
%
  M.~S.~Sloth,
  1403.8051 [hep-ph];
%
  T.~Fujita, M.~Kawasaki and S.~Yokoyama,
  1404.0951 [astro-ph.CO].
  
\bibitem{Ma:2014vua} 
  Y.~-Z.~Ma and Y.~Wang,
  1403.4585 [astro-ph.CO];
%
  M.~Czerny, T.~Kobayashi and F.~Takahashi,
  Phys.\ Lett.\ B {\bf 735}, 176 (2014),
  1403.4589 [astro-ph.CO];
%
  K.~M.~Smith, C.~Dvorkin, L.~Boyle, N.~Turok, M.~Halpern, G.~Hinshaw and B.~Gold,
  1404.0373 [astro-ph.CO].

\bibitem{Contaldi:2014zua} 
  C.~R.~Contaldi, M.~Peloso and L.~Sorbo,
  1403.4596 [astro-ph.CO];
%
  V.~íc.~Miranda, W.~Hu and P.~Adshead,
  1403.5231 [astro-ph.CO];
%
  K.~N.~Abazajian, G.~Aslanyan, R.~Easther and L.~C.~Price,
  1403.5922 [astro-ph.CO].

\bibitem{Giusarma:2014zza} 
  E.~Giusarma, E.~Di Valentino, M.~Lattanzi, A.~Melchiorri and O.~Mena,
  1403.4852 [astro-ph.CO];
%
  J.~-F.~Zhang, Y.~-H.~Li and X.~Zhang,
  1403.7028 [astro-ph.CO];
%
  C.~Dvorkin, M.~Wyman, D.~H.~Rudd and W.~Hu,
  1403.8049 [astro-ph.CO];
%
  L.~A.~Anchordoqui, H.~Goldberg, X.~Huang and B.~J.~Vlcek,
  JCAP {\bf 1406}, 042 (2014),
  1404.1825 [hep-ph].

\bibitem{Kawasaki:2014lqa} 
  M.~Kawasaki and S.~Yokoyama,
  JCAP {\bf 1405}, 046 (2014),
  1403.5823 [astro-ph.CO].
%
  M.~Kawasaki, T.~Sekiguchi, T.~Takahashi and S.~Yokoyama,
  1404.2175 [astro-ph.CO].
  
\bibitem{Intriligator:2006dd} 
  K.~A.~Intriligator, N.~Seiberg and D.~Shih,
  JHEP {\bf 0604}, 021 (2006),
  hep-th/0602239.
  
\bibitem{Izawa:2009nz} 
  K.~-I.~Izawa, F.~Takahashi, T.~T.~Yanagida and K.~Yonekura,
  Phys.\ Rev.\ D {\bf 80}, 085017 (2009),
  0905.1764 [hep-th].

\bibitem{Giddings:1988cx} 
  S.~B.~Giddings and A.~Strominger,
  Nucl.\ Phys.\ B {\bf 307}, 854 (1988);
%
  S.~R.~Coleman,
  Nucl.\ Phys.\ B {\bf 310}, 643 (1988);
%
  G.~Gilbert,
  Nucl.\ Phys.\ B {\bf 328}, 159 (1989);
%
  T.~Banks and N.~Seiberg,
  Phys.\ Rev.\ D {\bf 83}, 084019 (2011),
  1011.5120 [hep-th].
  
\end{thebibliography}
\end{document}